\documentclass[preprint2,emulateapj,natbib,iop]{aastex}

\usepackage{epstopdf,color}
\newcommand{\so}{$\sigma$\,Orio\-nis}
\newcommand{\mjup}{M$_{\rm Jup}$}

\slugcomment{Accepted for publication in ApJ}

\shorttitle{Optical and Near-Infrared Spectra of $\sigma$\,Orionis Isolated Planetary-mass Objects}
\shortauthors{Zapatero Osorio et al.}

\begin{document}

\title{Optical and Near-Infrared Spectra of $\sigma$\,Orionis Isolated Planetary-mass Objects}

\author{M$.$ R$.$ Zapatero Osorio}
\affil{Centro de Astrobiolog\'\i a (CSIC-INTA), Crta$.$ Ajalvir km 4, E-28850 Torrej\'on de Ardoz, Madrid, Spain.}
\email{mosorio@cab.inta-csic.es}

\author{V$.$ J$.$ S$.$ B\'ejar\altaffilmark{1}}
\affil{Instituto de Astrof\'\i sica de Canarias, C/$.$ V\'\i a L\'actea s/n, E-38205 La Laguna, Tenerife, Spain.}
\email{vbejar@iac.es}

\and

\author{K$.$ Pe\~na Ram\'\i rez}
\affil{Unidad de Astronom\'ia de la Universidad de Antofagasta, Av$.$ U$.$ de Antofagasta. 02800 Antofagasta, Chile.}
\email{karla.pena@uantof.cl}

\altaffiltext{1}{Also at Universidad de La Laguna, Tenerife, Spain.}

\begin{abstract}
We have obtained low-resolution optical (0.7--0.98 $\mu$m) and near-infrared (1.11--1.34 $\mu$m and 0.8--2.5 $\mu$m) spectra of twelve isolated planetary-mass candidates ($J$\,=\,18.2--19.9 mag) of the 3-Myr \so~star cluster with a view to  determining the spectroscopic properties of very young, substellar dwarfs and assembling a complete cluster  mass function. We have classified our targets by visual comparison with high- and low-gravity standards and by measuring newly defined spectroscopic indices. We derived L0--L4.5 and M9--L2.5 using high- and low-gravity standards, respectively. Our targets reveal clear signposts of youth, thus corroborating their cluster membership and planetary masses (6--13 M$_{\rm Jup}$). These observations complete the \so~mass function by spectroscopically confirming  the planetary-mass domain to a confidence level of $\sim$75 percent. The comparison of our spectra with BT-Settl solar metallicity model atmospheres yields a temperature scale of 2350--1800 K and a low surface gravity of  log\,$g$\,$\approx$\,4.0 [cm\,s$^{-2}$], as would be expected for young planetary-mass objects. We discuss the properties of the cluster least-massive population as a function of spectral type. We have also obtained the first  optical spectrum of S\,Ori\,70, a T dwarf in the direction of \so. Our data provide reference optical and near-infrared spectra of very young L dwarfs and a mass function that may be used as templates for future studies of low-mass substellar objects and exoplanets. The extrapolation of the \so~mass function to the solar neighborhood  may indicate that isolated planetary-mass objects with temperatures of $\sim$200--300 K and masses in the interval 6--13-M$_{\rm Jup}$ may be as numerous as very low-mass stars.
\end{abstract}



\section{Introduction \label{intro}}
A robust determination of the initial mass function (distribution of objects per unit mass) represents a 
step forward in our understanding of the formation processes giving rise to stars, brown dwarfs, and
planets. There are two basic approaches: i) studying the mass function of field objects and ii) building
the mass function of star-forming regions and young star clusters. Neither option is trouble-free. The latter approach has the advantage that the age and metallicity are fixed since all cluster 
members are believed to be coeval and formed from the same molecular cloud. Additionally, assembling mass
functions of star clusters of different ages potentially answers questions concerning what young stars, brown
dwarfs, and planets look like and how they evolve. \citet{luhman12} has reviewed various surveys for low-mass
members of nearby stellar associations, open clusters, star-forming regions, and the methods used to
characterize their properties and build the corresponding mass functions. 

Of the more than 1200 spectroscopically confirmed field L and T dwarfs, only several tens of 
the L-type objects show low gravity features that are indicative of young ages: weak alkali lines, strong 
TiO and VO bands, and a triangular $H$-band pseudo-continuum resembling the properties of directly imaged
exoplanetary companions to stars  (e.g., \citealt{kirk06,cruz09,liu13,schneider14,gauza15,gizis15}). 
They also have very red near- and mid-infrared colors (typically $J-K \ge 2.0$ mag) and, as discussed by 
some groups, dimmer absolute $J$- and brighter $W2$-band magnitudes than their old field spectroscopic 
counterparts (e.g., \citealt{faherty13}) while maintaining about the same bolometric 
luminosity \citep{filippazzo15}.  \citet{faherty16} present a compilation of all known nearby 
young L dwarfs in moving groups and the field; these authors discuss their properties in 
comparison to high-gravity dwarfs of similar spectral type. The trigonometric parallaxes of very 
red L dwarfs indicate that they are indeed young substellar objects of the solar neighborhood with 
ages in the interval 10--500 Myr and masses ranging from $\sim$10 to 45 M$_{\rm Jup}$ \citep{osorio14a}. 
There have been valuable attempts to classify and order these objects according to their increasing 
surface gravity \citep{allers13,faherty13,gagne15a}. However, a full interpretation of the 
spectroscopic and photometric data and a proper understanding of the dependence of the various atmospheric 
properties on gravity require samples of numerous objects with well-constrained age, distance, 
and metallicity, desiderata that are not always attainable when investigating young field L dwarfs. The members 
of star-forming regions and open clusters are located in small regions of the sky, which makes it an 
advantage for discovering new members, but the large distances of these clusters compared to nearby 
stellar moving groups makes it more challenging to follow up the candidates with high quality 
spectroscopy. Even so, cool substellar members of young open clusters offer a unique opportunity 
to establish a proper spectral sequence and to define the attributes of young brown dwarfs
and planetary-mass objects at given ages. 

Here, we present the spectroscopic follow-up of a significant fraction of isolated planetary-mass 
candidates (i.e., objects with masses below the deuterium burning mass limit at 13 M$_{\rm Jup}$,
\citealt{saumon96}) recently discovered in the young \so~cluster by \citet{pena12}. Our goals are 
twofold: first, we address the cluster membership of our candidates and describe their photometric 
and spectroscopic properties at optical and near-infrared wavelengths for the age of the cluster; 
second, we address the mass function in the planetary-mass domain. The \so~cluster has an age 
of $\sim$3 Myr \citep{osorio02a,sherry08}, solar metallicity \citep{gonzalez08}, low internal 
extinction ($E(B-V) = 0.05$ mag, \citealt{lee68}), and is located at a distance of 385$\pm$19 pc 
(determined by \citealt{hummel13} from interferometric data and adopted by \citealt{simon15}; the 
latter authors assigned a conservative error of 5\%). More recently, \citet{schaefer16} have measured a 
precise distance of 387.5\,$\pm$\,1.3 pc to the massive triple star $\sigma$\,Ori. The cluster of 
the same name is acknowledged as one suitable choice for observational efforts seeking to establish 
the continuity of the mass function from the O-type stars to the planetary-mass regime. See 
\citet{caballero08}, \citet{lodieu09}, \citet{bejar11}, and \citet[and references therein]{pena12} 
for a further description of the \so~cluster. 

This paper is organized as follows. In Section~\ref{targets}, we define the targets. The optical
and near-infrared spectroscopic observations are described in Section~\ref{obs}. Cluster membership
based on the detection of spectral signatures of low-gravity atmospheres is analyzed in 
Section~\ref{membership}. In Section~\ref{spt}, once  young cool dwarfs were confirmed, we assigned spectral types to the
low-gravity targets by defining spectral indices and  comparing the observations to spectra 
from the literature. This classification is compared to those of the field and 
members of young stellar moving groups in Sections~\ref{spt} and~\ref{properties}, In the latter
section, we also discuss the photometric and spectroscopic properties of the very young L dwarfs. 
In Section~\ref{model}, theoretical models are compared with observations to obtain the 
effective temperature and surface gravity of our targets. We discuss the masses of the cluster 
members (according to evolutionary models) and the planetary-mass function of \so~in Section~\ref{mf}. 
Our conclusions are presented in Section~\ref{conclusions}. Finally, in Section~\ref{so70} we present 
the first optical spectrum of S\,Ori\,70, a T dwarf found in the direction of the cluster.


\section{Target selection \label{targets}}
We selected twelve  planetary-mass candidates, recently discovered by \citet{pena12}, in the \so~cluster in the extensive 
$ZYJHK$ Visible and Infrared Survey Telescope for Astronomy ({\sc vista}).
None has spectra available in the literature. All except two of the candidates 
(S\,Ori\,J05382650$-$0209257 and S\,Ori\,J0538\-0323$-$0226568) lie outside the region where \citet[see
their Figure~2]{jeffries06} reported on a second, significantly less numerous population of young dwarfs 
($\le$10 Myr, although possibly slightly older than \so) with kinematic properties that differ from 
those of the cluster. Based on the findings of \citet{jeffries06}, we estimate that about one or two objects in 
our sample, most probably among the targets mentioned above, may belong to this second population. As 
pointed out by these authors, only accurate radial velocity measurements can distinguish the two kinematic
groups. This expected low contamination will not have an impact on the results of this paper since the 
ages of the two populations are very close (it is even possible that there is considerable overlap) and are lower than 10 Myr. 

\begin{figure}        
\plotone{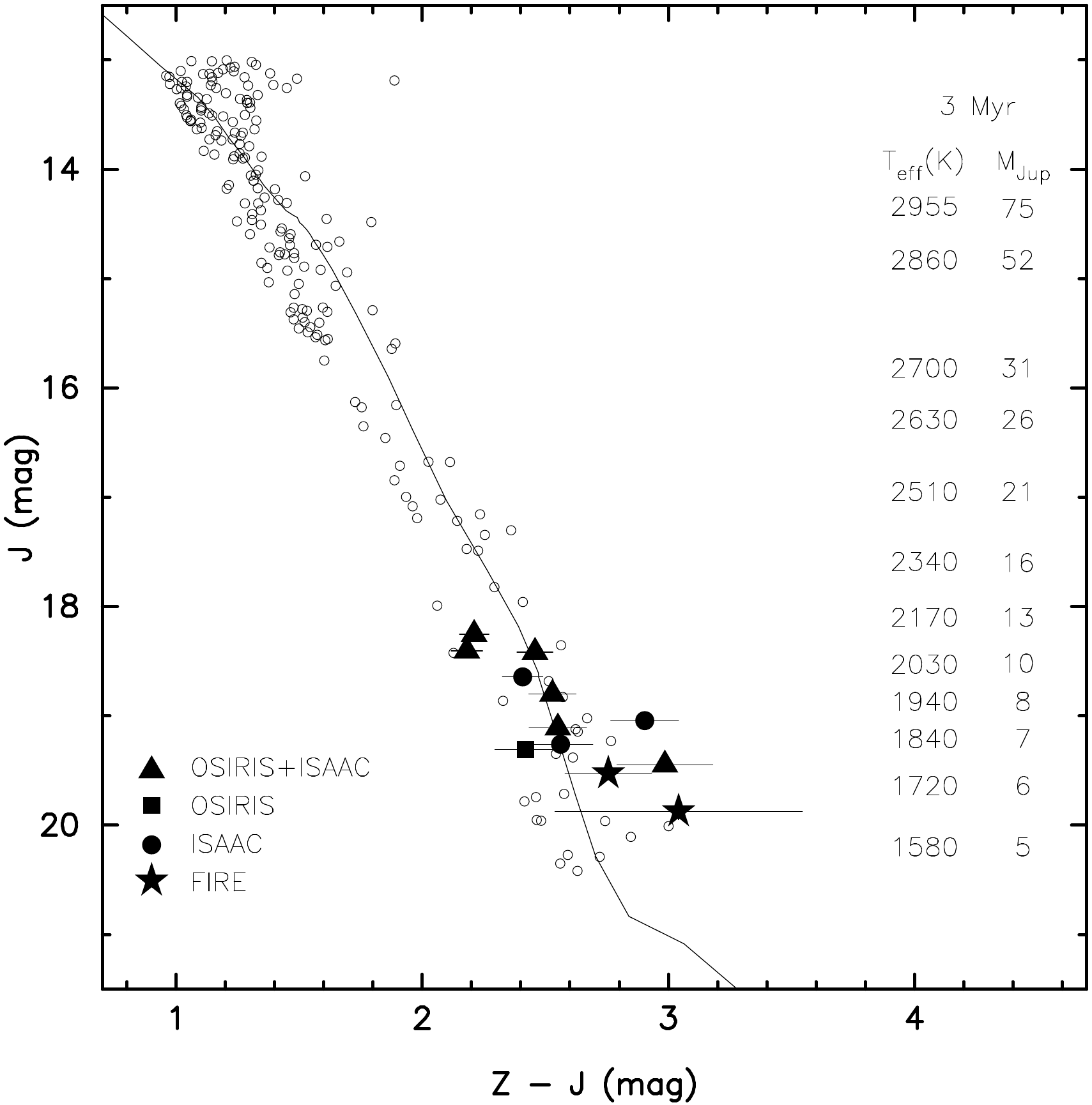}
\caption{Color--magnitude diagram of \so~low-mass star, brown dwarf, and planetary-mass 
candidates taken from \citet[open circles, {\sc vista} photometry]{pena12}. The 12 spectroscopic
targets of this paper are indicated by large solid symbols according to the observing instruments
(see the legend in the bottom left corner). The 3-Myr isochrone \citep{chabrier00} is shown 
with a solid line (conversion of theoretical luminosities and temperatures into observables
is explained in the text). Masses and $T_{\rm eff}$'s predicted by the 3-Myr model are given
in Jovian and kelvin units on the right  of the panel. For the clarity, only 
photometric error bars of the spectroscopic targets are depicted. The additional spectroscopic
target S\,Ori\,70 is not plotted because it has no $Z$-band photometry. \label{zj}}
\end{figure}

The twelve targets, along with $\sim$200 \so~low-mass candidate members, are shown in the $J$ 
versus $Z-J$ color--magnitude diagram of Figure~\ref{zj}. The vista photometry  
 can be translated to the 2MASS system \citep{skrutskie06} through color equations provided
by the Cambridge Astronomy Survey Unit\footnote{http://casu.ast.cam.ac.uk/surveys-projects/vista/technical/photometric-properties} 
(CASU). Superimposed on the photometry is the theoretical, solar metallicity isochrone 
corresponding to an age of 3 Myr \citep{chabrier00}. Predicted luminosities and effective temperatures
($T_{\rm eff}$) were transformed into {\sc vista} magnitudes and colors using  published relations between 
$J$-band bolometric corrections, observed $Z-J$ colors, measured surface temperatures, and spectral types
\citep{dahn02,golimowski04,stephens09,hewett06}. These relations are valid for temperatures between 
3300 K and $\sim$1000 K or spectral types M1--T6. We adopted the distance of 385 pc to the \so~cluster 
(see Section~\ref{intro}), which is slightly larger than the Hipparcos measurement \citep{perryman97}, 
although both distances are consistent to within 1 $\sigma$. Our program objects have $J$ and $Z-J$ colors 
in the intervals 18.2--19.9 mag and 2.2--3.1 mag, respectively. Because of their colors, and after 
comparison with the field, we expect them to display spectral types between late-M and late-L. As seen 
in Figure~\ref{zj}, they belong to the faint end of the tail of \so~candidate members known to date. If 
confirmed to be true cluster members, evolutionary models \citep{baraffe98,chabrier00,burrows03} predict 
that their surface temperatures and gravities will lie in the ranges $T_{\rm eff}$\,=\,2200--1700 K and
log\,$g$\,=\,3.5--4.0 [cm\,s$^{-2}$]; their masses would span from the deuterium burning-mass limit at 
about 13 M$_{\rm Jup}$ \citep{saumon96} down to 5--6 M$_{\rm Jup}$, i.e., within the planetary-mass domain.
Table~\ref{obslog} provides the targets' full names and their {\sc vista} $J$-band magnitudes. In what
follows, we shall use abridged names for the program objects. In the sample of \so~candidates, only two
(J05382951$-$0259591 and J05400004$-$0240331) were reported to show some mid-infrared flux excesses at
4.5 $\mu$m compatible with the presence of surrounding disks \citep{pena12}. Five targets (J05382951$-$0259591, 
J05382952$-$02\-29370, J05385751$-$0229055, J05380323$-$0226568, and J05380006$-$0247066) do not show
obvious mid-infrared flux excesses up to 4.5 $\mu$m, and the remaining five sources  have insufficient 
data in the mid-infrared to explore the existence of any flux excess. 

One additional object, S\,Ori\,70 \citep{osorio02b}, was included in the target list of optical 
observations. S\,Ori\,70 was spectroscopically classified in the near infrared as a T5.5\,$\pm$\,1.0 dwarf
($HK$-bands) and as a T6--T7 dwarf ($J$-band) by \citet{osorio02b} and \citet{burgasser04}, respectively. 
 S\,Ori\,70 was originally identified as a \so~candidate member; however, \citet{burgasser04} suggested, 
based on the similarity 
of its near-infrared spectra with the data of field dwarfs of related classification
 the possibility that  it
might instead be an old, massive field brown dwarf lying in the foreground of the \so~cluster. Using proper
motion studies and taking into account the observed colors, \citet{pena11} and \citet{pena15} indicated 
that S\,Ori\,70 is not a probable member of the \so~cluster but a foreground field  T dwarf with peculiar
colors, or a planet escaping from the Orion region. \citet{scholz08}, \citet{osorio08}, and \citet{luhman08}
noted that S\,Ori\,70 appears redder in the {\sl Spitzer} $[3.6]-[4.5]$ color than the reddest field
dwarfs of similar spectral classification, which may be owing to a low-gravity atmosphere or to the presence 
of a dusty disk. Here, we provide the first optical spectrum of S\,Ori\,70.


\section{Observations and data reduction\label{obs}}
\subsection{Optical spectroscopy}
We have carried out low-resolution spectroscopic observations at visible wavelengths of seven cluster 
planetary-mass candidates and S\,Ori\,70 using the Optical System for Imaging and low-Resolution 
Integrated Spectroscopy ({\sc osiris}) spectrograph \citep{cepa98} located at one of the Nasmyth 
foci of the Gran Telescopio de Canarias (GTC) at the Roque de los Muchachos Observatory (La Palma, Spain) 
on 2012 December 13--16 and 2014 November 25--26. {\sc osiris} consists of a mosaic of two Marconi 
CCD42-82 (2048\,$\times$\,4096 pix). All of our spectra were registered on the second detector, which 
is the default detector for long-slit spectroscopy because of its better cosmetics. We used a 2\,$\times$\,2 binning,
providing a pixel size of 0\farcs254, and the R300R grism, yielding a spectral 
coverage of 500--1200 nm and a nominal dispersion of 7.68 \AA\,pix$^{-1}$. The slit widths were 0\farcs8, 
1\farcs0, and 1\farcs23, depending on the seeing conditions. Typical seeing during the observations taken 
on 2012 December 13--16 was 0\farcs7--1\farcs0 at optical wavelengths and the weather
was mainly 
clear and spectroscopic. The seeing during the campaign on 2014 November 25--26 was 1\farcs0--1\farcs3. 
The binning of the pixels along the spectral direction and the projection of the slits onto the detector 
yielded spectral resolutions of $R$\,=\,200--310 (widest to narrowest slit) at 750 nm. An order blocking 
filter blueward of 450 nm was used; however, there exists a  second-order contribution redward of 950 nm 
(particularly important for blue objects), which was accounted for by observing spectrophotometric 
standard stars using the broad $z'$-band filter and the same spectroscopic configuration as that of the 
science targets. 

The {\sc osiris} spectra were acquired at two nodding positions along the slit separated by 8\arcsec--10\arcsec~for 
a proper subtraction of the Earth's sky emission contribution. Both the target and a reference bright star 
within typically 1\arcmin-distance from the program object were centered on the slit. Since our targets are very 
faint sources at optical wavelengths ($Z \ge 20.5$ mag) and the trace of their spectra is barely seen 
in individual exposures, we used the observations of the reference stars to  register precisely the 
two-dimensional spectra by computing relative object shifts. The registered frames were stacked together 
before optimally extracting the spectra of the targets. Data were not obtained at parallactic angle. Our
program objects were acquired on the slit using the $z'$ filter with a passband of 825--1000 nm, i.e., we
ensured that most of the red flux passed through the slit. In addition, the objects do not contribute significantly 
at blue wavelengths ($\le$650 nm), where the flux loss due Earth's atmospheric refraction would be more
noticeable. By observing the reference bright stars once at parallactic angle immediately after the targets 
we checked that no corrections due to blue-photon losses are required for our program sample. In 
Table~\ref{obslog}, we provide the journal of the {\sc osiris} observations, which include Universal Time 
(UT) observing dates, gratings, exposure times, slit widths, and air masses. Typical exposure times of single 
frames ranged from 1200 to 1800 s. The nodding observations were repeated several times to yield total 
on-source integrations of between 1 h and 4 h. 

Raw images were reduced with standard procedures including bias subtraction and flat-fielding within
{\sc iraf.\footnote{{\sc iraf} is distributed by the National Optical Astronomy Observatories, which are 
operated by the Association of Universities for Research in Astronomy, Inc., under cooperative agreement 
with the National Science Foundation.}} A full wavelength solution from calibration lamps taken during the
observing nights was applied to the spectra. The error associated with the fifth-order Legendre polynomial
fit to the wavelength calibration is typically 10\%~the nominal dispersion. We corrected the extracted spectra 
for instrumental response using data of the spectrophotometric standard star G\,158$-$100 and G\,191-B2B 
(white dwarfs) observed at parallactic angle on the same nights and with the same instrumental configuration 
as our targets. These standard stars have fluxes 
over the range $\sim$330--1000 nm in 
\citet{filippenko84} and \citet{massey90}. To complete the data reduction, target spectra were divided by the 
standard stars to remove the contribution of telluric absorption; the intrinsic features of the white dwarfs
were previously interpolated and their spectra were normalized to the continuum before using them for division 
into the science data. The spectrophotometric standard stars were observed a few hours before or after the
targets; therefore, some telluric residuals may be present in the corrected spectra, particularly the strong 
O$_2$ band at 760.5 nm.

The resulting reduced {\sc osiris} spectra of \so~candidates are depicted in Figure~\ref{opt}. They 
are ordered by increasing $J$-band magnitude and shifted by a constant for clarity. The top panel of 
Figure~\ref{opt} displays the wavelength range 700--980 nm because at shorter and longer wavelengths the
data have very poor signal-to-noise (S/N) ratio, and the instrumental response correction and wavelength 
calibration are not reliable at values beyond $\sim$1 $\mu$m. There was no signal detected below $\sim$800
nm from the T dwarf S\,Ori\,70. However, it is a very red source at optical wavelengths and the signal at around 
1 $\mu$m is still significant. Therefore, we illustrate the {\sc osiris} spectrum of S\,Ori\,70 between 0.78 
and 1.02 $\mu$m in the bottom panel of Figure~\ref{opt}.

Together with the science targets, we also observed the spectroscopic standard field dwarf 
2MASS\,J10584787$-$1548172 \citep{delfosse97}, which has an  L3 spectral type and a high-gravity atmosphere. 
The optical observations of the standard were taken at parallactic angle and were reduced and calibrated
in the same manner as the \so~candidates. We provide the journal of these observations at the bottom of 
Table~\ref{obslog}. The {\sc osiris} spectrum of 2MASS\,J10584787$-$1548172 was used for the spectral 
classification of the targets and the definition of the spectral properties of the young \so~objects in 
following Sections. The same standard was also observed at near-infrared wavelengths (see the following
subsection).

\subsection{Near-infrared spectroscopy}
\subsubsection{{\sc isaac} data}
$J$-band near-infrared spectroscopy was obtained for nine \so~planetary-mass candidates (six of
them in common with the optical targets) using the {\sc isaac} instrument that \citep{moorwood98} installed 
on the Nasmyth A focus of the third unit of the Very Large Telescope (VLT) array at Cerro Paranal 
(Chile). {\sc isaac} is an imager and spectrograph that covers the wavelength interval 1--5 $\mu$m. For our 
observations we used the short wavelength arm that is equipped with a 1024\,$\times$\,1024 Hawaii Rockwell 
detector with a pixel size of 0\farcs147, covering 1--2.5 $\mu$m. All of our data were obtained in the 
low-resolution mode with a slit width of 1\farcs0, centered at 1.25 $\mu$m. This instrumental configuration 
yielded a nominal spectral  dispersion of 3.49 \AA\,pix$^{-1}$, a resolving power of about 500 at the central 
frequency, and a wavelength coverage of 1.09--1.42 $\mu$m. Because of the faint nature of our targets and the 
fact that Earth's atmospheric water vapor absorption is very strong redward of 1.34 $\mu$m, the useful wavelength 
coverage is 1.11--1.34 $\mu$m. All of the {\sc isaac} observations were collected with a typical seeing of
0\farcs7--1\farcs0 on 2012 December 1--4 and 2013 January 26--27. 

In the near-infrared we employed the same observing strategy as in the optical. Both the targets and their 
bright reference stars were acquired through the $J$-band filter and simultaneously aligned in the 
120\farcs0-length slit. Individual 600 s exposures were obtained with the sources at two nod positions 
separated by typically 10\farcs0. All targets were observed in an ABBA nodding pattern giving total on-source
integration times of 1.3--3.3 h. In Table~\ref{obslog} we provide UT observing dates, exposure times, and air 
masses. To account for absorption by the Earth's atmosphere, calibration stars of warm spectral types B were 
observed immediately after the targets as close as possible to the same air mass, typically within
$\pm$0.1 air masses. Reduction of the raw data was accomplished using {\sc iraf} and in a manner similar
to that of the optical wavelengths. Pairs of nodded frames were subtracted to remove the background emission
contribution and then divided by the corresponding flat field. For each target, individual frames were
registered using the bright reference stars and stacked together to produce deeper data. The spectra were optimally
extracted and wavelength-calibrated using the terrestrial sky emission lines to a precision of about 
10--20\%~of the nominal dispersion. After removal of the intrinsic features (typically hydrogen lines) of 
the B-type stars, the calibration spectra were divided into the corresponding target spectra to remove
telluric absorptions and instrumental spectral response. Finally, the data were multiplied by a black body
curve of 32000 K to restore the spectral slope. 

Figure~\ref{nir} illustrates the final {\sc isaac} data ordered by increasing spectral type
(see Section~\ref{spt}). Together with the science targets there are three field dwarfs that were observed
with the same instrumental setup. They are classified in the optical and near-infrared as 
L1 (2MASS\,J10170754\-$+$1308398, \citealt{cruz03}), L3 (2MASS\,J10584787\-$-$1548172, \citealt{delfosse97}),
and L7.5 (2MASS J04234858$-$0414035, \citealt{geballe02}). In Table~\ref{obslog} (bottom), we
provide the journal of the observations of the spectroscopic standards, whose raw spectra were reduced 
in the same manner as the main program objects. The final data are depicted in Figure~\ref{nir} and 
will be employed for comparison with the \so~candidates.

\subsubsection{{\sc fire} data}
Two of the faintest targets were observed with the Folded-port Infrared Echellette ({\sc fire}) 
spectrometer \citep{simcoe08,simcoe10} on the 6.5-m Magellan telescope (Las Campanas Observatory, 
Chile) on 2014 December 01. This is a single object, prism cross-dispersed infrared instrument that 
covers the full 0.8--2.5 $\mu$m band in one shot at a spectral resolution of 45--60 \AA~for the long slit
mode and nominal slit width of 0\farcs6. We selected the high-throughput prism mode because it is more efficient
given the faint nature of the program objects and the long slit length of 40\arcsec, which is appropriate 
for the observing strategy previously described. The pixel size of the 2048\,$\times$\,2048 pix HAWAII-2RG
detector was 0\farcs15. Table~\ref{obslog} provides the journal of the {\sc fire} observations. Individual 
exposure times were 180 and 300 s and the nodding offset of the ABBA observing pattern was 3\arcsec, which
was convenient given the 0\farcs6 seeing conditions during the observations. Data reduction was carried out 
in a manner similar to that for the {\sc isaac} and {\sc osiris} spectra. Wavelength solution with a typical
precision of $\pm$1.4 \AA~was performed using arcs of internal Ne and Ar calibration lamps. To correct for
the telluric absorption we observed A-type stars, one for each target. 

The final {\sc fire} spectra are shown in Figure~\ref{fire}. To improve the S/N ratio of the data, we 
smoothed the spectra with a boxcar of 9 pixels (J054037382$-$0240011) and 11 pixels (J05401734\-$-$0236266). 
Both the original and smoothed data are illustrated in Figure~\ref{fire}.  This will be useful for analyzing 
the $YJHK$ pseudo-continuum for an spectral type assignment.


\section{Cluster membership \label{membership}}
The membership of our targets in \so~was addressed in terms of the red and late-type nature of the optical
and near-infrared spectra, and the evidence of low-gravity atmospheres using the near-infrared data.
In what follows, S\,Ori\,70 is not included because its cluster membership is clearly dubious (see 
\citealt{burgasser04,pena15}). It will be discussed separately in the Appendix. The detection of water 
vapor absorption in the near-infrared spectra of Figures~\ref{nir} and~\ref{fire} provides robust confirmation
of the cool nature of the candidates. Additionally, the rising slope of the optical spectra, together with
the detection of the TiO and VO absorption bands shown in Figure~\ref{opt}, denotes low temperature atmospheres.
Therefore, we concluded that all program sources are Galactic objects in origin (non-extragalactic sources), 
and that they most likely have cool temperatures.

Before making any attempt to determine spectral types, we studied whether the near-infrared spectra show signs 
of low gravity atmospheres, which in turn would lead to cluster membership confirmation. From the {\sc fire} 
data of Figure~\ref{fire}, it becomes apparent that the $H$-band pseudo-continua of J054037382$-$0240011 and
J05401734$-$0236266 have a peaked shape, in clear contrast to the $H$-band plateau spectrum of a high-gravity 
dwarf of similar energy distribution also shown in the Figure. This feature, where the $H$-band is dominated 
by strong water absorption bands to either side of the sharp peak located between 1.68 and 1.70 $\mu$m, was 
first reported  by \citet{lucas01} for very young ($\sim$1 Myr), cool Trapezium objects. It was ascribed to 
the effects of low pressure and low gravity atmospheres typical of self-gravitational collapsing objects with
$T_{\rm eff} \le 2700$ K. The triangle-shaped profile of the $H$-band has been accepted as a consistent 
signature of youth ever since (e.g.,  \citealt{allers07,rice10,allers13,liu13,schneider14,gagne14a}). It 
is also present in some 120-Myr Pleiades L dwarfs \citep{bihain10,osorio14b} and appears to discriminate 
low gravity atmospheres up to ages of about 300--500 Myr \citep{cruz09,osorio14b,gauza15}.  Nevertheless, 
it is debated whether this feature is also present in peculiar-metallicity dwarfs (see discussion in 
\citealt{looper08}, \citealt{allers13}, and \citealt{marocco14a}).

To further constrain the surface gravity, and in turn the age, of the {\sc fire} spectra, we used the 
spectroscopic $H_2(K)$ index defined by \citet{canty13}, which measures the flux ratio and the spectral 
slope between 2.15 and 2.24 $\mu$m in the $K$-band due to the absorption of the continuum by H$_2$. 
According to those authors, this index is more gravity-sensitive than other features, such as the strength of
the 2.21 $\mu$m Na\,{\sc i} doublet (also in the $K$-band) and the triangular $H$-band peak. For very young 
objects ($\le$10 Myr) the index returns a value $\le$1. For older objects, which have a negative slope 
between 2.15 and 2.24 $\mu$m, the index is $>$1. We measured $H_2(K) = 0.93$ for J054037382$-$0240011, 
which is compatible with a very young age. As for J05401734$-$0236266, the index returned a value of $H_2(K) 
\approx 1$; this object has a poorer S/N ratio spectrum making the index determination less reliable. 
Nevertheless, both objects display very similar {\sc fire} data and share related spectroscopic slopes, 
including the peaked $H$-band pseudo-continuum. The criteria of \citet{canty13} added additional proof of the low
gravity nature of the {\sc fire} spectra. 

We also investigated other gravity-sensitive features in the near-infrared spectra, in particular the
strength of the K\,{\sc i} absorption lines at 1.169, 1.177, 1.243, and 1.252 $\mu$m. The four 
near-infrared K\,{\sc i} lines produced by high-gravity atmospheres are clearly registered at the resolution of
our data, as  is demonstrated by the {\sc isaac} spectra of the field L1--L7.5 dwarfs shown in 
Figure~\ref{nir}. It is well known that these lines are linked to both temperature and gravity dependences 
\citep{mclean03,mclean07,mcgovern04}. Low gravity (or low pressure) atmospheres are characterized by
$J$-band spectra with alkali line strengths effectively weaker than ``normal''. From  visual inspection 
of the panels depicted in Figure~\ref{nir}, it becomes apparent that the \so~candidates typically show 
less intense K\,{\sc i} lines  compared to the field dwarfs of similar energy distribution and colors.
We interpreted this property as a sign of youth. 

To test the gravity indication from the strength and depth of the alkali lines we measured the
pseudo-equivalent width of the neutral K\,{\sc i} at 1.252 $\mu$m for the 11 candidates with
{\sc isaac} and {\sc fire} spectra. We chose the near-infrared K\,{\sc i} line with the longest 
wavelength to avoid contamination by FeH and other molecular bands; additionally, this line is less 
affected by telluric absorption and lies in a spectral region with relatively better S/N ratio. 
Pseudo-equivalent widths\footnote{Given the cool nature of our sample, equivalent widths are generally 
measured relative to the observed local pseudo-continuum formed by molecular absorptions. We refer to 
these equivalent widths as ``pseudo-equivalent widths''.} (pEWs) were obtained via direct integration of
the line profile with the {\sc splot} task in {\sc iraf}. The results of our measurements, given in 
Table~\ref{pew}, were extracted by adopting the base of the line as the pseudo-continuum. The error 
bars were determined after integrating over a reasonable range of possible pseudo-continua. With the 
exception of two candidates, J05382951$-$0259591 and J05382952$-$0229370, the K\,{\sc i} 
$\lambda$1.252-$\mu$m line is not obviously detected and we imposed upper limits on the line pEWs by 
measuring the strength of other spectroscopic features seen at nearby wavelengths. Our K\,{\sc i} pEWs 
agree with those measured for Taurus ($\sim$1 Myr) and Upper Scorpius ($\sim$5--10 Myr) members by 
\citet{bonnefoy14a} within 1-$\sigma$ the quoted uncertainties, and typically lie near and below the
boundary for objects to be assigned the highest score, indicating low gravity according to Figure~23 of  
\citet{allers13}. The K\,{\sc i}$_J$ index defined by these authors, which relies on the intensity of 
the $J$-band K\,{\sc i} lines, is also useful in determining the gravity score of our data. We measured 
K\,{\sc i}$_J$ values in the interval 0.87--1.11 for the {\sc isaac} and {\sc fire} spectra of
\so~candidates, from which we assigned very low (VL-G) and intermediate (INT-G) gravity flags to seven 
and four candidates, respectively. The resulting flags, or ``gravity classes'' according to the 
terminology of \citet{allers13}, are given in Table~\ref{pew}.

In Table~\ref{pew}, we also include the confirmed \so~member S\,Ori\,51 \citep{osorio00}, whose $J$-band
spectrum was published by \citet{mcgovern04}. These data have a spectral resolution of $R$\,$\sim$\,2000, 
four times higher than the {\sc isaac} data. We determined the K\,{\sc i} $\lambda$1.252-$\mu$m pEW to be 
2.0\,$\pm$\,0.5 \AA~from the spectrum of \citet{mcgovern04}, thus confirming the weak K\,{\sc i} lines,
and that good S/N, moderate-to-high resolution spectra are required to study the fine details of the 
alkali absorptions of low gravity atmospheres. S\,Ori\,51 ($J$ = 17.19 mag) was originally classified
as an M9$\pm$0.5 young dwarf at optical wavelengths \citep{barrado01}; using near-infrared spectra; its 
classification was recently revised to L0$\pm$0.5 by \citet{canty13}. The \so~candidates of this paper 
(Table~\ref{obslog}) are more than 1 mag fainter in the $J$-band; we thus expect them to have similar 
and later spectral types to S\,Ori\,51, as suggested by the redder near- and mid-infrared colors of 
our candidates (see Table~\ref{pew}). 

Figure~\ref{pewk} illustrates the K\,{\sc i} $\lambda$1.252-$\mu$m pEWs as a function of the $J-K$ and 
$K-[4.5]$ colors. These colors have a marked pattern with decreasing $T_{\rm eff}$ or cooler spectral
type \citep[and references therein]{leggett10a}, thus leaving the pEWs as the tracers of surface gravity
in the two panels of Figure~\ref{pewk}. Significant mid-infrared flux excesses may contribute to blur the
diagrams. However, as indicated in Section~\ref{targets}, only two \so~targets show rather moderate 
(even marginal) mid-infrared flux excesses, one of which (J05382951$-$0259591) has a near-infrared 
spectrum and is included in Figure~\ref{pewk} (the other one, J05400004$-$0240331, has  only an optical
spectrum). To put our K\,{\sc i} pEW measurements into context, we also plotted the pEWs obtained for 
field M, L, and T-type dwarfs reported in the literature \citep{mclean03,allers13,lodieu15}. The required
photometry of the field dwarfs was collected from the above-mentioned spectroscopic papers and \citet{patten06}, 
or retrieved from the {\sl WISE} public archive\footnote{We exchanged $[4.5]$ and $W2$ by assuming that
the $W2$ magnitude is similar to the {\sl Spitzer} $[4.5]$ data. This is a reasonable assumption for the
L dwarfs: the widths of the {\sl WISE} $W2$ and {\sl Spitzer} $[4.5]$ bands are similar and, although the
{\sl WISE} filter is slightly shifted to redder wavelegnths, there is no strong molecular absorption at 
these frequencies in the spectra of M and L dwarfs. Furthermore, \citet{wright11} also noted that the 
color term between {\sl Spitzer} $[4.5]$ and {\sl WISE} $W2$ is rather small ($[4.5]-W2 = 0.054$ mag)
and seems to have no trend with spectral type, even for the T dwarfs.}. Different spectral types and field 
dwarfs with spectroscopic evidence of low- and intermediate-gravity atmospheres
\citep{rebolo98,cruz03,faherty13,allers13,gauza15} are distinguished in the panels of Figure~\ref{pewk}. 
Although there is considerable scatter within the field objects, general trends can be seen in the figure. 
The K\,{\sc i} $\lambda$1.252-$\mu$m line strengthens from the M through the L types (i.e., with redder 
colors) and declines toward the coolest T dwarfs. Interestingly, the \so~candidates (including S\,Ori\,51) 
have pEWs and pEW upper limits that systematically lie below the values of high-gravity field dwarfs of
similar colors. They occupy regions populated by low-gravity objects, thus confirming a young age for all 
{\sc isaac} and {\sc fire} targets, except for J05385751$-$0229055, whose high value of pEW upper limit
(poor S/N ratio at around 1.252 $\mu$m) prevented us from unambiguously establishing its low-gravity 
atmosphere. Inspection of the other three K\,{\sc i} lines and the FeH absorption band at 1.20 $\mu$m 
(see Section~\ref{properties}) in the spectrum of this particular source yielded a similar result. 
Unfortunately, this object does not have an optical spectrum in our work for a further assessment of
its true atmospheric nature. Interestingly, the K\,{\sc i}$_J$ gravity-sensitive index of \citet{allers13}
yielded a value of 0.97 for J05385751$-$0229055, as expected for a very low gravity dwarf. 

From the above discussion, we conclude that 10 out of 11 \so~candidates with {\sc isaac} and {\sc fire} 
near-infrared spectra show evidence of low-gravity atmospheres reinforcing their membership of the cluster. 
A better quality near-infrared spectrum would be required for the remaining candidate before reaching any 
solid conclusion on its cluster membership. Additional membership criteria, such as accurate proper motion and 
radial velocity measurements, would be useful for strong cluster membership confirmation. Given the faint
nature of our targets, these measurements await the arrival of the next generation of space- and ground-based telescopes. 
The properties of the optical spectra are discussed in the context of low gravities and spectral types 
in Sections~\ref{spt} and~\ref{properties}. 

\section{Spectral type determination \label{spt}}
We assigned spectral types by visual comparison of the observed spectra with data from the 
literature and by measuring spectral indices previously defined and newly defined in this paper. 
As reference data, we employed spectral standards with high and low-to-intermediate gravity atmospheres 
from \citet{kirk10}, \citet{bonnefoy14a}, \citet{tinney98}, \citet{kirk99}, \citet{leggett01,leggett02}, 
\citet{reid01}, \citet{bejar08}, \citet{lodieu08}, \citet{lafreniere10}, and \citet{gagne14c}. Our
objective is to provide a classification based on the global morphology of the data. In this regard, 
the pseudo-continuum slope, rather than the individual atomic and molecular features, plays a key role 
in defining the spectral sequence. Because there is a classification scheme elaborated particularly  
for low-to-intermediate gravity L dwarfs (\citealt{allers13}; see also \citealt{faherty16} and references 
therein), we provided spectral types for the \so~candidates based on the comparison with the so-called 
young, low-to-intermediate gravity L dwarfs and field high-gravity dwarfs separately. The results are 
summarized in Table~\ref{spectypes}.

\subsection{Classification from visual comparison}
We compared all the spectra with one another to infer a relative spectral sequence within our sample and
concluded that J05403782$-$0240011, J05401734$-$0236226, and J05380006$-$0247066 show the steepest 
pseudo-continuum slopes, consistent with their being the latest-type objects in the target list. The 
three sources are also the faintest objects in the sample ($J$\,=\,19.5--19.9 mag). The dwarfs 
J05382650$-$0209257 and J05382471$-$0300283 appear to be the earliest-type sources with less steep 
spectral slopes and $J$-band magnitudes $\simeq$1.5 mag brighter. Any spectral type derivation obtained 
from comparison with reference spectra and/or spectral indices  should reflect this feature.

The spectral energy distribution of our targets does not differ significantly from those of normal ultra-cool 
field dwarfs with high- and low-gravity atmospheres (see also Section~\ref{properties}), thus allowing us a
direct analogy with the global spectral energy distributions. \citet{allers13} discussed, for the L 
types, whether there are spectral intervals where the pseudo-continuum shape is sensitive to spectral morphology 
with little dependence on gravity. Specifically, the regions are 1.07--1.40 $\mu$m ($J$-band) and 
 1.90--2.20 $\mu$m ($K$-band). We qualitatively determined which standard(s) provided the best matches to the 
 pseudo-continuum shape of our targets. From visual inspection of the spectral slopes, we concluded that the
 \so~sample has spectral types between $\sim$M9 and mid-L when compared to high-gravity standards, and late-M
 and early-L when matched to low-to-intermediate gravity dwarfs. As an example, the two {\sc fire} spectra 
 are classified as later than the young L1 member of the TW Hydrae object \citep{gagne14c} but as similar to, or
 slightly earlier than, the young L3 G\,196-3B because both {\sc fire} spectra display a pseudo-continuum 
 slope from blue to red wavelengths that is intermediate between that of the young L1 and L3 objects 
 (Figures~\ref{fire} and~\ref{fire2}). At the same time, the global energy distribution of the {\sc fire}
 data is reasonably reproduced (except for low-gravity features) by the overall shape of an L4.5 high-gravity 
 dwarf (DENIS\,J122815.2$-$154733, \citealt{leggett01}; shown in Figure~\ref{fire2}).

\subsection{Spectral indices}
For the six \so~candidates with optical and near-infrared data, both spectra were joined by using the
$Z-J$ colors (Table~\ref{pew}) published in \citet{pena12} and the zero-point fluxes and filter transmission 
curves of \citet{hewett06}. The resulting spectral energy distributions from 0.7 $\mu$m through 1.34 $\mu$m
are illustrated in Figure~\ref{optnir}. These combined data and the {\sc fire} spectra were the easiest to
classify since they cover a significant wavelength interval; more uncertain is establishing the 
typology of the four
objects with {\sc isaac} or {\sc osiris} spectra only. To aid and quantify the visual comparison exercise
we defined three spectral indices that measure the slope between the optical and the $J$-band and are
sensitive to the spectral type. They were useful for establishing an order among the spectra. The 
$S1$, $S2$, and $S3$ indices yield the flux ratios between 1.28--1.32 $\mu$m and various bluer wavelength intervals
that have relatively good quality in our {\sc osiris} and {\sc isaac} data as follows:
\begin{eqnarray}
S1 &=& \frac{f_{1.28-1.32}}{f_{0.8015-0.8415}}, \\
S2 &=& \frac{f_{1.28-1.32}}{f_{0.88-0.92}}, \\
S3 &= &\frac{f_{1.28-1.32}}{f_{1.175-1.215}}, 
\end{eqnarray}
where $f_{\lambda_1-\lambda_2}$ stands for relative mean fluxes in the intervals $\lambda_1-\lambda_2$ of
the same length, and $\lambda$ is given in $\mu$m. These indices were first determined for the high and 
low-to-intermediate gravity spectral standards given in \citet{kirk10} and the aforementioned references.
The behavior of these indices is not always linear with spectral type. In Table~\ref{coef} we provide 
the coefficients of the polynomial fits to the indices of the spectral standards according to the following expression:
\begin{eqnarray}
{\rm SpT} = c_0 + c_1\,S + c_2\,S^2 + c_3\,S^3 \label{eqspt}
\end{eqnarray}
where SpT values of $-1$, 0, 1, ... 7 correspond to spectral types M9, L0, L1, ... L7, respectively.
In Table~\ref{coef}, we also provide the validity regimes and the root mean square (rms) of the fits for 
each spectral index and for the high-gravity (``high-g'') and low-to-intermediate gravity (``young'') cases.

We used equation~\ref{eqspt} to determine  quantitatively the spectral types of the \so~candidates to within 
1--2 subtypes. The measured $S$ indices over the spectra of the \so~candidates are given in Table~\ref{spectypes}. 
Also provided in Table~\ref{spectypes} are the derived spectral types in the ``SpT young optNIR'' and
``SpT high-g optNIR'' columns; these were computed as the averages of all three measurements (when they existed) 
that include the optical and near-infrared spectra. The numbers were rounded to the closest integer or half 
subtype. With the $S1$, $S2$ and $S3$ indices we ensured that the optical-to-near-infrared and near-infrared
slopes of the \so~candidates werere well reproduced by the spectral standards, which agrees with the basis of our 
type determination approach. The least massive population of \so~known to date with magnitudes spanning the range 
$J \approx 17.2-19.9$ mag have a sequence of increasing spectral types between $\sim$L0 and $\sim$L4.5 
and between $\sim$M9 and $\sim$L2.5, according to the schemes based on high-gravity and low-to-intermediate gravity.

We also explored other near-infrared indices available in the literature based on the strength of H$_2$O 
absorption (e.g., \citealt{geballe02,gorlova03,slesnick04,allers13}). This signature shows an almost linear
behavior with spectral type from M6 through T6 as discussed by \citet{mclean03}. We found results consistent 
with our spectral type determinations to within $\pm$2.5 subtypes, except for J05370549$-$0251290, where the
typings differ by 4.5 subtypes (\citealt{geballe02} and \citealt{gorlova03} H$_2$O indices at 1.33 $\mu$m 
indicated $\ge$L6 versus the adopted L1.5). This relatively large difference is probably caused by large 
error bars associated with the indices of our data.

At optical wavelengths, the PC3 index described by \citet{martin99}, which measures the pseudo-continuum 
slope between two 40 \AA~bands centered at 0.825 and 0.756 $\mu$m, yielded a narrow spectral classification. 
The PC3 indices and the determined optical spectral types (``SpT opt'') are provided in Table~\ref{spectypes}. 
According to \citet{martin99}, the error associated with the spectral classification using the PC3 index is
typically $\pm$0.5 subtypes. All {\sc osiris} spectra share a similar typing of L1.5--L2.5, suggesting 
that the optical data are quite similar, despite covering nearly the same $J$-band magnitude interval as the
near-infrared spectra. This was already anticipated by the comparison of the optical data with the spectrum
of USco\,108\,B, a young Upper Scorpius member classified as an M9.5 source \citep{bejar08}, in 
Figure~\ref{opt}. Using the PC3 index, these authors determined a spectral type L1 for USco\,108\,B, in
better agreement with our derivations. The similarities between our data and USco\,108\,B are remarkable; 
they also provide additional proof of the low-gravity nature of the \so~candidates. 

In general, we found good agreement to within the quoted uncertainties between the  spectral types based on visual
observations, 
high-gravity and ``youth'' for the earliest sources. However, the high-gravity-based 
``optNIR'' spectral types are systematically later than those based on the \citet{allers13} system. 
This becomes apparent for the coolest types, where differences of up to 2.5 subtypes can be found. Also 
for the latest types, the disagreement between the optical-based and the high-gravity-based ``optNIR'' 
classification is 
noteworthy: J05380006$-$0247066 is classified as an L1.5 from the optical PC3 index, and 
as an L4.0 from the $S$ indices and the visual comparison to field data. Interestingly, the PC3-based typing 
and the ``youth''-based classification agree to within the error bars. Similar disagreements between optical
and near-infrared classifications are observed in other young cool dwarfs of the Upper Sco association 
(Lodieu 2016, priv$.$ communication), and disagreement was also noted for other \so~candidates by \citet{martin01}.
Additionally, \citet{allers13} found that the $J$-band classification of 10-Myr old cool dwarfs yields 
near-infrared types that are on average 1.3 subtypes later than their corresponding optical types. It is 
also possible that for late-L types, the PC3 index delivers optical spectral types a few subtypes earlier
than other classifying methods (see \citealt{martin06}).

We note that using TiO- and VO-based indices at optical wavelengths would produce spectral types close
to M9--M9.5 for all of our targets, i.e., earlier types than those derived from the slopes of the spectra.
We expect cluster members (including S\,Ori\,51) that cover $\approx$2.5 mag in the $J$-band and
differ by about 0.8 and 0.4 mag in the $Z-J$ and $J-K$ colors to have different subtypes and to form a coherent 
sequence of objects. As an example, this magnitude interval and relative color coverages  account for 
three to seven different M and L subtypes in the field. We ascribe the optical and near-infrared typing
mismatch to the low gravity nature  (and therefore low pressure atmospheres) of these \so~objects. TiO and VO 
are well-known gravity-sensitive features (they have a strong opacity at low pressures). The TiO- and 
VO-based spectral indices in the literature are anchored on high-gravity field dwarfs. For our sample of
young dwarfs that have the same metallicity and age we prefer to define a relative spectral classification 
that relies on the general slope of the observed spectra, which, as will be discussed in
Sections~\ref{properties} and~\ref{model}, correlate with other properties of the \so~sequence of members.

In what follows we adopt the spectral types labeled with the flag ``optNIR'' in Table~\ref{spectypes}. 
The associated uncertainties were obtained as the scatter of the various measurements. We prefer
the combined optical and near-infrared types over the optical-only types because, as indicated by
\citet{allers13}, the $J$- and $K$-band pseudo-continua show little dependence on surface gravity and
because the energy distribution of L-type objects peaks in the near-infrared. The adopted classification 
(both ``youth-'' and high-gravity-based determinations) correlates with the increasingly red colors 
(e.g., $Z-J$ from $\sim$2.2 to $\sim$3.0 mag, see also Section~\ref{properties}) and the decreasing
$T_{\rm eff}$, as determined from  comparison with model atmospheres (Section~\ref{model}). For
J05400004$-$0240331, which has an optical spectrum only, we have adopted an intermediate classification
with an error bar that covers the full range of spectral types for this work. We are confident of
this procedure because the optical data and the near-infrared colors of this object are similar 
to those of the rest of the target sample.

\section{Properties of \so~L dwarfs \label{properties}}
\subsection{Optical and near-infrared spectra \label{optnirall}}
Our targets reveal spectral types in the narrow interval L0--L4.5 (near-infrared, high-gravity-based),
M9--L2.5 (near-infrared, low-gravity-based), and L1.5--L2.5 (optical), where only a smooth behavior and
no major changes in the spectral features are expected \citep{kirk10}. To improve the S/N of the data,
we produced ``master'' spectra by combining all of the L1.5--L2.5 {\sc osiris} (optical) and
L0--L5 {\sc isaac} ($J$-band) data for confirmed \so~members. The noisy {\sc isaac} spectrum of 
J05370549$-$0251290 was not considered for producing the master near-infrared spectrum. We checked
the feasibility of this procedure by merging groups of three to four spectra ramdomly; we always
obtained similar outcomes from all combinations. The resulting master spectra are depicted in 
Figure~\ref{combspec}; we used them to establish the spectroscopic properties of solar metallicity 
L dwarfs with an age of $\sim$3 Myr, which corresponds to a surface gravity of log\,$g$ =~3.5--4.0 
(cm\,s$^{-2}$) according to various substellar evolutionary models \citep{baraffe98,chabrier00,burrows03}. 
Figure~\ref{combspec} also shows the spectra of field high-gravity dwarfs with spectral resolution 
similar to our data for a proper comparison (see Figure caption). The field standards are classified
as M9, L1 and L3 dwarfs, the last two types roughly correspond to the mean spectral type of the 
\so~master spectra. Note that the L3 standard (2MASS\,J10584787$-$1548172) is the same object for 
the optical and near-infrared panels of Figure~\ref{combspec} and was observed with the same 
instrumental configurations as the \so~targets. 

The improved S/N of the combined \so~near-infrared spectrum shown in the bottom panel of
Figure~\ref{combspec} allows us to confirm the presence of weak absorptions due to the Na\,{\sc i} 
doublet at $\sim$1.14 $\mu$m and the FeH bands at 1.20--1.24 $\mu$m. All the K\,{\sc i} lines are barely
seen with pEWs of 2.5 \AA~(1.169 $\mu$m), 3.7 \AA~(1.177 $\mu$m), 2.2 \AA~(1.243 $\mu$m), and
3.3 \AA~(1.252 $\mu$m), which have an associated uncertainty of $\pm$0.5 \AA. These pEWs should be 
interpreted as the intensity of the $J$-band K\,{\sc i} lines of low-gravity early-L dwarfs with an 
age of 3 Myr. They can be used as a reference for dating young objects of similar classification 
and metallicity in the field. Higher pEWs would indicate older ages. 

As seen from Table~\ref{pew}, a few \so~L0 dwarfs have smaller values and lower upper limits on the 
pEW of the reddest K\,{\sc i} line. This is explained by the fact that for a given surface gravity, 
the alkaline absorptions increase in strength with decreasing temperature and spectral type. The master 
spectrum depicted in the bottom panel of Figure~\ref{combspec} covers from L0 through mid-L. In any case, 
the so-obtained ``average'' pEW of the 1.252-$\mu$m line is smaller than the corresponding pEWs of 
high-gravity L0--L5 dwarfs (Figure~\ref{pewk}); this provides an additional test to confirm that the 
majority of our candidates are low-gravity sources (otherwise, the combined spectrum would have yielded 
stronger alkaline lines). 

As illustrated in Figure~\ref{combspec}, the \so~$J$-band master spectrum shows stronger water vapor 
absorption at 1.33 $\mu$m and a slightly redder pseudocontinuum slope than the field M9 spectrum, which 
indicates a later type on the basis of the spectroscopic classification criteria defined in the 
literature. This agrees with our spectral typing. The 1.33 $\mu$m-band appears marginally deeper in 
the \so~averaged spectrum than in the field L1 dwarf and is better reproduced by the field L3 dwarf. 
Despite the differences observed in the Na\,{\sc i}, K\,{\sc i} and FeH features, the pattern outlined
by the $J$-band pseudocontinuum of \so~L dwarfs and high-gravity objects of related types are 
remarkably similar, including the intensity of the water vapor absorption at 1.33 $\mu$m. Based on 
these properties and the discussion of previous sections, we concluded that the $J$-band wavelengths 
are appropriate for spectral classification within $\pm$2 subtypes (by using the pseudocontinuum 
slope) and for discriminating low-gravity from high-gravity dwarfs of similar metallicity (by
examining the intensity of the atomic and FeH features).

We followed a similar prescription of spectral comparison for the optical wavelengths as done above for 
the near-infrared. In the top panel of Figure~\ref{combspec}, the \so~master spectrum (red) is
compared with high-gravity M9, L1 and L3 standards (black). The optical data of \so~are characterized 
by the presence of weak Na\,{\sc i} absorption at $\sim$819.5 nm (a sign of youth, e.g.,
\citealt{steele95}), and strong TiO and VO bands. The latter, although also indicating a low-gravity
nature \citep{martin99}, would also suggest dwarf-based spectral types earlier than M9.5--L0 
\citep{luhman07a}, in contrast with the L classification derived from the near-infrared.
Interestingly, the \so~optical spectrum also reveals a broad K\,{\sc i} resonance doublet absorption 
at $\sim$766.5 nm. This line profile is quite sensitive to gravity and temperature conditions 
(e.g., \citealt[and references therein]{pavlenko00,kirk08}). Although the shape of this resonance 
doublet mimics that of the early-L types, the  K\,{\sc i} core appears to be weaker in the combined 
\so~spectrum than in the field, high-gravity L1 and L3 dwarfs in the top panel of 
Figure~\ref{combspec}, thus supporting low-gravity atmospheres. The red slope of the \so~combined 
optical spectrum is better reproduced by the L3 standard, which is consistent with the near-infrared 
high-gravity-based classification. As widely discussed in the literature, the transition from M to 
L types for high-gravity dwarfs is driven by the disappearance of TiO absorption at optical 
wavelengths \citep{kirk99,martin99}. However, atmospheres of very low gravity show strong TiO and 
VO features through the mid-L types (see next). As also observed at near-infrared wavelengths, 
the strength of the optical FeH bands is weaker than normal (in part owing to the intense TiO and
H$_2$O absorptions), while the water vapor feature at $\sim$935 nm appears to be stronger in 
the low-gravity spectrum. We caution that the telluric stars used to correct the optical science 
data were observed at smaller air masses than our \so~targets, and some telluric residuals might
be present in the visible spectra.

In Figure~\ref{allopt_with_young}, the optical master spectrum of \so~was also compared with the 
data of three field L dwarfs with spectroscopic evidence of low-to-intermediate gravity atmospheres
\citep{cruz09}. These data were collected with the same instrumental setup and were processed in
the same manner as the \so~spectra (thereby avoiding any systematic effects); the field objects' spectra
were published by \citet{osorio14a}. The comparison dwarfs were spectroscopically classified as L1
(2MASS\,J02411151$-$0326587), L2 (2MASS J00452143$+$1634446), and L3  (G\,196$-$3\,B), and were
given the status of very low-gravity (VL-G) sources by \citet{allers13}. Their likely ages are 
estimated at $\sim$500, $\sim$30--250, and $\sim$20--85 Myr, respectively, on the basis of the 
presence of lithium in their atmospheres, their location in the Herzsprung--Russell diagram, and 
their membership of young stellar moving groups \citep{osorio14a,gagne14b}. Hence, despite the
presence of youthful features in the optical spectra, these three comparison sources are significantly 
older than \so~by at least a factor of 10 and their gravities (a proxy of age) may differ. They cover 
the possible optical and ``optNIR'' low-gravity-based spectral types of our targets. As seen in 
Figure~\ref{allopt_with_young}, the 3-Myr spectrum shows stronger oxides (TiO, VO, and H$_2$O) at
all visible wavelengths than any of the other supposedly VL-G, but older (therefore, higher gravity)
dwarfs. We interpret this as evidence of the effects of the lower gravity atmospheres (younger
age) of the \so~objects. Note that the pseudo-continuum slope in the optical is quite similar for
all three spectral types (young L1, L2, and L3) of Figure~\ref{allopt_with_young}; only the
molecular absorptions due to TiO, VO, and H$_2$O seem to change significantly from one spectral type
to another. Because these oxides are known to depend on both atmospheric gravity and 
temperature (they become stronger with decreasing gravity and increasing temperature in the 
M--L transition), it might be possible that, at the resolution of our data, a low-gravity, low-temperature 
spectrum mimics a slightly higher gravity, higher temperature spectrum in the visible. Data at 
longer wavelengths would be required to solve this ambiguity. 

\subsection{Magnitude-spectral type diagram}

The spectral sequence of \so~low-mass stars, brown dwarfs, and planetary-mass dwarfs is illustrated
in Figure~\ref{jspt}: there are two panels depicting the high-gravity- and low-gravity-based spectral
types derived in this work. This is one of the first planetary-mass sequences defined for very young 
ages where the great majority of the low-mass objects are confirmed to be coeval via the diagnosis of 
spectroscopic features indicative of youth. We adopted the photometry given by the UKIRT Infrared Deep 
Sky Survey (UKIDSS) data release 10, particularly the Galactic Clusters Survey. The UKIDSS project is 
defined in \citet{lawrence07}. UKIDSS uses the UKIRT Wide Field Camera (WFCAM; \citealt{casali07})
and a photometric system described in \citet{hewett06}. The pipeline processing and science archive
are described in \citet[][and references therein]{hambly08}. The UKIDSS $J$-band photometry and the 
data published by \citet{pena12} are consistent typically to within 1\,$\sigma$ the quoted uncertainties 
at bright and intermediate magnitudes, and to within $\sim$2\,$\sigma$ at the faintest magnitudes. 
However, the UKIDSS $J-K$ color is systematically redded by about 0.05 mag than the $J-K$ index based 
on the Visible and Infrared Survey Telescope for Astronomy (VISTA) photometric system (see
\citealt{cross12}) employed in \citet{pena12}. A further discussion of the comparison between the VISTA 
and UKIDSS photometry of \so~objects will be provided in a forthcoming paper. In this paper,
the UKIDSS $J$- and $K$-band photometry of our spectroscopic targets and the photometry of
\citet{pena12} are listed in Table~\ref{ukidssphot}. In Figure~\ref{jspt}, the average location of field
dwarfs taken to the distance of the cluster is also shown, together with the \so~sequence of 
spectroscopically-confirmed members \citep{barrado01,martin01,sacco07,pena12}. For spectral types 
$\ge$M6, we used the field sequence of \citet[their Table~19]{faherty16}, which provides a sequence 
of high-gravity dwarfs by removing peculiar objects, binaries, low-metallicty dwarfs, and young 
field sources. Furthermore, it was constructed by using a relatively high number of dwarfs with known
trigonometric parallax. The photometric dispersion of the high-gravity field sequence is also 
depicted in both panels of Figure~\ref{jspt}.

The \so~sequence is well defined and extrapolates the previously known cluster sequence of more 
luminous members. It runs parallel to the field sequence of high-gravity dwarfs when considering 
the high-gravity-based spectral types (top panel of Figure~\ref{jspt}), as expected for very 
young objects that are still undergoing self-gravitational contraction. However, the cluster 
sequence defined by the ``youth''-based classification (bottom panel) approaches the high-gravity 
field sequence, thus indicating that the slope of the $J$-band magnitude--spectral type relationship
may be different for low- and high-gravity cool objects. This feature has been extensively 
discussed in the literature (see \citealt{faherty16} and \citealt{liu16}, and references therein).
Note that this is not the case for color--magnitude diagrams such as that shown in Figure~\ref{zj}, 
where the \so~members show a clearly defined sequence.

\subsection{Photometric colors}
In  recent years, the very red near- and mid-infrared colors of cool dwarfs, particularly
the $J-K$ index, have been widely associated with a young age (e.g., \citealt[and references 
therein]{cruz09,kirk10,faherty12,liu13,schneider14,marocco14b,gizis15}). This understanding is
driven by the fact that young-aged L-type members of nearby, solar metallicity stellar moving 
groups (such as $\beta$ Pictoris, Tucana-Horologium, AB Doradus, etc.) show a general flux excess 
towards longer wavelengths accompanied by weak alkali lines, peaked $H$-band pseudocontinuum,
and enhanced metal oxide absorption bands in their optical and infrared spectra. The latter 
properties are accepted spectroscopic evidence of a low-gravity nature (as discussed in 
Sections~\ref{membership} and~\ref{optnirall}). The reddening characteristic has thus been 
interpreted as mostly due to the low gravity of the atmospheres and to an excess of dust 
(condensates) in the photospheres, which heavily influence the photometric and spectroscopic
properties of the young L dwarfs. Interestingly, there are a few very dusty L dwarfs with no 
evidence of youth or low gravity (see \citealt{liu16} and references therein). It is also
acknowledged that the very red colors might be caused by a high metallic content of the 
low-temperature atmospheres (e.g., \citealt[and references therein]{looper08,leggett10b}).

We confirmed the membership of most of our targets in the 3-Myr \so~cluster. They all display
spectroscopic features ascribed to low-gravity atmospheres and delineate a photometric and 
spectroscopic progression that is over-luminous with respect to the field sequence 
(Figure~\ref{jspt}), as expected for the age of the cluster. In what follows, we discuss 
the color properties of our targets as a function of their spectral types.

\subsubsection{High-gravity-based spectral types}
Figure~\ref{jkspt} illustrates UKIDSS $J-K$ colors against spectral types  from 
early-M through  mid-L. In the top panel of the Figure, the high-gravity-based types of 
the \so~sources of this work are shown. For the two targets lacking UKIDSS $J$-band magnitudes,
we adopted the VISTA $J$ and UKIDSS $K$ values based on the similarity of the $J$-filter
measurements between these two photometric systems (Section~\ref{optnirall}). As depicted 
in the top panel of Figure~\ref{jkspt}, the L-type \so~objects do not present obvious near-infrared
colors in clear excess of the field dwarfs with similar spectral classification. 

The mean location of the field (i.e., high gravity atmospheres) for types $\ge$M6 is indicated 
by the solid line (2MASS data were taken from \citealt{faherty16}, and colors were conveniently 
transformed into the UKIDSS photometric system. The color dispersion of the field is also shown).
For each spectral subtype, the field sequence has a color dispersion of typically 
$\pm$0.15--0.25 mag. The \so~members of this paper and those from the compilation made by 
\citet{pena12} and \citet{sacco07} follow the trend described by the field. Among the 
M dwarfs, there are several \so~outliers, which show mid-infrared flux excesses typical of
surrounding disks \citep{pena12}. From Figure~\ref{jkspt} it becomes apparent that the M- 
and L-type \so~and field dwarfs have similar $J-K$ colors to within the quoted dispersion. 
Furthermore, other L dwarfs with ages $\le$10 Myr, such as additional \so~members from
\citet{osorio00}, \citet{barrado01}, and \citet{martin01}, Upper Scorpius members recently 
discovered by \citet{lodieu08}, \citet{bejar08}, \citet{lafreniere10}, \citet{pena16}, 
and members of the TW Hydrae (TWA) moving group compiled by \citet{chauvin04,chauvin05a},
\citet{gagne15a}, \citet{kellogg15,kellogg16}, and \citet{schneider16a} also have colors 
resembling the indices of the \so~L0--L4.5 objects of this paper. Figure~\ref{jkspt}
shows the location of the \so, Upper Scorpius and TWA sequences using different symbols. 
Only a few Upper Sco L1-type members and 2MASSW\,J1207334$-$393254b \citep[TWA member,][]{chauvin04} 
appear redder than the TWA and \so~sequences. The similarity of the very young L4--L5 dwarfs 
is illustrated in Figure~\ref{fire2}, where the L4 spectrum of an Upper Scorpius member 
is shown in comparison with two of our \so~targets: all three spectra are well matched.

Other young cool dwarfs with known age and spectroscopic signposts of low gravity atmospheres
are also depicted in Figure~\ref{jkspt}. The figure shows,  in  order of increasing age,  
the location of members of the $\sim$25-Myr $\beta$~Pictoris moving group taken from the 
compilation made by \citet{lagrange09}, \citet{liu13}, \citet{schneider14}, \citet{gagne15a},
and \citet{deacon16}, members of stellar moving groups with ages in the interval 20--50 Myr 
(see \citealt{gagne15a} for the lists of objects and moving groups), and members of the
120-Myr Pleiades cluster \citep{bihain10,osorio14b,osorio14c}. All of these objects are
supposed to share solar metallicity. For a given spectral type, the similarity of the $J-K$ 
colors for all ages is consistent with the observed photometric dispersion from the mid-Ms
through $\sim$L0. However, the situation changes at cooler types. The L0--L5 dwarfs with 
intermediate ages ($\beta$~Pic and particularly the 20--50 Myr moving group members) 
tend to have on average redder $J-K$ colors than their younger and older (the field) 
spectral couterparts. The impact of the very red $J-K$ colors in the near-infrared spectra 
can be seen in Figure~\ref{fire2}, where the spectra of 2MASSW\,J1207334$-$393254b \citep[TWA 
member, spectrum from][]{patience12} and G\,196$-$3\,B \citep[20--85 Myr, spectrum from][]{bihain10} 
are shown in comparison with our {\sc fire} data.  As discussed in \citet{bihain10}, 
Pleiades M and early-L members show slightly redder $J-K$ colors than the field by 0.11 mag, 
which the authors ascribe to both reddening and youth. \citet{osorio14b,osorio14c} argued
that the Pleiades sequence continues towards even redder $J-K$ colors for cooler types, matching 
the colors observed from giant planets around stars (e.g., planets HR\,8799bcde, 
\citealt{marois08,marois10}) and other extremely red field dwarfs with likely planetary 
masses. As illustrated in Figure~\ref{jkspt}, dwarfs of young stellar moving groups and 
Pleiades members with spectral types $\ge$L5 and ages in the interval 10--500 Myr have 
$J-K$ colors 0.2--1.0 mag redder than the field ($\sim$1--10 Gyr) spectral counterparts. 

The top panel of Figure~\ref{jkspt} suggests that the L dwarfs become redder than the field,
achieving up to $J-K \approx 3$ mag at ages typically $\ge$10 Myr and then turn back to
``normal'' colors at an age (gravity) that is not well constrained yet. This timing appears 
to be older than 120--300 Myr  for the latest L types since Pleiades late-L dwarfs (120 Myr) 
and the L7 dwarf (150--300 Myr) of \citet{gauza15} still display very red atmospheres.
Regarding the early-L types, as is seen in Figure~\ref{jkspt}, both the 120-Myr 
Pleiades cluster and the field have colors that are consistent to within the uncertainties. \citet{aller16} 
also noted that the L0--L4 members of the AB~Doradus moving group, which has an age similar
to that of the Pleiades, do not have such distinct near-infrared colors from field ultracool 
dwarfs. The authors ascribed this effect to the higher-gravity atmospheres of AB~Doradus 
members than those of the the TWA and $\beta$~Pic moving groups members. By adopting the
high-gravity-based spectral types, it seems that the near-infrared colors of the L dwarfs
behave nonlinearly with low surface gravity, and atmospheric gravity alone cannot 
account for the reddening and subsequent blue turn-over of the colors with age (increasing 
gravity). Given the young-to-intermediate ages of the very red L dwarfs (most appear to have 
ages in the interval 10--500 Myr), other scenarios may be at play such as an enhanced dust
production at certain atmospheric pressures, the presence of warm debris disks, which might 
contribute to the signficant long wavelength reddening (e.g., \citealt{zakhozhay16}), and/or 
thermo-chemical instabilities that produce changes in the temperature structure of cool atmospheres \citep{tremblin16}.

\subsubsection{Youth-based spectral types}
The bottom panel of Figure~\ref{jkspt} shows the low-to-intermediate gravity-based types of 
the \so~objects. Under this classification scheme, the great majority of the cluster members 
appear to be very red late-M and early-L dwarfs, as compared to the high-gravity dwarfs of related
types. This scenario suggests that the ``red nature'' of L dwarfs develop very early in the 
evolution of the low-mass objects with cool atmospheres (as opposed to the scenario described in
the previous section), and that this property disappears at an age $\ge$120 Myr for the latest
L types, and around 120--150 Myr for the early-to-mid L sources. 

\section{Comparison with model atmospheres \label{model}}
We used the latest generation of the BT-Settl synthetic spectra\footnote{http://perso.ens-lyon.fr/france.allard/}
created with the {\sc phoenix} code (version 15.5, \citealt{hauschildt97,allard01}) for solar 
metallicity. These models aim to describe the atmospheres of low-mass stars, brown dwarfs, 
and planets without irradiation. They include a cloud model by accounting for the formation and 
gravitational settling of dust grains for $T_{\rm eff} \le 2700$ K following the approach of \citet{rossow78}. 
Further details on  BT-Settl theoretical atmospheres are given in \citet{allard12,allard14} and
\citet{manjavacas15}. For our purposes, we employed a subsample of the BT-Settl solar metallicity 
models that extend from $T_{\rm eff}$ = 1200 to 3000 K in steps of 50--100 K and log\,$g$ = 2.5 
to 5.5 (cm\,s$^{-2}$) in steps of 0.5 dex. We performed a comparison between observed and BT-Settl 
synthetic spectra to derive the physical parameters of our \so~objects with {\sc fire} and 
{\sc osiris}$+${\sc isaac} data because these sources have a long wavelength coverage that may provide 
well-constrained parameters. The resolution of the original BT-Settl computations was degraded with a
Gaussian profile to match that of the optical and near-infrared observations, and all theoretical and 
observed spectra were normalized to unity in the interval 1.28--1.32 $\mu$m, which does not include 
strong spectral features (the observations are not flux calibrated). This normalization allowed us 
to compare the observed and predicted molecular and atomic signatures, as well as the pseudo-continuum 
behavior, and provided an additional check on the veracity of the synthetic spectra. Given the 
low resolution of our data, we neither broadened the BT-Settl models to account for rotation nor 
corrected for radial velocity shifts. 

Best fits to the observed spectra were found by minimizing the following merit function:
\begin{equation}
\Delta_f = \sqrt{\sum_{i=1}^{2}(f_{\rm obs}-f_{\rm syn})^2_{\lambda_i}}
\label{eq}
\end{equation}
where $f_{\rm obs}$ and $f_{\rm syn}$ stand for the observed and theoretical flux densities, 
and $\lambda_i$ accounts for the wavelength intervals 0.700--0.9815 and 1.091--1.345 $\mu$m 
({\sc osiris} and {\sc isaac} data), and 0.90--1.32 and 2.00--2.36 $\mu$m ({\sc fire} data). 
The $H$-band was excluded from the comparisons (see below). The regions of strong telluric water vapor 
absorption in the near-infrared were also excluded. We fixed the chemical abundance to the solar value,
which is the metallicity of the \so~cluster \citep{gonzalez08}, and kept $T_{\rm eff}$ and log\,$g$ 
free. The merit functions were obtained for each {\sc fire} and pair of {\sc osiris}$+${\sc isaac}
spectra. Examples are shown in Figure~\ref{goodness}. The merit functions show a marked decrease 
towards a minimum from the warm to the cool temperatures. We assigned errors of $\pm$100 K and $\pm$0.5 
dex to the derived $T_{\rm eff}$ and log\,$g$ (corresponding to the sampling of the models). The 
acceptable temperature and gravity parameters were visually inspected by comparing them with the observed spectra. 

We found that the best-fit gravities providing the smallest values of $\Delta_f$ are in the interval 
log\,$g$ = 3.5--4.5 dex. For all the observed spectra used we consistently derived such low gravities
as are to be expected for low-mass objects with an age of a few Myr. They  agree quantitatively to within
$\pm$0.5 dex with the predictions made for \so~by  evolutionary models (e.g., 
\citealt{baraffe98,chabrier00,burrows03}). The best-fit $T_{\rm eff}$s are given in Table~\ref{pew}; 
these correspond to log\,$g$ = 4.0 dex. The measured spectral types and the $T_{\rm eff}$s obtained from 
the minimization technique successfully correlate, i.e., the cooler the temperature, the later the 
spectral type, thus supporting our relative ordering of the data and the spectral classification. 

For the low gravities, such as those of the \so~low mass objects, our analysis shows that the
$T_{\rm eff}$ scale of Table~\ref{pew}  differs slightly from that of the field high-gravity dwarfs
of related types. Our determinations are systematically warmer by 100--200 K than recent 
$T_{\rm eff}$--spectral type relations valid for field L0--L5 dwarfs \citep{stephens09}. We note,
however, that this difference is close to the temperature uncertainty associated with the model 
fitting technique, and that there might be some uncontrolled systematics in the models. A similar 
effect was discussed for the M dwarfs by \citet{luhman99} and \citet{luhman03}. These authors developed 
a temperature scale for young, low-gravity M dwarfs that is intermediate between the dwarf (cooler)
and giant (warmer) scales. This scale has been used by various groups to study the low-mass
population of young star clusters (e.g., \citealt{briceno02}). Our temperature scale for the young 
early-L dwarfs seems to follow the tendency reported by \citet[see their Table~8]{luhman03}. A better 
agreement is found with the $T_{\rm eff}$ predictions made by the evolutionary models, yet our
derivations from the spectral fitting analysis are $\sim$100 K warmer (albeit close to the quoted
uncertainty in $T_{\rm eff}$). 

Figures~\ref{models1} and~\ref{models2} illustrate the comparison between the observed spectra 
and the best-fit model atmospheres. In Figure~\ref{models1}, we also included the {\sc vista} 
photometry (appropriately normalized to the observed spectra) and the theoretical photometry computed by 
integrating the best-fit models using the {\sc vista} filter passband profiles and flux densities of
\citet{hewett06}. The overall aspect of the optical and near-infrared data is successfully reproduced
by the theoretical spectra, implying that no relevant molecule or atomic element is missing from the
atmospheric computations. However, not all the fine details of the observations are well matched by 
the models. At optical wavelengths, the various TiO, VO, FeH and CrH absorptions and the red optical
slope are nicely fitted using  solar metallicity, log\,$g$ = 4.0 dex, models with the temperatures of
Table~\ref{pew} (Figure~\ref{models1}). However, the water vapor absorption at $\sim$935 nm appears to be 
slightly overestimated by the model atmospheres. As for the alkaline lines, the theoretical spectra
degraded to the low resolution of our data exhibit no measurable Na\,{\sc i} (at 
$\sim$819.5 nm), Cs\,{\sc i}, and Rb\,{\sc i}; this agrees with the properties of the \so~spectra.
The resonance doublet of K\,{\sc i} in the visible, although one of the strongest features at low 
temperatures, depends strongly on atmospheric pressure; the weak intensity of this feature present 
in the low-gravity \so~spectra appears to be reasonably well matched by the theory. 

The success of the BT-Settl low-gravity models at  near-infrared wavelengths is not as remarkable 
as it is in the optical. While the strength of the alkaline near-infrared lines are acceptably
predicted by the theory (at the resolution of our data), the broad features due to VO at 1.06 
$\mu$m and VO and TiO between 1.17 and 1.29 $\mu$m \citep{mcgovern04} are typically less intense
in the observed data. Not even the true shape of these bands is  fitted by the theory, probably because
of the lack of accurate molecular data to describe the VO A--X ($\Delta\nu = -1$) and TiO $\Phi$
($\Delta\nu = -1$) band heads. \citet{manjavacas15} have discussed the possibility that the pEWs of the K\,{\sc i} 
$\lambda$1.252 $\mu$m of high-gravity models overestimate those measured from the observed spectra 
of field early-L dwarfs (an opposite effect was observed for the late-L and T dwarfs). This does 
not occur at the low gravities of \so, perhaps because of our poor understanding of the 
opacities of the dust being formed in high pressure (gravity) atmospheric layers (the 
pseudocontinuum around 1.25 $\mu$m is dominated by condensate grain opaicities, \citealt{ackerman01,faherty14}).
The model $H$-band delivers a poor fit to the observed triangular shape of the data 
(Figure~\ref{models2}), probably because of the poor description of the water opacities at 
these wavelengths \citep{allard12}. The small S/N of the $K$-band spectra at around 2.3 $\mu$m 
(CO bands) prevented us from giving a qualitative despcription of the observed--theoretical comparison results. 
However, we note that the models attractively reproduce the relative fluxes between the optical, 
$J$-, $H$-, and $K$-bands, which can be understood as evidence in favour of a proper treatment of the 
gas and dust condensation chemistry and the dust opacity at these low temperatures and surface gravities. 

\section{Masses and \so~mass function\label{mf}}
Given the similarity of the spectral energy distributions of the \so~members with those of the 
field objects of related classification, we used the $J$- and $K$-band bolometric corrections (BCs) 
recently determined for M and L dwarfs by \citet{filippazzo15}, and the magnitudes and spectral types
of our targets to determine absolute luminosities. The luminosities provided in Table~\ref{pew} 
correspond to the mean values between the $J$- and $K$-band derivations, both of which agree to within 
$\pm$0.1 dex. For the Sun, we used $M_{\rm bol} = 4.73$ mag. The associated uncertainty in luminosity
was obtained by considering the photometric errors and by adopting error bars of $\pm$0.20 mag in the 
BCs, and $\pm$0.25 mag in the distance modulus of the cluster ($m-M = 7.93 \pm 0.10$ mag). 
\citet{filippazzo15} measured a dispersion of $\pm$0.16 and $\pm$0.24 mag for the $J$- and $K$-band
BCs. We did not consider the spectral type uncertainties because they do not have a significant effect
on the BC values beyond the adopted BC error; from early- through mid-Ls, the $J$- and $K$-band BCs vary
little to within a few subtypes. This behavior is also present in other BC determinations available in the 
literature (e.g., \citealt{leggett01,dahn02,golimowski04,schmidt14}).

The luminosities were then compared to the predictions made by the 3-Myr isochrone from \citet{chabrier00} 
to determine the masses\footnote{We employed the following mass conversion: 1 M$_\odot$ = 1047.56
M$_{\rm Jup}$.} listed in the last column of Table~\ref{pew}. The quoted mass uncertainties take into 
account the errors in luminosity. If the cluster were younger (1 Myr) or older (5 Myr), the masses
would be smaller by a factor of 1.5--2.0 or greater by a factor of 1.3, respectively. At the age of 
3 Myr, our targets have masses from $\approx$6 M$_{\rm Jup}$ up to the deuterium burning-mass limit at
$\approx$13 M$_{\rm Jup}$. They can serve as very low-gravity templates to study other young 
isolated and planetary companion sources in the field, and star-forming regions and star clusters. According
to the theory of substellar evolution, they mature towards fainter luminosities and cooler temperatures,
reaching $L$\,$\sim$\,10$^{-7}$\,L$_\odot$ and $T_{\rm eff}$\,$\sim$\,240--400 K at the age of a few Gyr.
They would belong to the recently defined Y spectral type \citep{cushing11} and would have prohibitively 
dim optical and near-infrared magnitudes for current ground-based and space telescopes. 
Most of their flux is expected to be emitted at $\ge$3 $\mu$m, an example of which could be
WISE\,J085510.83$-$071442.5 \citep{luhman14}, a 3--10 M$_{\rm Jup}$ isolated planetary-mass dwarf 
located at 2 pc from the Sun. To date, this object remains undetected at optical wavelengths and has
detections in a few near-infrared wavelengths, and mid-infrared {\sl Spitzer} and {\sl WISE} short-wavelength 
bands \citep{schneider16b,luhman16,osorio16}; indeed, it is the reddest and coolest dwarf ($\le$250 K) 
ever discovered \citep{beamin14}. WISE\,J085510.83$-$071442.5 may represent the ``old'' face of the 
\so~isolated planetary-mass objects studied here. 

The validation of a significant number of \so~planetary-mass members adds solid support to the cluster 
luminosity and mass functions presented in Figures~13 and 15 of \citet{pena12}, particularly in the 
substellar domain. In the mass interval 6--13 M$_{\rm Jup}$, these authors counted 22 cluster candidates 
in a circular area of radius of $30'$ around the massive, multiple star $\sigma$\,Ori. Here, we 
established the cluster membership of 10 (possibly 11) of these objects using a gravity-sensitive 
spectroscopic diagnosis. Another two L dwarfs (S\,Ori\,56 and 60) within the aforementioned mass range 
have mid-infrared flux excesses compatible with surrounding disks \citep[and references therein]{luhman08}, 
thus confirming their likely cluster membership. Therefore, 55--60\%~of the original 6--13
M$_{\rm Jup}$ sample is corroborated by well-accepted signs of youth. The statistics increases 
up to 77\%~by considering another 4 sources (S\,Ori\,58, 62, 66, and 68) that have L spectral types 
\citep{martin01,barrado01} and magnitudes consistent with the \so~spectroscopic sequence (Figure~\ref{jkspt}). 
The \so~mass function assembled by \citet{pena12} is thus verified from the high-mass stars 
through the planetary-mass domain. The mass function can be used as a reference for future investigations. According to the \so~mass function (in its linear form), there are as many 6--13 M$_{\rm Jup}$ members as 0.075--0.15 M$_\odot$ low-mass stars in the cluster. The observed density of isolated planetary-mass objects accounts for one such object for every $\sim 8 \pm 2$ stars in \so. This ratio of planetary-mass to stellar
cluster members agrees with recent estimates based on the least massive population of other young 
clusters and stellar moving groups (e.g., \citealt{gagne15a}), but is smaller than that determined
for the TWA moving group by \citet{gagne17}. By extrapolating the rising behavior of the \so~mass 
function to the field, we would expect as many 6--13 M$_{\rm Jup}$ isolated planetary-mass objects with 
properties similar to WISE\,J085510.83$-$071442.5 \citep{luhman14} as 0.075--0.15 M$_\odot$ late-M and
early-L stars in the solar neighborhood. This estimate is in line with the results of \citet{sumi11} 
based on microlensing events, although we caution that these authors were looking at even less 
massive objects ($\sim$1 M$_{\rm Jup}$). Their discovery is indeed challenging.

\section{Conclusions and final remarks\label{conclusions}}
We have collected low-resolution optical and near-infrared spectra of twelve planetary-mass candidates
of the young ($\sim$3 Myr), solar metallicity \so~cluster using {\sc osiris} (0.70--0.98
$\micron$, $R \sim 300$) on the GTC, {\sc isaac} (1.11--1.34 $\mu$m, $R\sim500$) on the VLT, and 
{\sc fire} (0.8--2.5 $\micron$, $R\sim300$) on the Magellan Telescope. These sources represent about half of
the sample populating the planetary-mass interval (6--13 M$_{\rm Jup}$) of the cluster mass function 
defined in \citet{pena12}. Confirming their low-gravity nature and in turn, their membership in
\so, is critical for building a reference mass function. All twelve targets have $J$-band magnitudes
in the range 18.2--19.9 mag and define the faint end of the photometric cluster sequence. 

By visual inspection of the comparisons of our data with spectroscopic reference high- and low-gravity 
dwarfs, and by measuring newly-defined indices and known indices from the literature that quantify 
the spectral slopes and water vapor absorption, we determined that the \so~candidates have spectral 
types in the interval L0--L4.5 (high-gravity-based) and M9--L2.5 (low-gravity-based) from the
near-infrared spectra and L1.5--L2.5 from the visible data. The observed spectra revealed features
of low-gravity atmospheres for nearly all of the targets, such as weak alkali lines due to Na\,{\sc i} 
and K\,{\sc i}, peak-shaped $H$-band pseudo-continua, weak CrH and FeH absorptions, and strong
molecular absorptions due to oxides (TiO and VO). This confirms them as bona fide members of \so.
Contrary to the definition of the dwarf-based L classification for field high-gravity objects (where
oxides vanish from the optical spectra), the TiO and VO signatures of young L dwarfs do not disappear 
from the visible data and maintain their strength through the mid-L types. We caution that using the
gravity-sensitive TiO- and VO-based indices in the optical,  would lead one to derive earlier spectral types
for the \so~objects. The individual data were combined to produce acceptable S/N ``master'' spectra 
at optical and near-infrared wavelengths that can be used as templates for future studies of young
L dwarfs. As a reference, our measurements showed that the pEWs of the near-infrared K\,{\sc i} lines are 
2.5$\pm$0.5 \AA~(1.169 $\mu$m), 3.7$\pm$0.5 \AA~(1.177 $\mu$m), 2.2$\pm$0.5 \AA~(1.243 $\mu$m), and
3.3$\pm$0.5 \AA~(1.252 $\mu$m) for young early-L dwarfs with an age of $\sim$3 Myr. Higher pEWs 
would indicate older ages. 

Considering the low-to-intermediate gra\-vity-based spectral classification, nearly all  \so~objects 
present infrared colors in excess of those of the majority of the known dwarfs of related types, in 
line with  observations of other young L dwarfs in the field (e.g., 
\citealt{chauvin04,cruz09,kirk10,faherty12,liu13,schneider14,marocco14b,osorio14b,gizis15,gauza15,kellogg15}). 
This implies that the very red nature of L dwarfs develops early in the evolution of cool, 
low-mass objects and lasts up to 120--150 Myr for the early-L types. The reddening effect has
been explained in the literature in terms of the presence of dusty atmospheres at low gravities. The
classification based on the similarity of the global spectral energy distribution of \so~sources 
with high-gravity references leads to the conclusion that  very young dwarfs  have no obvious color excesses 
and suggests that the reddening feature appears at ages $\ge$10 Myr. There might be other scenarios
at play to account for the very red colors of the young L dwarfs, such as the presence of warm
debris disks, which may contribute to the long wavelength reddening (e.g., \citealt{zakhozhay16}) 
and/or instabilities that produce changes in the temperature structure of the atmospheres \citep{tremblin16}.

We have compared the \so~observations with BT-Settl, solar metallicity model atmospheres deriving low
gravities, log\,$g \approx 4$ [cm\,s$^{-2}$], as expected for 3-Myr low-mass sources. We also 
obtained a $T_{\rm eff}$ scale that correlates with the near-infrared spectral types and the observed
colors: 2350--1800 K for the L0--L4.5 (high-gravity-based) and M9--L2.5 (low-gravity-based)
\so~dwarfs. At the resolution of our data, the synthetic spectra appear to reproduce reasonably
well the spectral slope and some atomic and molecular features of the observations. However, the 
theory predicts stronger VO absorption in the near-infrared than is observed. Water vapor bands,
particularly in the $H$ band, are  another pending issue for the models. 

Through comparison with evolutionary tracks, we determined the masses of the \so~L dwarfs to be in 
the interval 6--13 M$_{\rm Jup}$ and confirm them as planetary-mass members of the cluster. 
The \so~mass function presented in \citet{pena12} is now confirmed with the spectroscopic 
observations of up to $\sim$75\%~of the cluster members from the high-mass O-type stars through 
the planetary-mass domain at 6 M$_{\rm Jup}$. It is expected that the \so~isolated planetary-mass 
objects evolve and look like the 2-pc distant WISE\,J085510.83$-$071442.5 \citep{luhman14} at the 
age of a few Gyr, of which there might be as many as late-M and early-L dwarfs in the solar neighborhood.

We also included the first optical spectrum of S\,Ori\,70 (T5--T7) obtained with {\sc osiris} on 
the GTC aimed at identifying possible features indicative of a low-gravity atmosphere. Although the 
spectrum has poor S/N ($Z \ge 24$ mag) and no photons were registered blueward of $\simeq$800 nm, 
a very red slope is clearly detected that is compatible with a T6--T7 spectral type, which agrees 
with the near-infrared typing of S\,Ori\,70. A strong water vapor absorption is detected at 935 nm. 
Better quality data and spectra with sufficient S/N in the blue wing of the K\,{\sc i} doublet
are required for a solid characterization of the surface gravity of this particular object.


\acknowledgments
We thank the anonymous referee for useful comments. We also thank
Terry Mahoney for revising the English of this paper. This work is
based on observations made with the Gran Telescopio Canarias (GTC),
installed at the Spanish Observatorio del Roque de los Muchachos of
the Instituto de Astrof\'\i sica de Canarias, on the island of La
Palma. This paper is also based on observations made with ESO
Telescopes at the Paranal Observatory under program ID 090.C-0766 and
also includes data gathered with the 6.5 meter Magellan Telescopes at
Las Campanas Observatory, Chile. This research has benefitted from the
SpeX Prism Spectral Libraries\footnote{http://pono.ucsd.edu/~adam/browndwarfs/spexprism} and the Montreal Brown Dwarf and Exoplanet Spectral Library\footnote{https://jgagneastro.wordpress.com/the-montreal-spectral-library/}, maintained by Adam Burgasser and by Jonathan Gagn\'e, respectively.
This research is
partly financed by the Spanish Ministry of Economy and Competitivity
through projects AYA2014-54348-C3-2-R, AYA2015-69350-C3-2-P, AYA2016-79425-C3-2-P, and the
Chilean FONDECYT Postdoctoral grant 3140351. KPR acknowledges CONICYT PAI Concurso Nacional Inserci\'on en la Academia, Convocatoria 2016 Folio PAI79160052.

{\it Facility:} \facility{VLT ({\sc isaac}); GTC ({\sc osiris}); Magellan ({\sc fire}).}

\bibliographystyle{aa} 
\bibliography{1.bib}

\begin{thebibliography}{152}
\expandafter\ifx\csname natexlab\endcsname\relax\def\natexlab#1{#1}\fi

\bibitem[{Ackerman \& Marley(2001)}]{ackerman01}
Ackerman, A.~S. \& Marley, M.~S. 2001, ApJ, 556, 872

\bibitem[{Allard(2014)}]{allard14}
Allard, F. 2014 (Presented at the workshop on Exploring the Formation and
  Evolution of Planetary Systems, IAU Symposium, 299, 271)

\bibitem[{Allard {et~al.}(2001)Allard, Hauschildt, Alexander, Tamanai, \&
  Schweitzer}]{allard01}
Allard, F., Hauschildt, P.~H., Alexander, D.~R., Tamanai, A., \& Schweitzer, A.
  2001, ApJ, 556, 357

\bibitem[{Allard {et~al.}(2012)Allard, Homeier, \& Freytag}]{allard12}
Allard, F., Homeier, D., \& Freytag, B. 2012, Royal Society of London
  Philosophical Transactions Series A, 370, 2765

\bibitem[{Aller {et~al.}(2016)Aller, Liu, Magnier, Best, Kotson, Burgett,
  Chambers, \& et~al.}]{aller16}
Aller, K.~M., Liu, M.~C., Magnier, E.~A., {et~al.} 2016, ApJ, 821, 120

\bibitem[{Allers {et~al.}(2007)Allers, Jaffe, Luhman, Liu, Wilson, Skrustskie,
  Nelson, Peterson, Smith, \& Cushing}]{allers07}
Allers, K.~N., Jaffe, D.~T., Luhman, K.~L., {et~al.} 2007, ApJ, 657, 511

\bibitem[{Allers \& Liu(2013)}]{allers13}
Allers, K.~N. \& Liu, M.~C. 2013, ApJ, 772, 79

\bibitem[{Baraffe {et~al.}(1998)Baraffe, Chabrier, Allard, \&
  Hauschildt}]{baraffe98}
Baraffe, I., Chabrier, G., Allard, F., \& Hauschildt, P.~H. 1998, A\&A, 337,
  403

\bibitem[{Barrado~y Navascu\'es {et~al.}(2001)Barrado~y Navascu\'es,
  Zapatero~Osorio, B\'ejar, Rebolo, Mart{\'i}n, Mundt, \&
  Bailer-Jones}]{barrado01}
Barrado~y Navascu\'es, D., Zapatero~Osorio, M.~R., B\'ejar, V. J.~S., {et~al.}
  2001, A\&A, 377, L9

\bibitem[{Beam\'in {et~al.}(2014)Beam\'in, Ivanov, Bayo, Mu\^zi\'c, Boffin,
  Allard, Homeier, Minniti, Gromadzki, Kurtev, Lodieu, Mart\'in, \&
  M\'endez}]{beamin14}
Beam\'in, J.~C., Ivanov, V.~D., Bayo, A., {et~al.} 2014, A\&A, 570L, 8

\bibitem[{B\'ejar {et~al.}(2008)B\'ejar, Zapatero~Osorio, P\'erez-Garrido,
  \'Alvarez, Mart{\'i}n, Rebolo, Vill\'o-P\'erez, \&
  D{\'}az-S\'anchez}]{bejar08}
B\'ejar, V. J.~S., Zapatero~Osorio, M.~R., P\'erez-Garrido, A., {et~al.} 2008,
  ApJ, 673, L185

\bibitem[{B\'ejar {et~al.}(2011)B\'ejar, Zapatero~Osorio, Rebolo, Caballero,
  Barrado~y Navascu\'es, Mart\'in, Mundt, \& Bailer-Jones}]{bejar11}
B\'ejar, V. J.~S., Zapatero~Osorio, M.~R., Rebolo, R., {et~al.} 2011, ApJ, 743,
  64

\bibitem[{Best {et~al.}(2015)Best, Liu, Magnier, Deacon, Aller, Redstone,
  Burgett, Chambers, \& et~al.}]{best15}
Best, W. M.~J., Liu, M.~C., Magnier, E.~A., {et~al.} 2015, ApJ, 814, 118

\bibitem[{Bihain {et~al.}(2010)Bihain, Rebolo, Zapatero~Osorio, B\'ejar, \&
  Caballero}]{bihain10}
Bihain, G., Rebolo, R., Zapatero~Osorio, M.~R., B\'ejar, V. J.~S., \&
  Caballero, J.~A. 2010, A\&A, 519, 93

\bibitem[{Bonnefoy {et~al.}(2014a)Bonnefoy, Chauvin, Lagrange, Rojo, Allard,
  Pinte, Dumas, \& Homeier}]{bonnefoy14a}
Bonnefoy, M., Chauvin, G., Lagrange, A.-M., {et~al.} 2014a, A\&A, 562, A127

\bibitem[{Brice\~no {et~al.}(2002)Brice\~no, Luhman, Hartmann, Stauffer, \&
  Kirkpatrick}]{briceno02}
Brice\~no, C., Luhman, K.~L., Hartmann, L., Stauffer, J.~R., \& Kirkpatrick,
  J.~D. 2002, ApJ, 580, 317

\bibitem[{Burgasser {et~al.}(2003)Burgasser, Kirkpatrick, Liebert, \&
  Burrows}]{burgasser03}
Burgasser, A.~J., Kirkpatrick, J.~D., Liebert, J., \& Burrows, A. 2003, ApJ,
  594, 510

\bibitem[{Burgasser {et~al.}(2004)Burgasser, Kirkpatrick, McGovern, McLean,
  Prato, \& Reid}]{burgasser04}
Burgasser, A.~J., Kirkpatrick, J.~D., McGovern, M.~R., {et~al.} 2004, ApJ, 604,
  827

\bibitem[{Burrows {et~al.}(2003)Burrows, Sudarsky, \& Lunine}]{burrows03}
Burrows, A., Sudarsky, D., \& Lunine, J.~I. 2003, ApJ, 596, 587

\bibitem[{Caballero(2008)}]{caballero08}
Caballero, J.~A. 2008, A\&A, 478, 667

\bibitem[{Canty {et~al.}(2013)Canty, Lucas, Roche, \& Pinfield}]{canty13}
Canty, J.~I., Lucas, P.~W., Roche, P.~F., \& Pinfield, D.~J. 2013, MNRAS, 435,
  2650

\bibitem[{Casali {et~al.}(2007)Casali, Adamson, Alves~de Oliveira, Almaini,
  Burch, Elliot, Folger, Foucaud, Hambly, \& et~al.}]{casali07}
Casali, M., Adamson, A., Alves~de Oliveira, C., {et~al.} 2007, A\&A, 467, 777

\bibitem[{Cepa(1998)}]{cepa98}
Cepa, J. 1998, Ap\&SS, 263, 369

\bibitem[{Chabrier {et~al.}(2000)Chabrier, Baraffe, Allard, \&
  Hauschildt}]{chabrier00}
Chabrier, G., Baraffe, I., Allard, F., \& Hauschildt, P.~H. 2000, \apj, 542,
  464

\bibitem[{Chauvin {et~al.}(2004)Chauvin, Lagrange, Dumas, Zuckerman, Mouillet,
  Song, Beuzit, \& Lowrance}]{chauvin04}
Chauvin, G., Lagrange, A.-M., Dumas, C., {et~al.} 2004, A\&A, 425, L29

\bibitem[{Chauvin {et~al.}(2005a)Chauvin, Lagrange, Dumas, Zuckerman, Mouillet,
  Song, Beuzit, \& Lowrance}]{chauvin05a}
Chauvin, G., Lagrange, A.-M., Dumas, C., {et~al.} 2005a, A\&A, 438, L25

\bibitem[{Chiang \& Chen(2015)}]{chiang15}
Chiang, P. \& Chen, W.~P. 2015, ApJ, 811, L16

\bibitem[{Cross {et~al.}(2012)Cross, Collins, Mann, Read, Sutorius, Blake,
  Holliman, Hambly, Emerson, Lawrence, \& Noddle}]{cross12}
Cross, N. J.~G., Collins, R.~S., Mann, R.~G., {et~al.} 2012, A\&A, 548, A119

\bibitem[{Cruz {et~al.}(2009)Cruz, Kirkpatrick, \& Burgasser}]{cruz09}
Cruz, K.~L., Kirkpatrick, J.~D., \& Burgasser, A.~J. 2009, AJ, 137, 3345

\bibitem[{Cruz {et~al.}(2003)Cruz, Reid, Liebert, Kirkpatrick, \&
  Lowrance}]{cruz03}
Cruz, K.~L., Reid, I.~N., Liebert, J., Kirkpatrick, J.~D., \& Lowrance, P.~J.
  2003, AJ, 126, 2421

\bibitem[{Cushing {et~al.}(2011)Cushing, Kirkpatrick, Gelino, Griffith,
  Skrutskie, Mainzer, Marsh, Beichman, \& et~al.}]{cushing11}
Cushing, M.~C., Kirkpatrick, J.~D., Gelino, C.~R., {et~al.} 2011, ApJ, 743, 50

\bibitem[{Cushing {et~al.}(2005)Cushing, Rayner, \& Vacca}]{cushing05}
Cushing, M.~C., Rayner, J.~T., \& Vacca, W.~D. 2005, ApJ, 623, 1115

\bibitem[{Dahn {et~al.}(2002)Dahn, Harris, Vrba, Guetter, Canzian, Henden,
  Levine, Luginbuhl, Monet, Monet, Pier, Stone, Walker, Burgasser, Gizis,
  Kirkpatrick, Liebert, \& Reid}]{dahn02}
Dahn, C.~C., Harris, H.~C., Vrba, F.~J., {et~al.} 2002, AJ, 124, 1170

\bibitem[{Deacon {et~al.}(2016)Deacon, Schlieder, \& Murphy}]{deacon16}
Deacon, N.~R., Schlieder, J.~E., \& Murphy, S.~J. 2016, MNRAS, 457, 3191

\bibitem[{Delfosse {et~al.}(1997)Delfosse, Tinney, Forveille, \&
  et~al.}]{delfosse97}
Delfosse, X., Tinney, C., Forveille, T., \& et~al. 1997, ApJ, 327, L25

\bibitem[{Faherty {et~al.}(2014)Faherty, Beletsky, Burgasser, Tinney, Osip,
  Filippazzo, \& Simcoe}]{faherty14}
Faherty, J.~K., Beletsky, Y., Burgasser, A.~J., {et~al.} 2014, ApJ, 790, 90

\bibitem[{Faherty {et~al.}(2012)Faherty, Burgasser, Walter, Van~de Bliek,
  Shara, Cruz, West, Vrba, \& Anglada-Escud\'e}]{faherty12}
Faherty, J.~K., Burgasser, A.~J., Walter, F.~M., {et~al.} 2012, ApJ, 752, 56

\bibitem[{Faherty {et~al.}(2013)Faherty, Rice, Cruz, Mamajek, \&
  N\'u\~nez}]{faherty13}
Faherty, J.~K., Rice, E.~L., Cruz, K.~L., Mamajek, E.~E., \& N\'u\~nez, A.
  2013, AJ, 145, 2

\bibitem[{Faherty {et~al.}(2016)Faherty, Riedel, Cruz, Gagne, Filippazzo,
  Lambrides, Fica, Weinberger, Thorstense, Tinney, Baldassare, Lemonier, \&
  Rice}]{faherty16}
Faherty, J.~K., Riedel, A.~R., Cruz, K.~L., {et~al.} 2016, ApJS, 225, 10

\bibitem[{Filippazzo {et~al.}(2015)Filippazzo, Rice, Faherty, Cruz, Van~Gordon,
  \& Looper}]{filippazzo15}
Filippazzo, J.~C., Rice, E.~L., Faherty, J., {et~al.} 2015, ApJ, 810, 158

\bibitem[{Filippenko \& Greenstein(1984)}]{filippenko84}
Filippenko, A., V. \& Greenstein, J.~L. 1984, PASP, 96, 530

\bibitem[{Gagn\'e {et~al.}(2015b)Gagn\'e, Burgasser, Faherty, Lafreni\'ere,
  Doyon, Filippazzo, Bowsher, \& Nicholls}]{gagne15b}
Gagn\'e, J., Burgasser, A.~K., Faherty, J.~K., {et~al.} 2015b, ApJ, 808, L20

\bibitem[{Gagn\'e {et~al.}(2014c)Gagn\'e, Faherty, Cruz, Lefreni\`ere, Doyon,
  Malo, \& Artigau}]{gagne14c}
Gagn\'e, J., Faherty, J.~K., Cruz, K., {et~al.} 2014c, ApJ, 785, 14

\bibitem[{Gagn\'e {et~al.}(2015a)Gagn\'e, Faherty, Cruz, Lafreni\'ere, Doyon,
  Malo, Burgasser, Naud, Artigau, Bouchard, \& et~al..}]{gagne15a}
Gagn\'e, J., Faherty, J.~K., Cruz, K.~L., {et~al.} 2015a, ApJS, 219, 33

\bibitem[{Gagn\'e {et~al.}(2017)Gagn\'e, Faherty, Mamajek, Malo, Doyon,
  Filippazzo, Weinberger, Donaldson, L\'epine, Lafren\`ere, \&
  et~al.}]{gagne17}
Gagn\'e, J., Faherty, J.~K., Mamajek, E., {et~al.} 2017, ApJS, 228, 18

\bibitem[{Gagn\'e {et~al.}(2014a)Gagn\'e, Lafreni\`ere, Doyon, Artigau, Malo,
  Robert, \& Nadeau}]{gagne14a}
Gagn\'e, J., Lafreni\`ere, D., Doyon, R., {et~al.} 2014a, ApJ, 792, L17

\bibitem[{Gagn\'e {et~al.}(2014b)Gagn\'e, Lafreni\`ere, Doyon, Malo, \&
  Artigau}]{gagne14b}
Gagn\'e, J., Lafreni\`ere, D., Doyon, R., Malo, L., \& Artigau, E. 2014b, ApJ,
  783, 121

\bibitem[{Gauza {et~al.}(2015)Gauza, B\'ejar, P\'erez~Garrido, Zapatero~Osorio,
  Lodieu, Rebolo, Pall\'e, \& Nowak}]{gauza15}
Gauza, B., B\'ejar, V. J.~S., P\'erez~Garrido, A., {et~al.} 2015, ApJ, 804, 96

\bibitem[{Geballe {et~al.}(2002)Geballe, Knapp, Leggett, Fan, Golimowski,
  Anderson, Brinkmann, Csabai, Gunn, Hawley, \& et~al.}]{geballe02}
Geballe, T.~R., Knapp, G.~R., Leggett, S.~K., {et~al.} 2002, ApJ, 564, 466

\bibitem[{Gizis {et~al.}(2015)Gizis, Allers, Liu, Harris, Faherty, Burgasser,
  \& Kirkpatrick}]{gizis15}
Gizis, J.~E., Allers, K.~N., Liu, M.~C., {et~al.} 2015, ApJ, 799, 203

\bibitem[{Goldman {et~al.}(2010)Goldman, Marsat, Henning, Clemens, \&
  Greiner}]{goldman10}
Goldman, B., Marsat, S., Henning, T., Clemens, C., \& Greiner, J. 2010, MNRAS,
  405, 1140

\bibitem[{Golimowski {et~al.}(2004)Golimowski, Leggett, Marley, Fan, Geballe,
  Knapp, Vrba, Henden, Luginbuhl, Guetter, Munn, Canzian, Zheng, Tsvetanov,
  Chiu, Glazebrook, Hoversten, Schneider, \& Brinkmann}]{golimowski04}
Golimowski, D.~A., Leggett, S.~K., Marley, M.~S., {et~al.} 2004, AJ, 127, 3516

\bibitem[{Gonz\'alez~Hern\'andez {et~al.}(2008)Gonz\'alez~Hern\'andez,
  Caballero, Rebolo, B\'ejar, Barrado~y Navascu\'es, Mart\'in, \&
  Zapatero~Osorio}]{gonzalez08}
Gonz\'alez~Hern\'andez, J.~I., Caballero, J.~A., Rebolo, R., {et~al.} 2008,
  A\&A, 490, 1135

\bibitem[{Gorlova {et~al.}(2003)Gorlova, Meyer, Rieke, \& Liebert}]{gorlova03}
Gorlova, N.~I., Meyer, M.~R., Rieke, G.~H., \& Liebert, J. 2003, ApJ, 593, 1074

\bibitem[{Hambly {et~al.}(2008)Hambly, Collins, Cross, Mann, Read, Sutorius,
  Bond, Bryant, Emerson, Lawrence, \& et~al.}]{hambly08}
Hambly, N.~C., Collins, R.~S., Cross, N. J.~G., {et~al.} 2008, MNRAS, 384, 637

\bibitem[{Hauschildt {et~al.}(1997)Hauschildt, Baron, \& Allard}]{hauschildt97}
Hauschildt, P.~H., Baron, E., \& Allard, F. 1997, ApJ, 483, 390

\bibitem[{Hewett {et~al.}(2006)Hewett, Warren, Leggett, \& Hodgkin}]{hewett06}
Hewett, P.~C., Warren, S.~J., Leggett, S.~K., \& Hodgkin, S.~T. 2006, \mnras,
  367, 454

\bibitem[{Hummel {et~al.}(2013)Hummel, Rivinius, Nieva, Stahl, van Belle, \&
  Zavala}]{hummel13}
Hummel, C.~A., Rivinius, T., Nieva, M.-F., {et~al.} 2013, A\&A, 554, A52

\bibitem[{Jeffries {et~al.}(2006)Jeffries, Maxted, Oliveira, \&
  Naylor}]{jeffries06}
Jeffries, R.~D., Maxted, P. F.~L., Oliveira, J.~., \& Naylor, T. 2006, MNRAS,
  371, L6

\bibitem[{Kellogg {et~al.}(2016)Kellogg, Metchev, Gagn\'e, \&
  Faherty}]{kellogg16}
Kellogg, K., Metchev, S., Gagn\'e, J., \& Faherty, J. 2016, ApJL, 821, L15

\bibitem[{Kellogg {et~al.}(2015)Kellogg, Metchev, Geissler, Hicks, Kirkpatrick,
  \& Kurtev}]{kellogg15}
Kellogg, K., Metchev, S., Geissler, K., {et~al.} 2015, AJ, 150, 182

\bibitem[{King {et~al.}(2010)King, McCaughrean, Homeier, Allard, \&
  Lodieu}]{king10}
King, R.~R., McCaughrean, M.~J., Homeier, D., Allard, F., a. S. R.-D., \&
  Lodieu, N. 2010, A\&A, 510, 99

\bibitem[{Kirkpatrick(2005)}]{kirk05}
Kirkpatrick, J.~D. 2005, ARA\&A, 43, 195

\bibitem[{Kirkpatrick {et~al.}(2006)Kirkpatrick, Barman, Burgasser, McGovern,
  McLean, Tinney, \& Lowrance}]{kirk06}
Kirkpatrick, J.~D., Barman, T.~S., Burgasser, A.~J., {et~al.} 2006, ApJ, 639,
  1120

\bibitem[{Kirkpatrick {et~al.}(2008)Kirkpatrick, Cruz, Barman, Burgasser,
  Looper, Gelino, Lowrance, Liebert, Carpenter, Hillenbrand, \&
  Stauffer}]{kirk08}
Kirkpatrick, J.~D., Cruz, K.~L., Barman, T.~S., {et~al.} 2008, ApJ, 689, 1295

\bibitem[{Kirkpatrick {et~al.}(2010)Kirkpatrick, Looper, Burgasser, Schurr,
  Cutri, Cushing, Cruz, Sweet, \& et~al.}]{kirk10}
Kirkpatrick, J.~D., Looper, D.~L., Burgasser, A.~J., {et~al.} 2010, ApJS, 190,
  100

\bibitem[{Kirkpatrick {et~al.}(1999)Kirkpatrick, Reid, Liebert, Cutri, Nelson,
  Beichman, Dahn, Monet, Gizis, \& Skrutskie}]{kirk99}
Kirkpatrick, J.~D., Reid, I.~N., Liebert, J., {et~al.} 1999, ApJ, 519, 802

\bibitem[{Knapp {et~al.}(2004)Knapp, Leggett, Fan, Marley, Geballe, Golimowski,
  Finkbeiner, Gunn, Hennawi, Ivezi\'c, \& et~al.}]{knapp04}
Knapp, G.~R., Leggett, S.~K., Fan, X., {et~al.} 2004, AJ, 127, 3553

\bibitem[{Lafreni\`ere {et~al.}(2010)Lafreni\`ere, Jayawardhana, \& van
  Kerkwijk}]{lafreniere10}
Lafreni\`ere, D., Jayawardhana, R., \& van Kerkwijk, M.~H. 2010, ApJ, 719, 497

\bibitem[{Lagrange {et~al.}(2009)Lagrange, Gratadour, Chauvin, Fusco,
  Ehrenreich, Mouillet, Rousset, Rouan, \& et~al.}]{lagrange09}
Lagrange, A.-M., Gratadour, D., Chauvin, G., {et~al.} 2009, A\&A, 493, L21

\bibitem[{Lawrence {et~al.}(2007)Lawrence, Warren, Almaini, Edge, Hambly,
  Jameson, Lucas, Casali, Adamson, Dye, \& et~al.}]{lawrence07}
Lawrence, A., Warren, S.~J., Almaini, O., {et~al.} 2007, MNRAS, 379, 1599

\bibitem[{Lee(1968)}]{lee68}
Lee, T.~A. 1968, ApJ, 152, 913

\bibitem[{Leggett {et~al.}(2001)Leggett, Allard, Geballe, Hauschildt, \&
  Schweitzer}]{leggett01}
Leggett, S.~K., Allard, F., Geballe, T.~R., Hauschildt, P.~H., \& Schweitzer,
  A. 2001, ApJ, 548, 908

\bibitem[{Leggett {et~al.}(2010b)Leggett, Burningham, Saumon, Marley, Warren,
  Smart, Jones, Luchas, Pinfield, \& Tamura}]{leggett10b}
Leggett, S.~K., Burningham, B., Saumon, D., {et~al.} 2010b, ApJ, 710, 1627

\bibitem[{Leggett {et~al.}(2002)Leggett, Golimowski, Fan, Geballe, Knapp,
  Brinkmann, Csabai, Gunn, Hawley, Henry, Hindsley, Ivezic, Lupton, Pier,
  Schneider, Smith, Strauss, Uomoto, \& York}]{leggett02}
Leggett, S.~K., Golimowski, D., Fan, X., {et~al.} 2002, ApJ, 564, 452

\bibitem[{Leggett {et~al.}(2007)Leggett, Saumon, Marley, Geballe, Golimowski,
  Stephens, \& Fan}]{leggett07}
Leggett, S.~K., Saumon, D., Marley, M.~S., {et~al.} 2007, ApJ, 655, 1079

\bibitem[{Leggett {et~al.}(2010a)Leggett, Burningham, Saumon, Marley, Warren,
  Smart, Jones, Lucas, Pinfield, \& Tamura}]{leggett10a}
Leggett, S.~L., Burningham, B., Saumon, D., {et~al.} 2010a, ApJ, 710, 1627

\bibitem[{Liebert \& Burgasser(2007)}]{liebert07}
Liebert, J. \& Burgasser, A.~J. 2007, ApJ, 655, 522

\bibitem[{Liu {et~al.}(2016)Liu, Dupuy, \& Allers}]{liu16}
Liu, M.~C., Dupuy, T.~J., \& Allers, K.~N. 2016, ApJ, 833, 96

\bibitem[{Liu {et~al.}(2013)Liu, Magnier, Deacon, Allers, Dupuy, Kotson, Aller,
  Burgett, Chambers, Draper, \& et~al.}]{liu13}
Liu, M.~C., Magnier, E.~A., Deacon, N.~R., {et~al.} 2013, ApJ, 777, L20

\bibitem[{Lodieu {et~al.}(2008)Lodieu, Hambly, Jameson, \& Hodgkin}]{lodieu08}
Lodieu, N., Hambly, N.~C., Jameson, R.~F., \& Hodgkin, S.~T. 2008, MNRAS, 383,
  1385

\bibitem[{Lodieu {et~al.}(2007)Lodieu, Hambly, Jameson, Hodgkin, Carraro, \&
  Kendall}]{lodieu07}
Lodieu, N., Hambly, N.~C., Jameson, R.~F., {et~al.} 2007, MNRAS, 374, 372

\bibitem[{Lodieu {et~al.}(2015)Lodieu, Zapatero~Osorio, Rebolo, B\'ejar,
  Pavlenko, \& P\'erez~Garrido}]{lodieu15}
Lodieu, N., Zapatero~Osorio, M.~R., Rebolo, R., {et~al.} 2015, A\&A, 581, 73

\bibitem[{Lodieu {et~al.}(2009)Lodieu, Zapatero~Osorio, Rebolo, Mart\'in, \&
  Hambly}]{lodieu09}
Lodieu, N., Zapatero~Osorio, M.~R., Rebolo, R., Mart\'in, E.~L., \& Hambly,
  N.~C. 2009, A\&A, 505, 1115

\bibitem[{Looper {et~al.}(2008)Looper, Kirkpatrick, Cutri, barman, Burgasser,
  Cushing, Roellig, McGovern, McLean, Rice, Swift, \& Schurr}]{looper08}
Looper, D.~L., Kirkpatrick, J.~D., Cutri, R.~M., {et~al.} 2008, ApJ, 686, 528

\bibitem[{Lucas {et~al.}(2001)Lucas, Roche, Allard, \& Hauschildt}]{lucas01}
Lucas, P.~W., Roche, P.~F., Allard, F., \& Hauschildt, P.~H. 2001, MNRAS, 326,
  695

\bibitem[{Luhman \& Esplin(2016)}]{luhman16}
Luhman, K. \& Esplin, T.~L. 2016, AJ, in press

\bibitem[{Luhman(1999)}]{luhman99}
Luhman, K.~L. 1999, ApJ, 525, 466

\bibitem[{Luhman(2012)}]{luhman12}
Luhman, K.~L. 2012, ARA\&A, 50, 65

\bibitem[{Luhman(2014)}]{luhman14}
Luhman, K.~L. 2014, ApJ, 786, L18

\bibitem[{Luhman {et~al.}(2007a)Luhman, Allers, Jaffe, Cushing, Williams,
  Slesnick, \& Vacca}]{luhman07a}
Luhman, K.~L., Allers, K.~N., Jaffe, D.~T., {et~al.} 2007a, ApJ, 659, 1629

\bibitem[{Luhman {et~al.}(2008)Luhman, Hern\'andez, Downes, Hartmann, \&
  Brice\~no}]{luhman08}
Luhman, K.~L., Hern\'andez, J., Downes, J.~J., Hartmann, L., \& Brice\~no, C.
  2008, ApJ, 688, 362

\bibitem[{Luhman {et~al.}(2007b)Luhman, Patten, Marengo, Schuster, Hora, Ellis,
  Stauffer, Sonnett, Winston, \& et~al.}]{luhman07b}
Luhman, K.~L., Patten, B.~M., Marengo, M., {et~al.} 2007b, ApJ, 654, 570

\bibitem[{Luhman {et~al.}(2003)Luhman, Stauffer, Muench, Rieke, Lada, Bouvier,
  \& Lada}]{luhman03}
Luhman, K.~L., Stauffer, J.~R., Muench, A.~A., {et~al.} 2003, ApJ, 593, 1093

\bibitem[{Macintosh {et~al.}(2015)Macintosh, Graham, Barman, De~Rosa,
  Konopacky, Marley, Marois, Nielsen, Pueyo, Rajan, Rameau, Saumon, Wang,
  Patience, Ammons, Arriaga, Artigau, Beckwith, Brewster, Bruzzone, Bulger,
  Burningham, Burrows, Chen, Chiang, Chilcote, Dawson, Dong, Doyon, Draper,
  Duchêne, Esposito, Fabrycky, Fitzgerald, Follette, Fortney, Gerard,
  Goodsell, Greenbaum, Hibon, Hinkley, Cotten, Hung, Ingraham, Johnson-Groh,
  Kalas, Lafreniere, Larkin, Lee, Line, Long, Maire, Marchis, Matthews, Max,
  Metchev, Millar-Blanchaer, Mittal, Morley, Morzinski, Murray-Clay,
  Oppenheimer, Palmer, Patel, Perrin, Poyneer, Rafikov, Rantakyrö, Rice, Rojo,
  Rudy, Ruffio, Ruiz, Sadakuni, Saddlemyer, Salama, Savransky, Schneider,
  Sivaramakrishnan, Song, Soummer, Thomas, Vasisht, Wallace, Ward-Duong,
  Wiktorowicz, Wolff, \& Zuckerman}]{macintosh15}
Macintosh, B., Graham, J.~R., Barman, T., {et~al.} 2015, Science, 350, 64

\bibitem[{Manjavacas {et~al.}(2015)Manjavacas, Goldman, Alcal\'a,
  Zapatero~Osorio, B\'ejar, Homeier, Bonnefoy, Smart, Henning, \&
  Allard}]{manjavacas15}
Manjavacas, E., Goldman, B., Alcal\'a, J.~M., {et~al.} 2015, MNRAS, 455, 1341

\bibitem[{Marocco {et~al.}(2014a)Marocco, Day-Jones, Lucas, Jones, Smart,
  Zhang, Gomes, Burningham, Pinfield, Raddi, \& Smith}]{marocco14a}
Marocco, F., Day-Jones, A.~C., Lucas, P.~W., {et~al.} 2014a, MNRAS, 439, 372

\bibitem[{Marocco {et~al.}(2014b)Marocco, Jones, Day-Jones, Pinfield, Lucas,
  Burningham, Zang, Smart, Gomes, \& Smith}]{marocco14b}
Marocco, F., Jones, H. R.~A., Day-Jones, A.~C., {et~al.} 2014b, MNRAS, 449, 365

\bibitem[{Marois {et~al.}(2008)Marois, Macintosh, Barman, Zuckerman, Song,
  Patience, Lafreni\`ere, \& Doyon}]{marois08}
Marois, C., Macintosh, B., Barman, T., {et~al.} 2008, Science, 322, 1348

\bibitem[{Marois {et~al.}(2010)Marois, Zuckerman, Konopacky, Macintosh, \&
  Barman}]{marois10}
Marois, C., Zuckerman, B., Konopacky, Q.~M., Macintosh, B., \& Barman, T. 2010,
  Nature, 468, 1080

\bibitem[{Mart{\'i}n {et~al.}(2006)Mart{\'i}n, Brandner, Bouy, Basri, Davis,
  Deshpande, \& Montgomery}]{martin06}
Mart{\'i}n, E.~L., Brandner, W., Bouy, H., {et~al.} 2006, A\&A, 456, 253

\bibitem[{Mart{\'i}n {et~al.}(1999)Mart{\'i}n, Delfosse, Basri, Goldman,
  Forveille, \& Zapatero~Osorio}]{martin99}
Mart{\'i}n, E.~L., Delfosse, X., Basri, G., {et~al.} 1999, AJ, 118, 2466

\bibitem[{Mart\'in {et~al.}(2001)Mart\'in, Zapatero~Osorio, Barrado~y
  Navascu\'es, B\'ejar, \& rebolo}]{martin01}
Mart\'in, E.~L., Zapatero~Osorio, M.~R., Barrado~y Navascu\'es, D., B\'ejar, V.
  J.~S., \& rebolo, R. 2001, ApJ, 558, L117

\bibitem[{Massey \& Gronwall(1990)}]{massey90}
Massey, P. \& Gronwall, C. 1990, ApJ, 358, 344

\bibitem[{McCaughrean {et~al.}(2004)McCaughrean, Close, Scholz, Lenzen, Biller,
  Brandner, Hartung, \& Lodieu}]{mccaughrean04}
McCaughrean, M.~J., Close, L.~M., Scholz, R.-D., {et~al.} 2004, A\&A, 413, 1029

\bibitem[{McGovern {et~al.}(2004)McGovern, Kirkpatrick, McLean, Burgasser,
  Prato, \& Lowrance}]{mcgovern04}
McGovern, M.~R., Kirkpatrick, J.~D., McLean, I.~S., {et~al.} 2004, ApJ, 600,
  1020

\bibitem[{McLean {et~al.}(2003)McLean, McGovern, Burgasser, Kirkpatrick, Prato,
  \& Kim}]{mclean03}
McLean, I.~S., McGovern, M.~R., Burgasser, A.~J., {et~al.} 2003, ApJ, 596, 561

\bibitem[{McLean {et~al.}(2007)McLean, Prato, McGovern, Burgasser, Kirkpatrick,
  Rice, \& Kim}]{mclean07}
McLean, I.~S., Prato, L., McGovern, M.~R., {et~al.} 2007, ApJ, 658, 1217

\bibitem[{Moorwood {et~al.}(1998)Moorwood, Cuby, Biereichel, Brynnel, Delabre,
  Devillard, van Dijsseldonk, Finger, Gemperlein, Gilmozzi, Herlin, Huster,
  Knudstrup, Lidman, Lizon, Mehrgan, Meyer, Nicolini, Petr, Spyromilio, \&
  Stegmeier}]{moorwood98}
Moorwood, A., Cuby, J.-G., Biereichel, P., {et~al.} 1998, The Messenger, 94, 7

\bibitem[{Naud {et~al.}(2014)Naud, Artigau, Malo, Albert, Doyon, Lafreni\`ere,
  Gagn\`e, Saumon, Morley, Allard, Homeier, \& et~al.}]{naud14}
Naud, M.-E., Artigau, E., Malo, L., {et~al.} 2014, ApJ, 787, 5

\bibitem[{Patience {et~al.}(2012)Patience, King, De~Rosa, Vigan, Witte, Rice,
  Helling, \& Hauschildt}]{patience12}
Patience, K., King, R.~R., De~Rosa, R.~J., {et~al.} 2012, A\&A, 540, 85

\bibitem[{Patten {et~al.}(2006)Patten, Stauffer, Burrows, Marengo, Hora,
  Luhman, Sonnett, Henry, \& et~al.}]{patten06}
Patten, B.~M., Stauffer, J.~R., Burrows, A., {et~al.} 2006, ApJ, 651, 502

\bibitem[{Pavlenko {et~al.}(2000)Pavlenko, Zapatero~Osorio, \&
  Rebolo}]{pavlenko00}
Pavlenko, Y., Zapatero~Osorio, M.~R., \& Rebolo, R. 2000, A\&A, 355, 245

\bibitem[{Pe\~na Ram\'irez {et~al.}(2016)Pe\~na Ram\'irez, B\'ejar, \&
  Zapatero~Osorio}]{pena16}
Pe\~na Ram\'irez, K., B\'ejar, V. J.~S., \& Zapatero~Osorio, M.~R. 2016, A\&A,
  586, A157

\bibitem[{Pe\~na Ram{\'i}rez {et~al.}(2012)Pe\~na Ram{\'i}rez, B\'ejar,
  Zapatero~Osorio, Petr-Gotzens, \& Mart{\'i}n}]{pena12}
Pe\~na Ram{\'i}rez, K., B\'ejar, V. J.~S., Zapatero~Osorio, M.~R.,
  Petr-Gotzens, M.~G., \& Mart{\'i}n, E.~L. 2012, ApJ, 754, 30

\bibitem[{Pe\~na Ram{\'i}rez {et~al.}(2015)Pe\~na Ram{\'i}rez, Zapatero~Osorio,
  \& B\'ejar}]{pena15}
Pe\~na Ram{\'i}rez, K., Zapatero~Osorio, M.~R., \& B\'ejar, V. J.~S. 2015,
  A\&A, 574, 118

\bibitem[{Pe\~na Ram\'irez {et~al.}(2011)Pe\~na Ram\'irez, Zapatero~Osorio,
  B\'ejar, Rebolo, \& Bihain}]{pena11}
Pe\~na Ram\'irez, K., Zapatero~Osorio, M.~R., B\'ejar, V. J.~S., Rebolo, R., \&
  Bihain, G. 2011, A\&A, 532, 42

\bibitem[{Perryman {et~al.}(1997)Perryman, Lindegren, Kovalevsky, Hoeg,
  Bastian, Bernacca, Cr\'ez\'e, Donati, Grenon, Grewing, van Leeuwen, van~der
  Marel, Mignard, Murray, Le~Poole, Schrijver, Turon, Arenou, Froeschlé, \&
  Petersen}]{perryman97}
Perryman, M. A.~C., Lindegren, L., Kovalevsky, J., {et~al.} 1997, A\&A, 323,
  L49

\bibitem[{Rebolo {et~al.}(1998)Rebolo, Zapatero~Osorio, Madruga, B\'ejar,
  Arribas, \& Licandro}]{rebolo98}
Rebolo, R., Zapatero~Osorio, M.~R., Madruga, S., {et~al.} 1998, Science, 282,
  1309

\bibitem[{Reid {et~al.}(2001)Reid, Burgasser, Cruz, Kirkpatrick, \&
  Gizis}]{reid01}
Reid, I.~N., Burgasser, A.~J., Cruz, K.~L., Kirkpatrick, J.~D., \& Gizis, J.~E.
  2001, AJ, 121, 1710

\bibitem[{Rice {et~al.}(2010)Rice, Barman, McLean, Prato, \&
  Kirkpatrick}]{rice10}
Rice, E.~L., Barman, T., McLean, I.~S., Prato, L., \& Kirkpatrick, J.~D. 2010,
  ApJS, 186, 63

\bibitem[{Rossow(1978)}]{rossow78}
Rossow, W.~B. 1978, Icarus, 36, 1

\bibitem[{Sacco {et~al.}(2007)Sacco, Randich, Franciosini, Pallavicini, \&
  Palla}]{sacco07}
Sacco, G.~G., Randich, S., Franciosini, E., Pallavicini, R., \& Palla, F. 2007,
  A\&A, 462, L23

\bibitem[{Saumon {et~al.}(1996)Saumon, Hubbard, Burrows, Guillot, Lunine, \&
  Chabrier}]{saumon96}
Saumon, D., Hubbard, W.~B., Burrows, A., {et~al.} 1996, ApJ, 460, 993

\bibitem[{Schaefer {et~al.}(2016)Schaefer, Hummel, Gies, Zavala, Monnier,
  Walter, Turner, Baron, ten Brummelaar, Che, Farrington, Kraus, Sturmann, \&
  Sturmann}]{schaefer16}
Schaefer, G.~H., Hummel, C.~A., Gies, D.~R., {et~al.} 2016, AJ, 152, 213

\bibitem[{Schmidt {et~al.}(2014)Schmidt, West, Bochanski, Hawley, \&
  Kielty}]{schmidt14}
Schmidt, S., J., West, A.~A., Bochanski, J.~J., Hawley, S.~L., \& Kielty, C.
  2014, PASP, 126, 642

\bibitem[{Schneider {et~al.}(2016b)Schneider, Cushing, Kirkpatrick, \&
  Gelino}]{schneider16b}
Schneider, A.~C., Cushing, M.~C., Kirkpatrick, J.~D., \& Gelino, C.~R. 2016b,
  ApJL, 823, L35

\bibitem[{Schneider {et~al.}(2014)Schneider, Cushing, Kirkpatrick, Mace,
  Gelino, Faherty, Fajardo-Acosta, \& Sheppard}]{schneider14}
Schneider, A.~C., Cushing, M.~C., Kirkpatrick, J.~D., {et~al.} 2014, AJ, 147,
  34

\bibitem[{Schneider {et~al.}(2016a)Schneider, Windsor, Cushing, Kirkpatrick, \&
  Wright}]{schneider16a}
Schneider, A.~C., Windsor, J., Cushing, M.~C., Kirkpatrick, J.~D., \& Wright,
  E.~L. 2016a, ApJL, 822, L1

\bibitem[{Scholz \& Jayawardhana(2008)}]{scholz08}
Scholz, A. \& Jayawardhana, R. 2008, ApJ, 672, L49

\bibitem[{Scholz {et~al.}(2003)Scholz, McCaughrean, Lodieu, \&
  Kuhlbrodt}]{scholz03}
Scholz, R.-D., McCaughrean, M.~J., Lodieu, N., \& Kuhlbrodt, B. 2003, A\&A,
  398, L29

\bibitem[{Sherry {et~al.}(2008)Sherry, Walter, Wolk, \& Adams}]{sherry08}
Sherry, W.~H., Walter, F.~M., Wolk, S.~J., \& Adams, N.~R. 2008, AJ, 135, 1616

\bibitem[{Simcoe {et~al.}(2008)Simcoe, Burgasser, Bernstein, \&
  et~al.}]{simcoe08}
Simcoe, R.~A., Burgasser, A.~J., Bernstein, R.~A., \& et~al. 2008, Proc. SPIE,
  7014, 70140U

\bibitem[{Simcoe {et~al.}(2010)Simcoe, Burgasser, Bochanski, \&
  et~al.}]{simcoe10}
Simcoe, R.~A., Burgasser, A.~J., Bochanski, J.~J., \& et~al. 2010, Proc. SPIE,
  7735, 773514

\bibitem[{Sim\'on-D\'iaz {et~al.}(2015)Sim\'on-D\'iaz, Caballero, Korenzo,
  Ma\'iz~Apell\'aniz, Schneider, Negueruela, Barb\'a, Dorda, Marco, Montes,
  Pellerin, S\'anchez-Berm\'udez, S\'odor, \& Sota}]{simon15}
Sim\'on-D\'iaz, S., Caballero, J.~A., Korenzo, J., {et~al.} 2015, ApJ, 799, 169

\bibitem[{Skrutskie {et~al.}(2006)Skrutskie, Cutri, Stiening, Weinberg,
  Schneider, Carpenter, Beichman, Capps, Chester, Huchra, Liebert, Lonsdale,
  Monet, Price, Seitzer, Jarrett, Kirkpatrick, Gizis, Howard, Evans, Fowler,
  Fullmer, Hurt, Light, Kopan, Marsh, McCallon, Tam, Van~Dyk, \&
  Wheelock}]{skrutskie06}
Skrutskie, M.~F., Cutri, R.~M., Stiening, R., {et~al.} 2006, AJ, 131, 1163

\bibitem[{Slesnick {et~al.}(2004)Slesnick, Hillenbrand, \&
  Carpenter}]{slesnick04}
Slesnick, C.~L., Hillenbrand, L.~A., \& Carpenter, J.~M. 2004, ApJ, 610, 1045

\bibitem[{Steele \& Jameson(1995)}]{steele95}
Steele, I.~A. \& Jameson, R.~F. 1995, MNRAS, 272, 630

\bibitem[{Stephens {et~al.}(2009)Stephens, Leggett, Cushing, Marley, Saumon,
  Geballe, Golimowski, Fan, \& Noll}]{stephens09}
Stephens, D.~C., Leggett, S.~K., Cushing, M.~C., {et~al.} 2009, ApJ, 702, 154

\bibitem[{Sumi {et~al.}(2011)Sumi, Kamiya, Bennett, Bond, Abe, Botzler, Fukui,
  Furusawa, Hearnshaw, Itow, Kilmartin, Korpela, Lin, Ling, Masuda, Matsubara,
  Miyake, Motomura, Muraki, Nagaya, Nakamura, Ohnishi, Okumura, Perrott,
  Rattenbury, Saito, Sako, Sullivan, Sweatman, Tristram, ;~Udalski, Szymanski,
  Kubiak, Pietrzyński, Poleski, Soszyński, Wyrzykowski, Ulaczyk, \&
  in~Astrophysics (MOA)~Collaboration}]{sumi11}
Sumi, T., Kamiya, K., Bennett, D.~P., {et~al.} 2011, Nature, 473, 349

\bibitem[{Tinney {et~al.}(1998)Tinney, Delfosse, Forveille, \&
  Allard}]{tinney98}
Tinney, C.~G., Delfosse, X., Forveille, T., \& Allard, F. 1998, A\&A, 338, 1066

\bibitem[{Tremblin {et~al.}(2016)Tremblin, Amundsen, Chabrier, Baraffe,
  Drummond, Hinkley, Mourier, \& Venot}]{tremblin16}
Tremblin, P., Amundsen, D.~S., Chabrier, G., {et~al.} 2016, ApJL, 817, L19

\bibitem[{Wright {et~al.}(2011)Wright, Mainzer, Gelino, \&
  Kirkpatrick}]{wright11}
Wright, E.~L., Mainzer, A., Gelino, C., \& Kirkpatrick, J.~D. 2011,
  arXiv:1104.2569

\bibitem[{Zakhozhay {et~al.}(2016)Zakhozhay, Zapatero~Osorio, \&
  B\'ejar}]{zakhozhay16}
Zakhozhay, O., Zapatero~Osorio, M.~R., \& B\'ejar, V. J.~S. 2016, MNRAS,
  submitted

\bibitem[{Zapatero~Osorio {et~al.}(2008)Zapatero~Osorio, B\'ejar, Bihain,
  Mart{\'i}n, Rebolo, Vill\'o-P\'erez, D{\'i}az-S\'anchez, P\'erez~Garrido,
  Caballero, Henning, Mundt, Barrado Y~Navascu\'es, \& Bailer-Jones}]{osorio08}
Zapatero~Osorio, M.~R., B\'ejar, V. J.~S., Bihain, G., {et~al.} 2008, A\&A,
  477, 895

\bibitem[{Zapatero~Osorio {et~al.}(2014{\natexlab{a}})Zapatero~Osorio, B\'ejar,
  Mart\'in, G\'alvez~Ortiz, Rebolo, Bihain, Henning, Boudreault, Goldman,
  Mundt, Caballero, \& Miles-P\'aez}]{osorio14b}
Zapatero~Osorio, M.~R., B\'ejar, V. J.~S., Mart\'in, E.~L., {et~al.}
  2014{\natexlab{a}}, A\&A, 572, 67

\bibitem[{Zapatero~Osorio {et~al.}(2000)Zapatero~Osorio, B\'ejar, Mart{\'i}n,
  Rebolo, Barrado~y Navascu\'es, Bailer-Jones, \& Mundt}]{osorio00}
Zapatero~Osorio, M.~R., B\'ejar, V. J.~S., Mart{\'i}n, E.~L., {et~al.} 2000,
  Science, 290, 103

\bibitem[{Zapatero~Osorio {et~al.}(2002)Zapatero~Osorio, B\'ejar, Mart{\'i}n,
  Rebolo, Barrado~y Navascu\'es, Mundt, Eislöffel, \& Caballero}]{osorio02b}
Zapatero~Osorio, M.~R., B\'ejar, V. J.~S., Mart{\'i}n, E.~L., {et~al.} 2002,
  \apj, 578, 536

\bibitem[{Zapatero~Osorio {et~al.}(2014{\natexlab{b}})Zapatero~Osorio, B\'ejar,
  Miles-P\'aez, Pe\~na Ram\'\i~rez, Rebolo, \& Pall\'e}]{osorio14c}
Zapatero~Osorio, M.~R., B\'ejar, V. J.~S., Miles-P\'aez, P.~A., {et~al.}
  2014{\natexlab{b}}, A\&A, 568, 6

\bibitem[{Zapatero~Osorio {et~al.}(2012a)Zapatero~Osorio, B\'ejar, Pavlenko,
  Rebolo, Allende~Prieto, Mart\'in, \& Garc\'ia~L\'opez}]{osorio02a}
Zapatero~Osorio, M.~R., B\'ejar, V. J.~S., Pavlenko, Y., {et~al.} 2012a, A\&A,
  384, 937

\bibitem[{Zapatero~Osorio {et~al.}(2014a)Zapatero~Osorio, G\'alvez~Ortiz,
  Bihain, Bailer-Jones, Rebolo, Henning, Boudreault, B\'ejar, Goldman, Mundt,
  \& Caballero}]{osorio14a}
Zapatero~Osorio, M.~R., G\'alvez~Ortiz, M.~C., Bihain, G., {et~al.} 2014a,
  A\&A, 568, 77

\bibitem[{Zapatero~Osorio {et~al.}(2016)Zapatero~Osorio, Lodieu, B\'ejar,
  Mart\'in, Ivanov, Bayo, Boffin, Mužić, Minniti, \& Beam\'in}]{osorio16}
Zapatero~Osorio, M.~R., Lodieu, N., B\'ejar, V. J.~S., {et~al.} 2016, A\&A,
  592, A80

\end{thebibliography}

\clearpage

\appendix

\section{S\,Ori\,70 \label{so70}}
\begin{figure}        
\epsscale{0.5}
\plotone{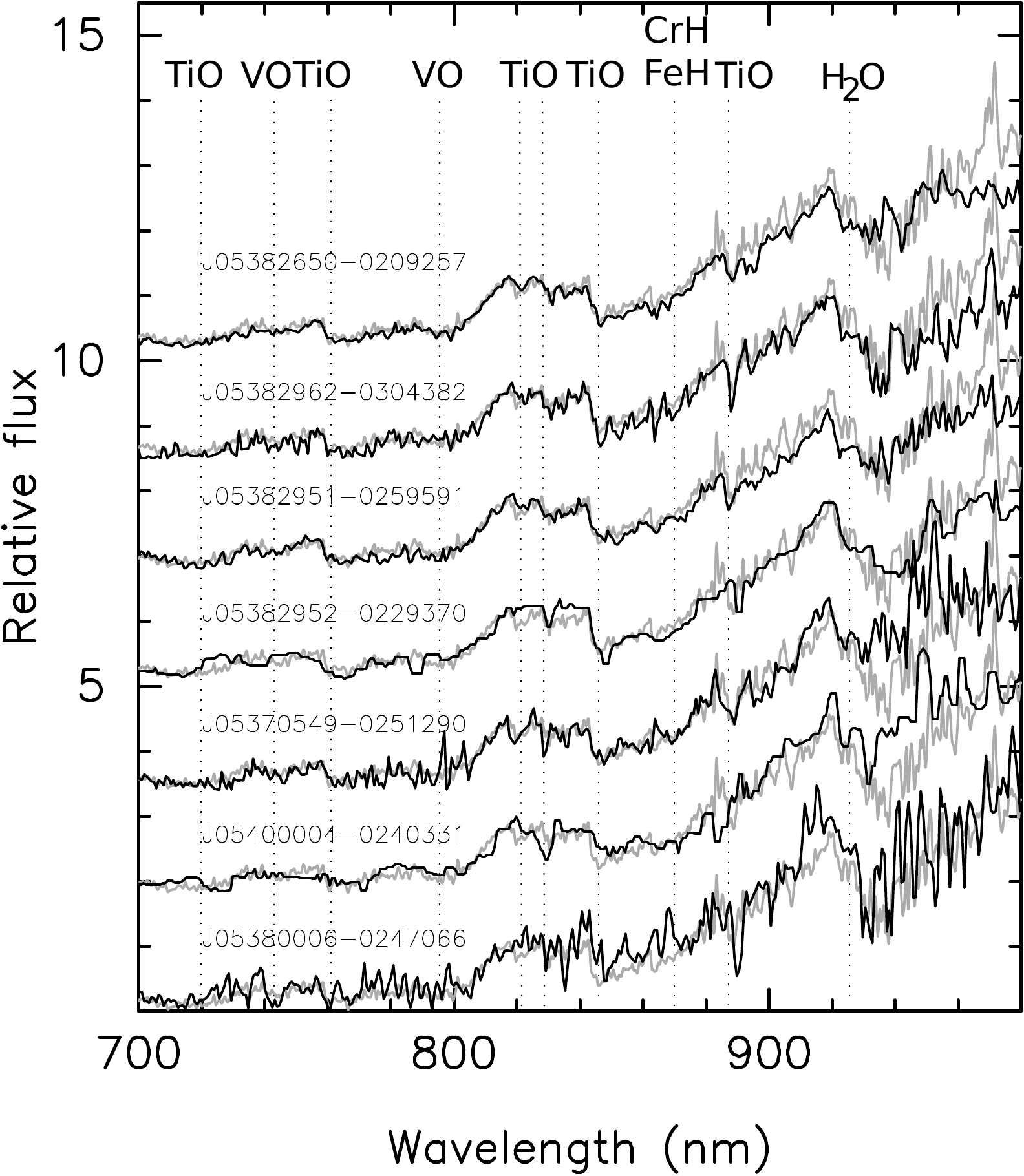}
\plotone{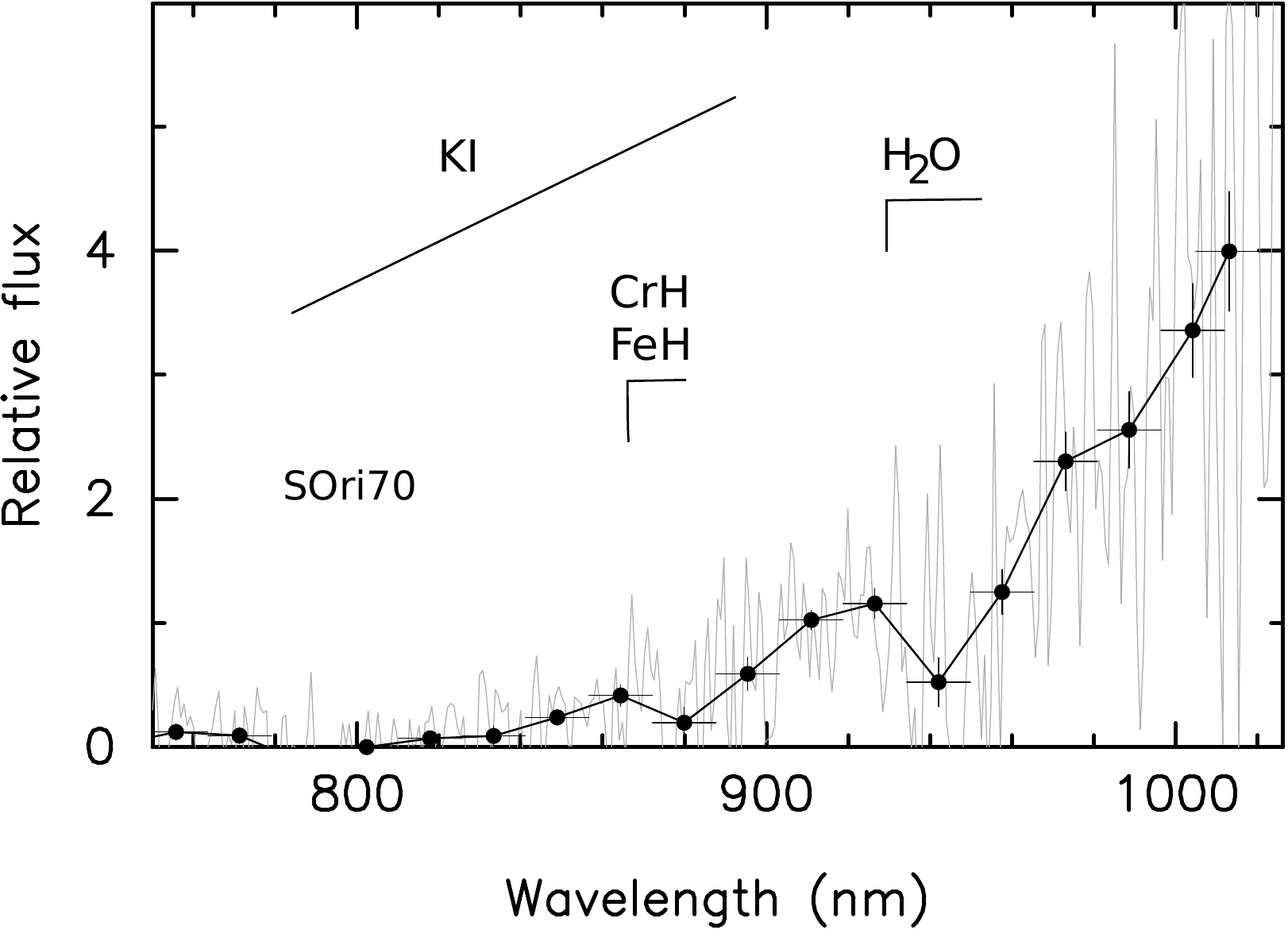}
\caption{{\sc osiris} optical spectra ($R \approx 250-310$ at 750 nm) free of telluric
absorption. {\sl (Top panel).} From top to bottom, the data are ordered by increasing $J$-band
magnitude. The optical spectrum (gray lines) of USco\,108\,B \citep{bejar08} is shown for
comparison. The data of J05400004$-$0240331 and J05382952$-$0229370 are median filtered with a 
box of 5 and 9 pixels. The spectrum of J05380006$-$0247066 corresponds to the average data 
acquired on two different occasions. All spectra are normalized to unity at 814--817.5 nm and are 
vertically offset for clarity. {\sl (Bottom panel).} The spectrum of S\,Ori\,70 is shown in gray 
and is normalized to unity at around 920 nm. The rebinned spectrum (with a boxcar of 20 pixels) is
depicted with  solid black circles, where the horizontal and vertical error bars account for the 
wavelength interval and flux errors of the rebinned flux values. Various molecular features are 
identified following \citet{martin99} and \citet{kirk05}.
\label{opt}}
\end{figure}

\begin{figure*}        
\epsscale{1}
\plotone{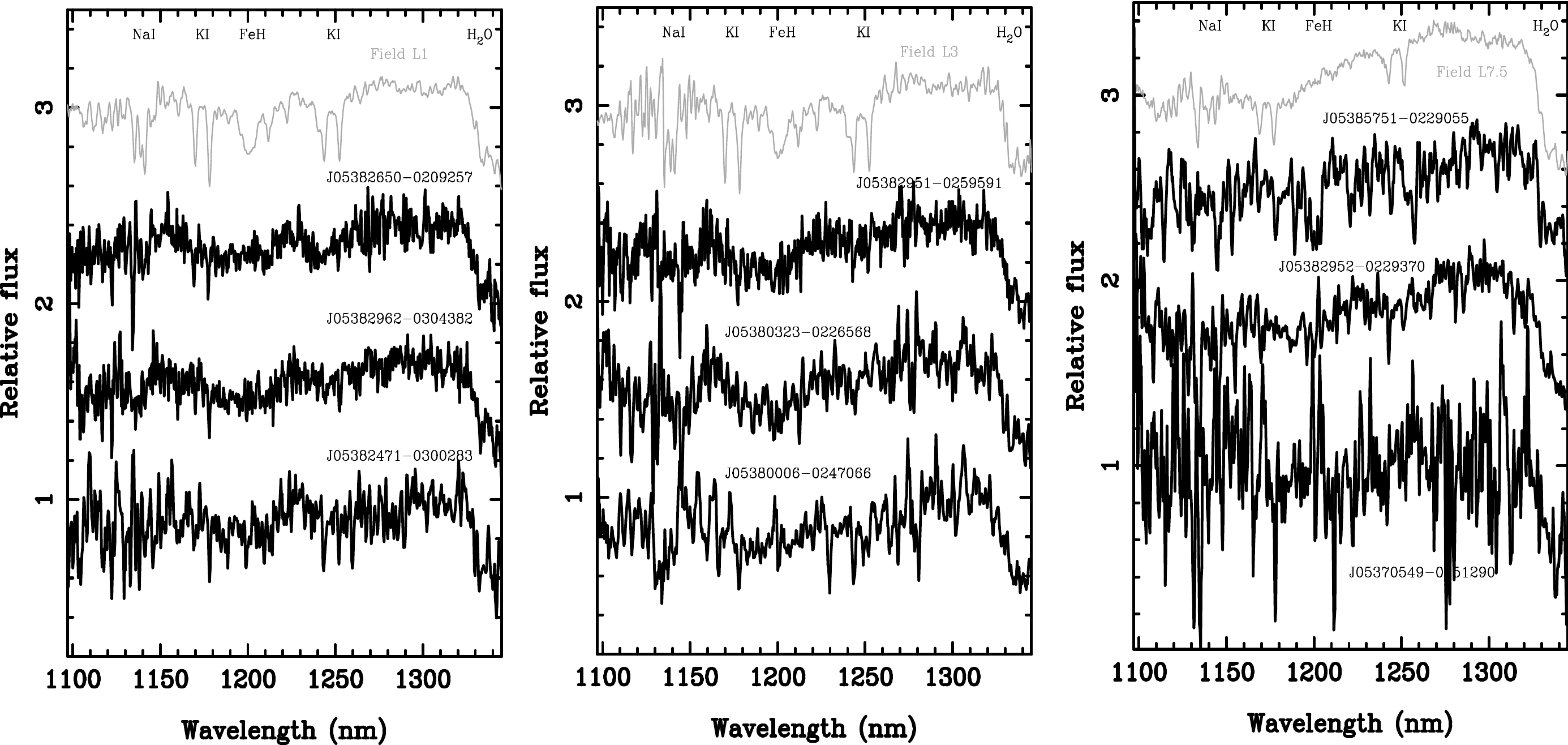}
\caption{{\sc isaac} spectra of \so~candidates (black) and comparison field dwarfs (gray, 
Table~\ref{obslog}) obtained with a resolution of $\sim$500 at 1.25 $\mu$m. All spectra are 
normalized to unity in the wavelength interval 1.28--1.32 $\mu$m and are offset for clarity. 
The data of the following objects are smoothed to increase the signal-to-noise ratio:
J05382471$-$0300283 (3 pix), J05382952$-$0229370 (3 pix), J05370549$-$0251290 (3 pix), 
J05380323$-$0226568 (3 pix), J05385751$-$0229055 (5 pix), and J05380006$-$0247066 (5 pix). 
Molecular and atomic features are identified according to \citet{cushing05}.
\label{nir}}
\end{figure*}

\begin{figure}        
\epsscale{0.8}
\plotone{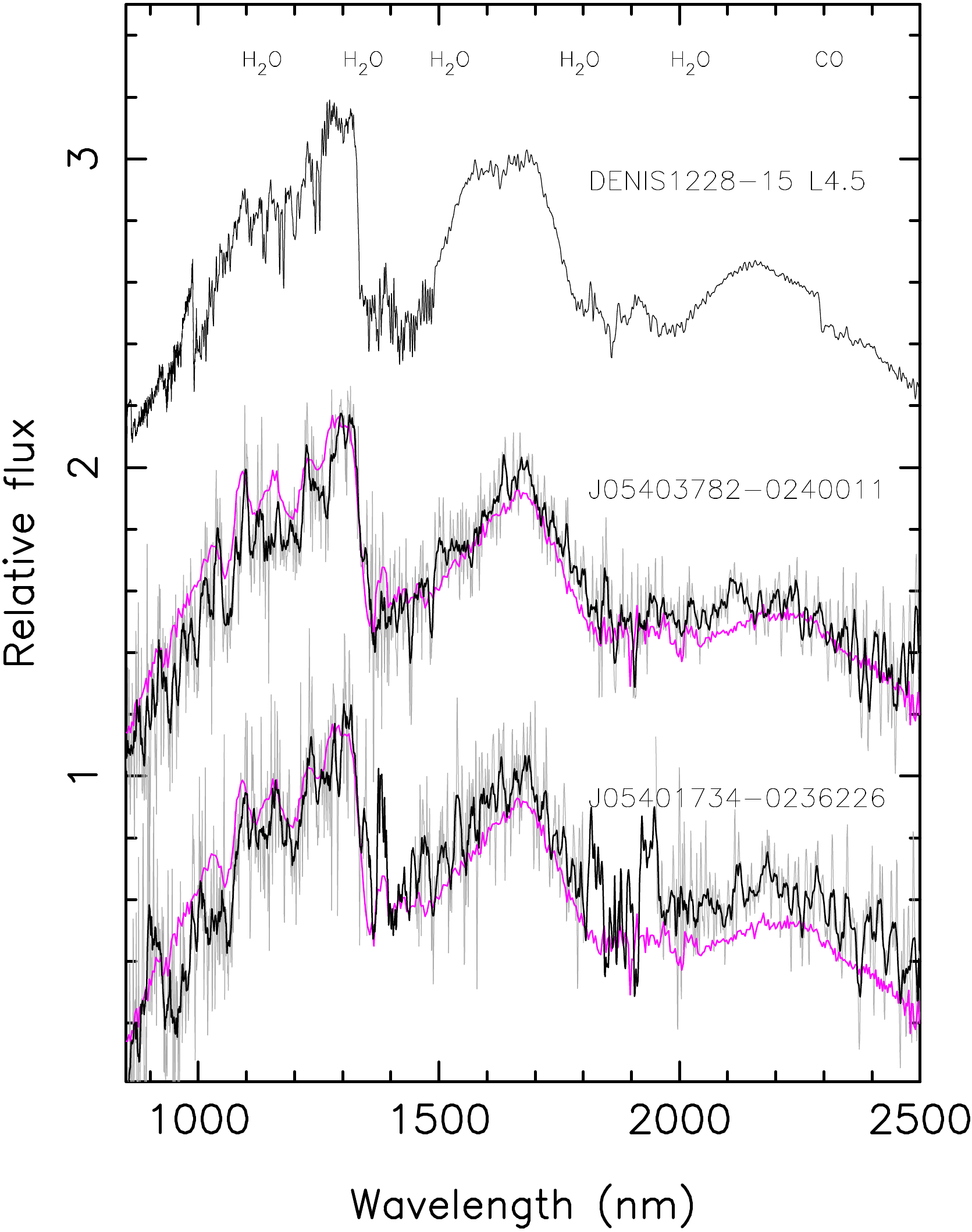}
\caption{The {\sc fire} spectra of two \so~candidates are shown with  gray (original) and 
black lines (smoothed by 9 pixels -- J054037382$-$0240011, and 11 pixels -- J05401734$-$0236266). 
For clarity, the regions of strong telluric absorption are removed from the original data of J05401734$-$0236266. 
For reference, the top spectrum corresponds to a field high-gravity L4.5 dwarf 
\citep{leggett01}. 2MASS J1207$-$3900, an L1 member of the TW Hydrae moving group \citep{gagne14c},
is depicted in magenta. This spectrum is normalized to the peak of the $J$-band of the \so~objects.
The remaining spectra are normalized to unity at the peak of the $H$-band and are offset for 
clarity. Some molecular species are identified. 
\label{fire}}
\end{figure}

\begin{figure*}        
\epsscale{1}
\plottwo{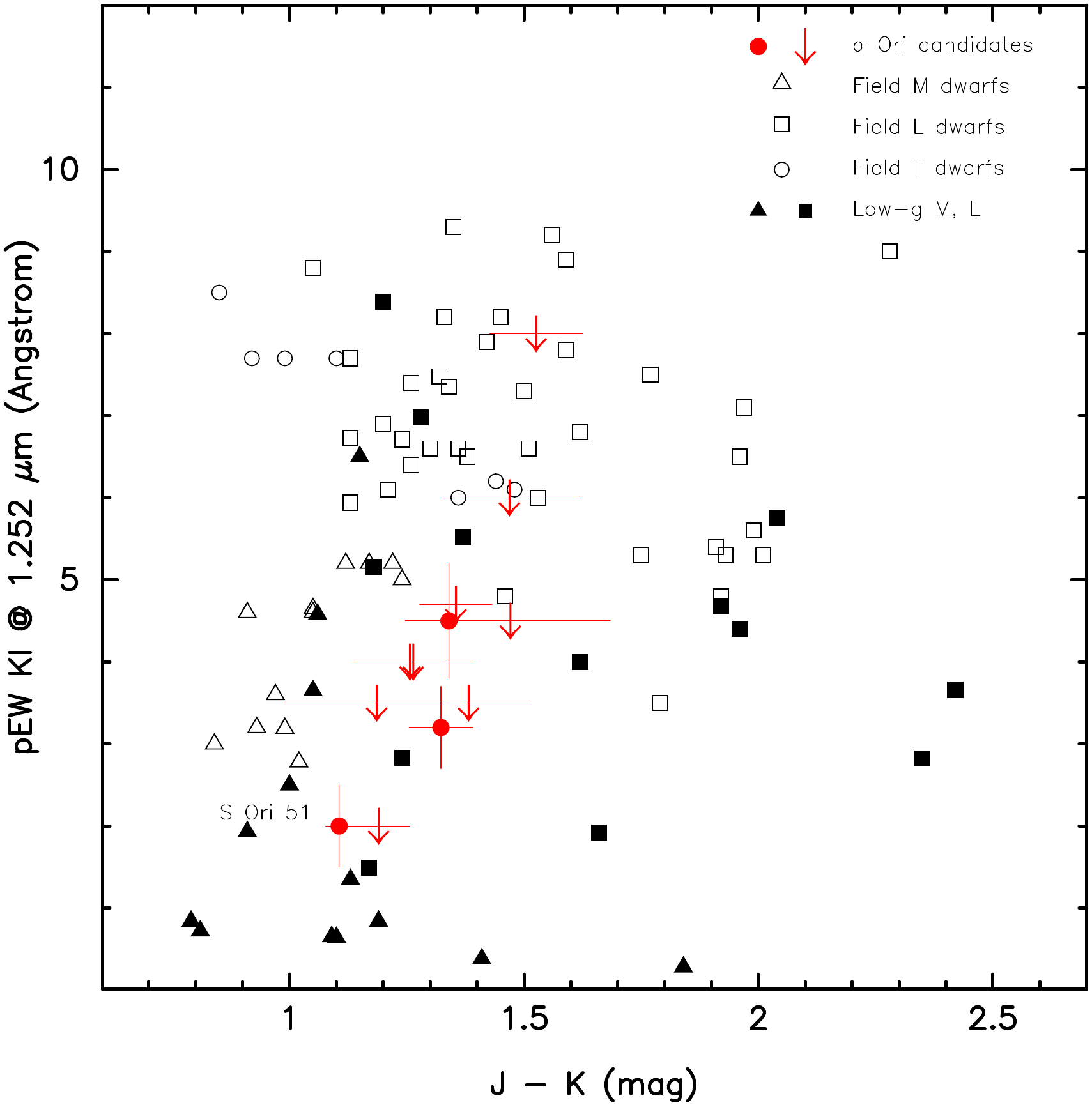}{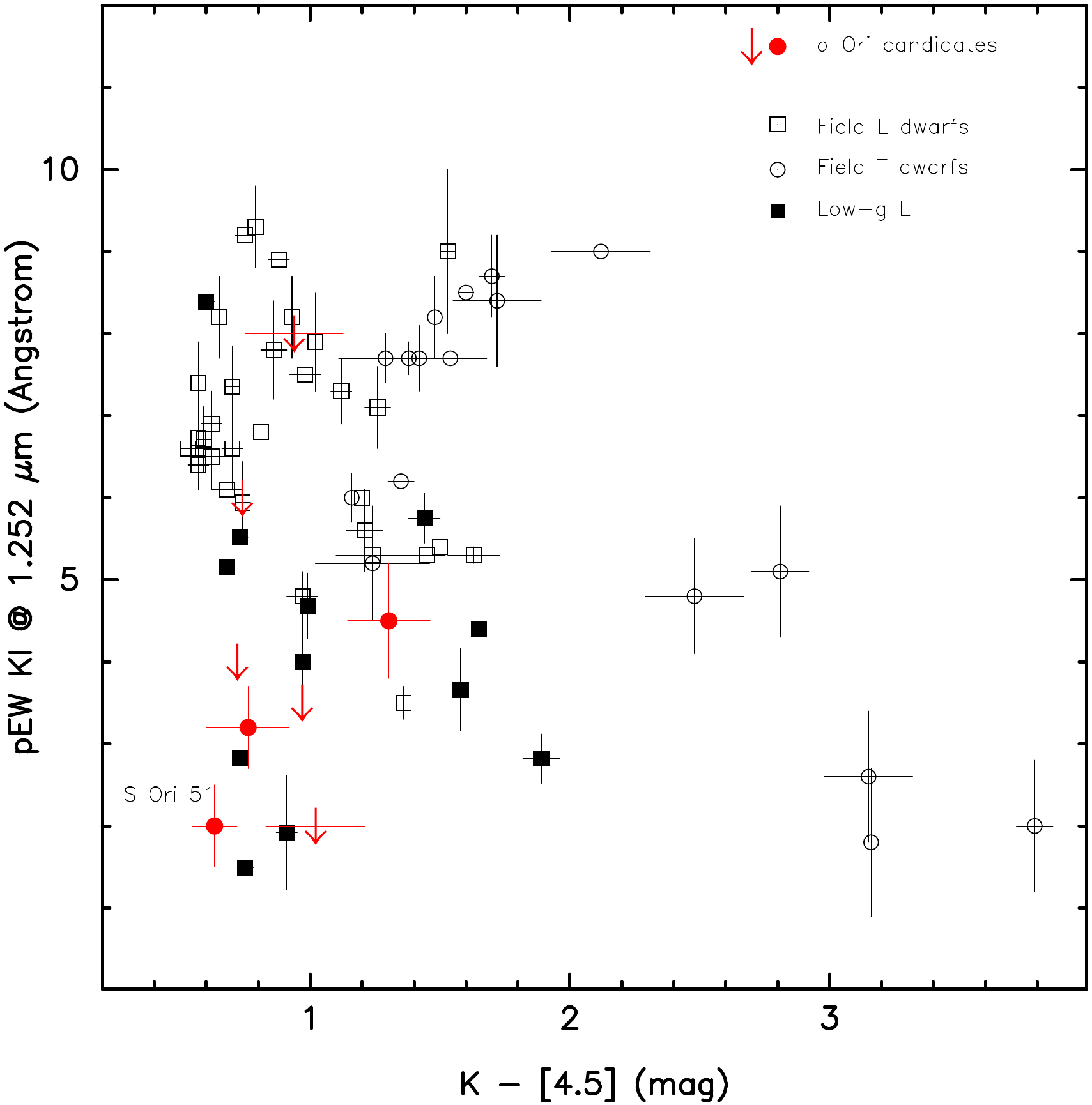}
\caption{Pseudo-equivalent widths (pEW) of the K\,{\sc i} line at 1.252 $\mu$m as a function of 
the $J-K$ (left) and $K-[4.5]$ (right) colors. The data corresponding to \so~candidates are shown 
in red (arrows denote upper limits). The pEWs of field M, L, and T dwarfs from the literature 
\citep{mclean03,allers13,lodieu15} are illustrated in black: open symbols stand for high-gravity 
atmospheres, and solid symbols correspond to dwarfs with reported spectroscopic features of low- 
and intermediate-gravity atmospheres.
\label{pewk}}
\end{figure*}

\begin{figure*}        
\epsscale{1}
\plottwo{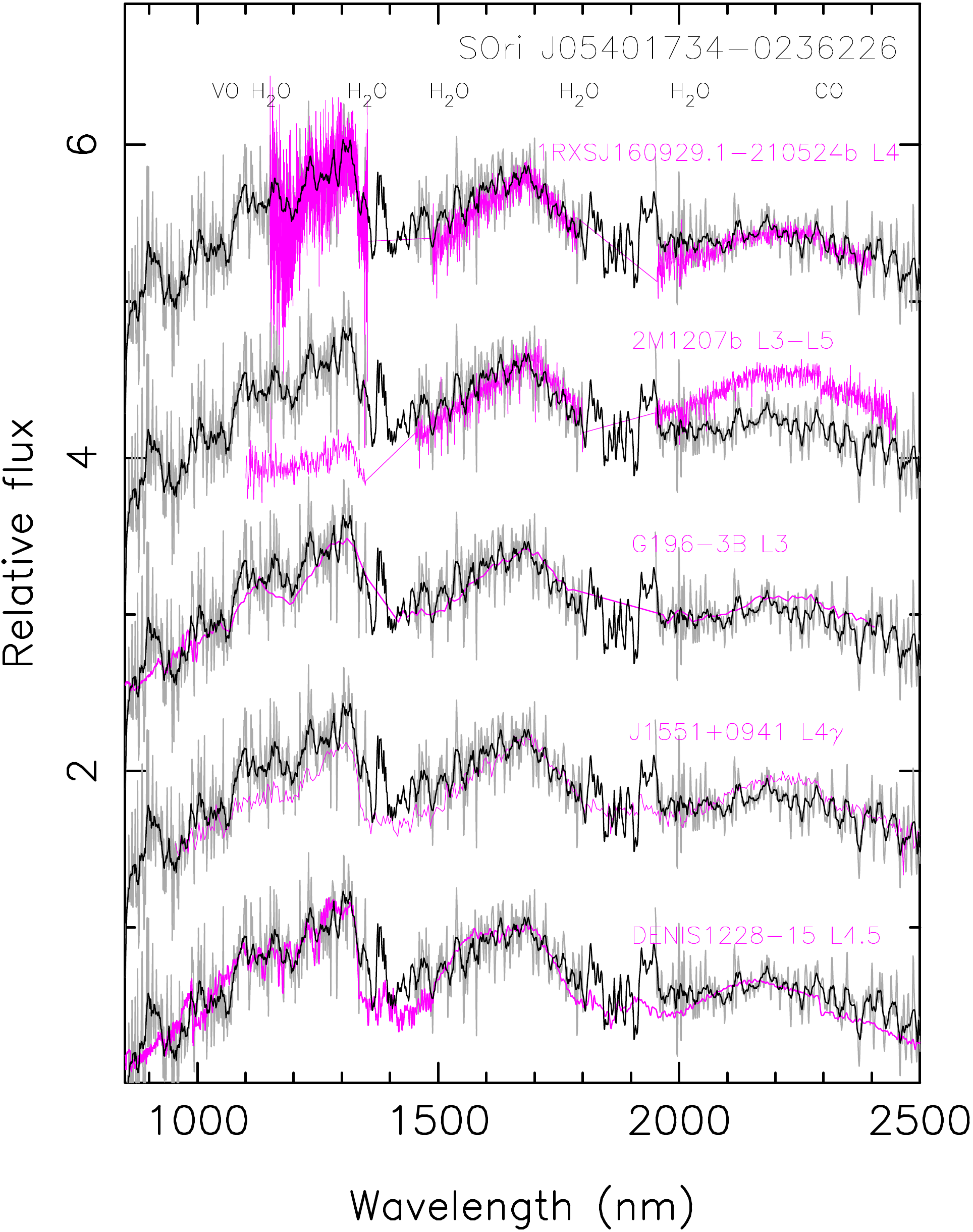}{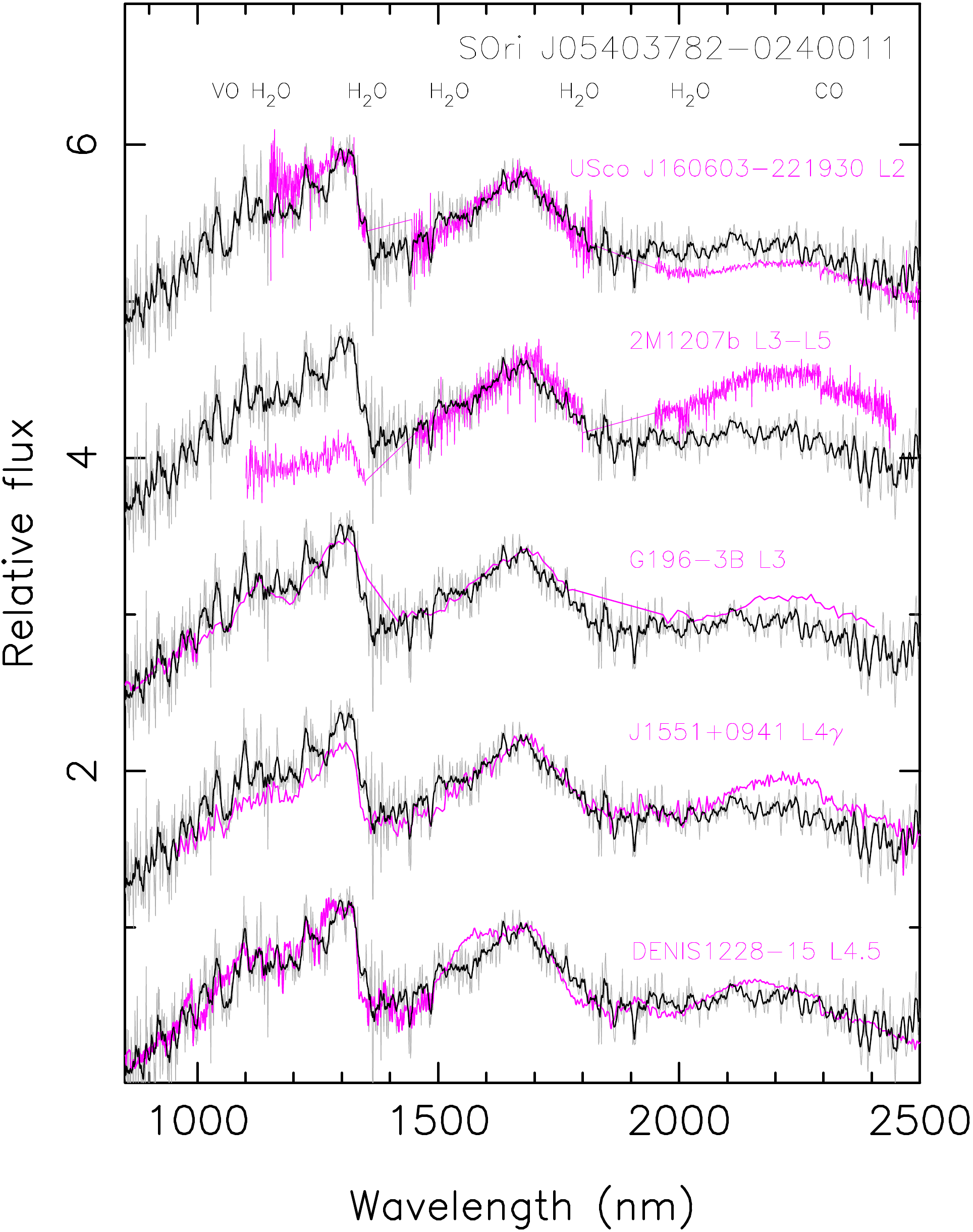}
\caption{The {\sc fire} spectra of J05401734$-$0236226 (left) and J05403782$-$0240011 (right) 
are plotted in gray (original) and black (smoothed by 11 and 9 pixels, respectively). They are 
compared with L2--L5 dwarfs of different ages (magenta lines). The top comparison spectra correspond
to the L4 1RXS\,J160929.1$-$210524b \citep[left panel]{lafreniere10} and the L2 USco 
J160603$-$221930 \citep[right panel]{lodieu07,lodieu08}, both are members of the 5--10-Myr Upper 
Scorpius association. In the middle of both panels, there are three redder than average young dwarfs: 
2M1207b \citep{chauvin04,chauvin05a}, an L3-L5 member of the $\sim$10-Myr TWA association (the
illustrated spectrum was published by \citealt{patience12}), G\,196$-$3\,B \citep{rebolo98}, an 
L3 source with a likely age of 20--85 Myr (plotted spectrum from \citealt{bihain10}), and 
2MASS J15515237$+$0941148 (very low gravity L4 according to \citealt{allers13}). The bottom 
comparison spectra belong to the field L4.5 dwarf DENIS\,J122815.2$-$154733 \citep{leggett01}, 
with a likely age of Gyr. The spectra are normalized to the peak of the $H$-band and are offset 
vertically. Some molecular features are indicated.
\label{fire2}}
\end{figure*}

\begin{figure}        
\epsscale{0.6}
\plotone{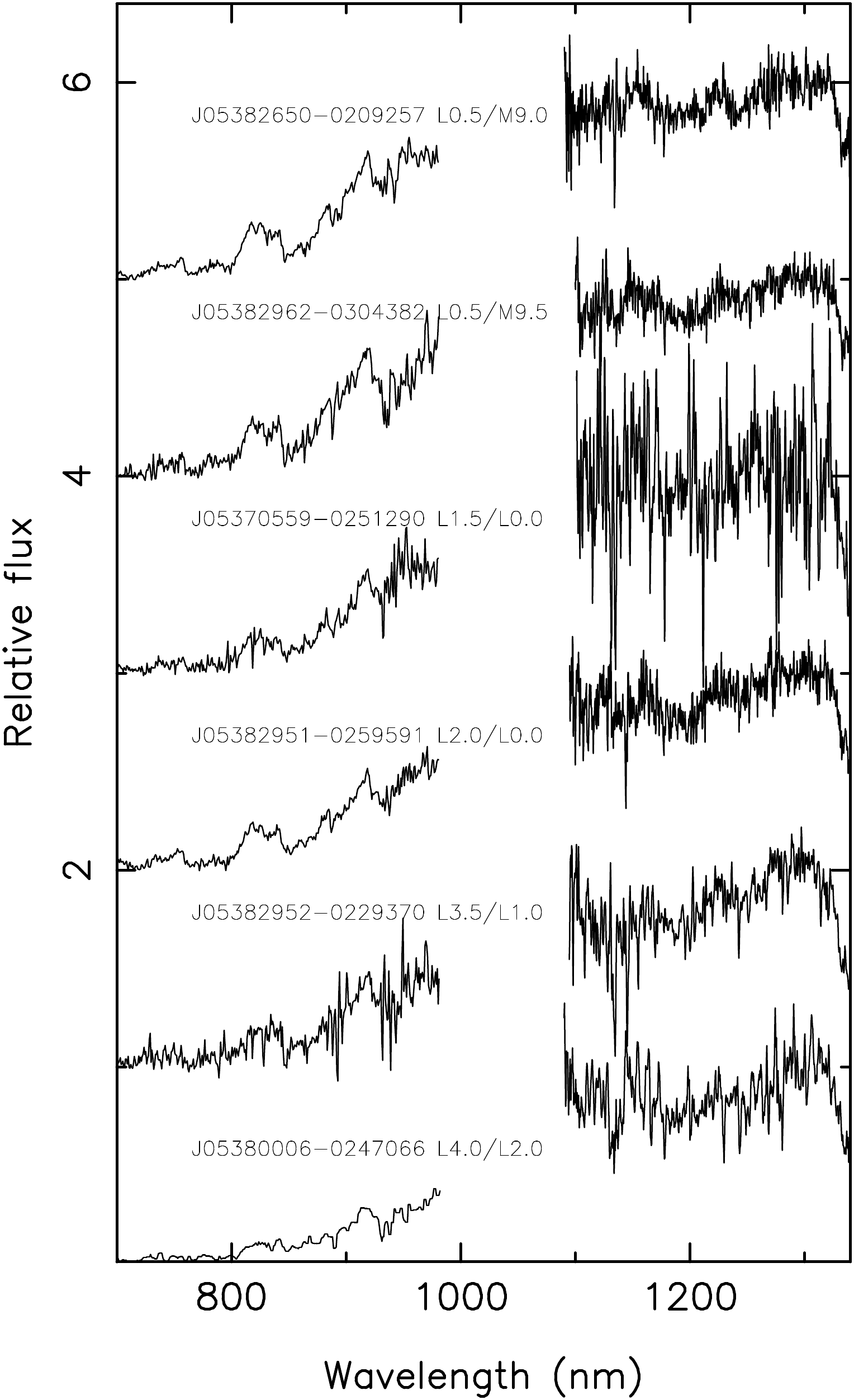}
\caption{{\sc osiris} and {\sc isaac} spectra are brought together for six \so~candidates (black lines).
The data are normalized to unity at 1.28--1.32 $\mu$m, and an integer offset is added to separate the 
spectra vertically. Spectral types are indicated (high-gravity-based/low-gravity-based).
\label{optnir}}
\end{figure}

\begin{figure}        
\epsscale{0.65}
\plotone{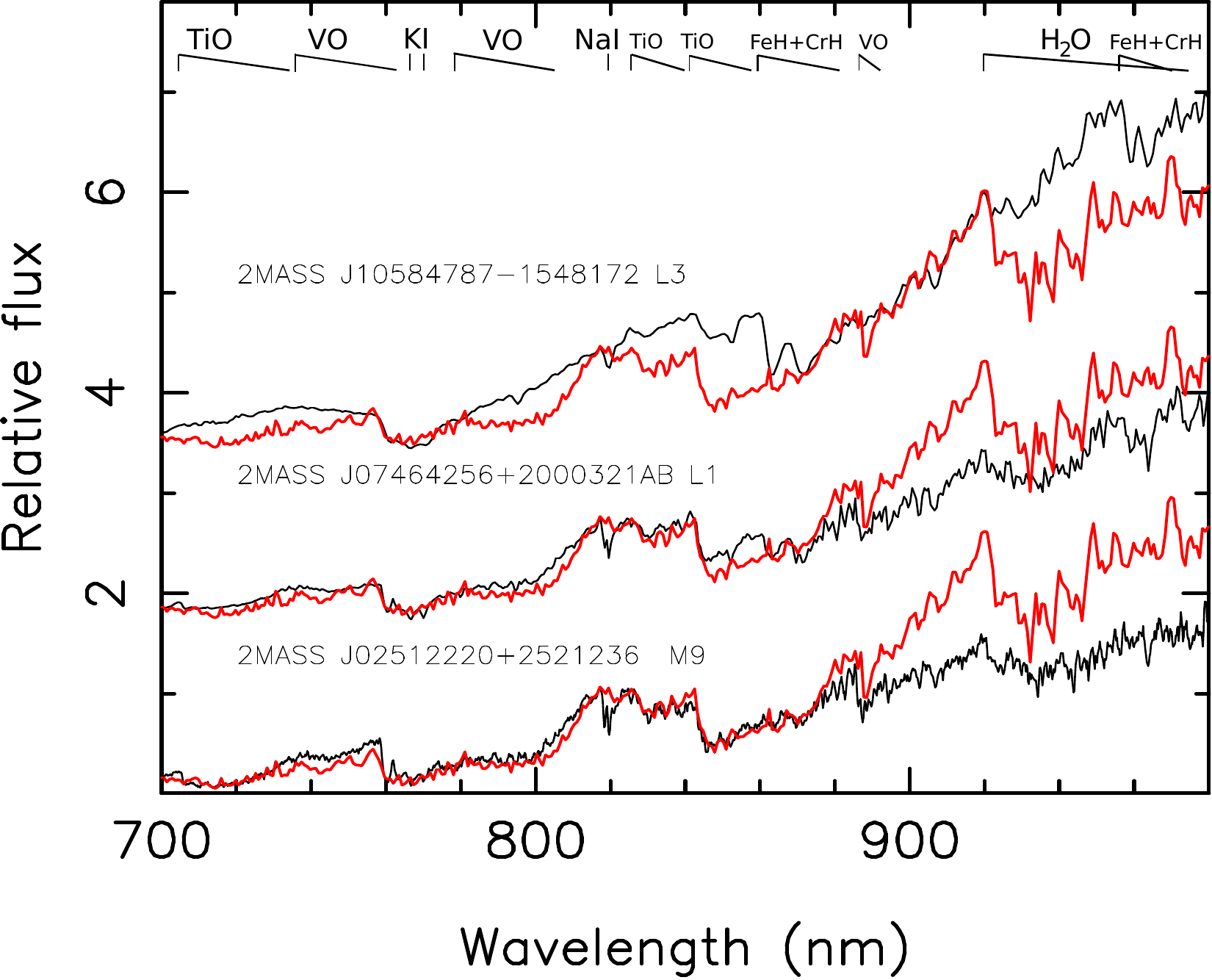}
\plotone{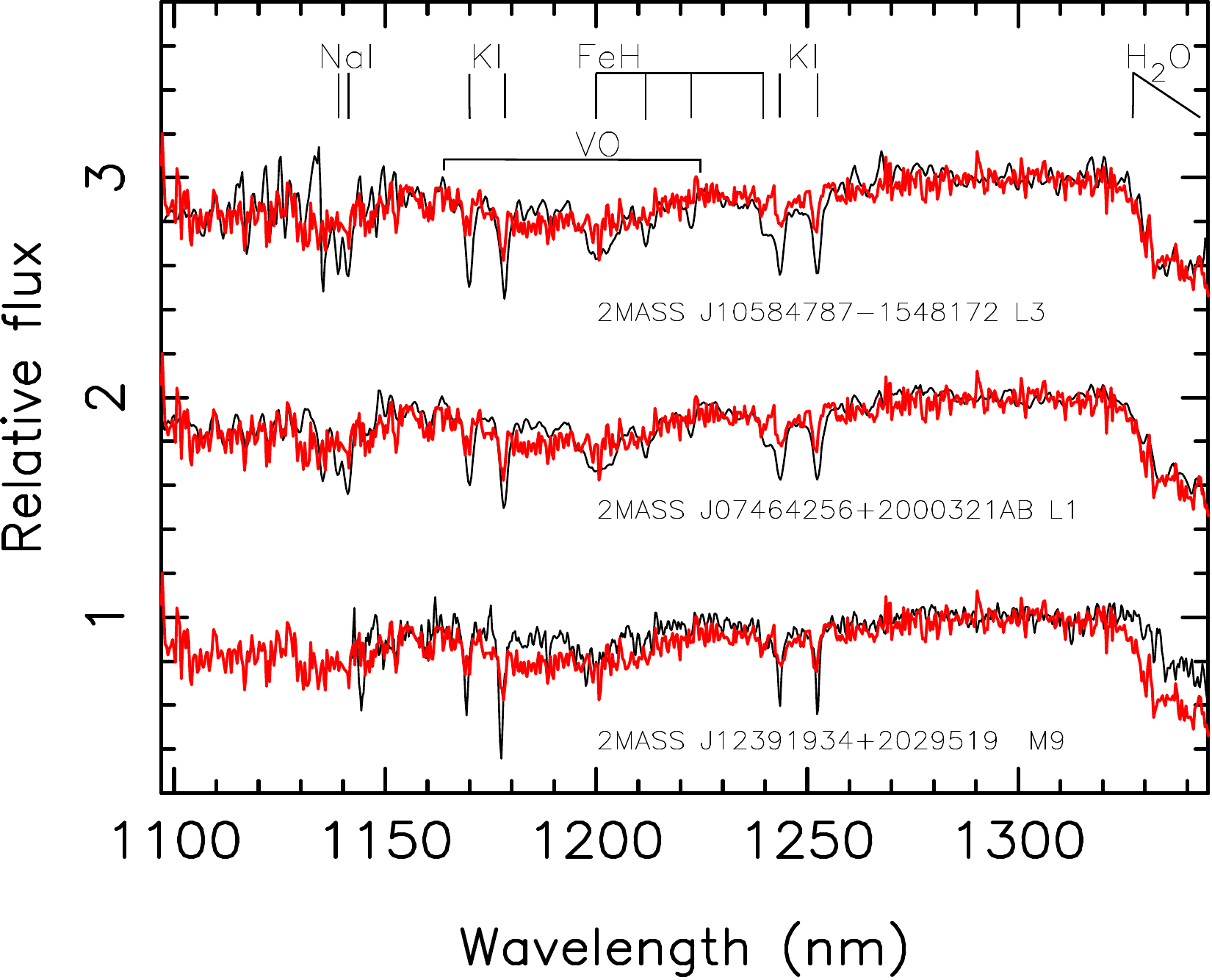}
\caption{The combination of all {\sc osiris} (top panel) and {\sc isaac} (bottom panel) spectra of
\so~members is shown in red. The field high-gravity spectral standards are plotted as black 
lines; in the top panel there are an M9 (2MASS\,J02512220$+$2521236, \citealt{kirk99}), an L1 
(2MASS\,J07464256$+$2000321AB, spectrum from \citealt{reid01}) and an L3 (2MASS\,J10584787$-$1548172, 
{\sc osiris} spectrum), and the bottom panel shows the {\sc isaac} data of an L1 
(2MASS\,J10170754$+$1308398) and the same L3 as for the optical, and the M9 dwarf 2MASS\,J12391934$+$2029519, 
spectrum taken from \citealt{mclean03,mclean07}). All spectra have the same wavelength resolution
for a proper comparison ($R \approx 250$ and $\approx 500$ for the optical and near-infrared). The 
most prominent features are labeled.  Optical and near-infrared spectra are normalized to unity at 
814--817.5 nm and 1.28--1.32 $\mu$m. The data are vertically shifted by a constant.
\label{combspec}}
\end{figure}

\begin{figure}        
\epsscale{1}
\plotone{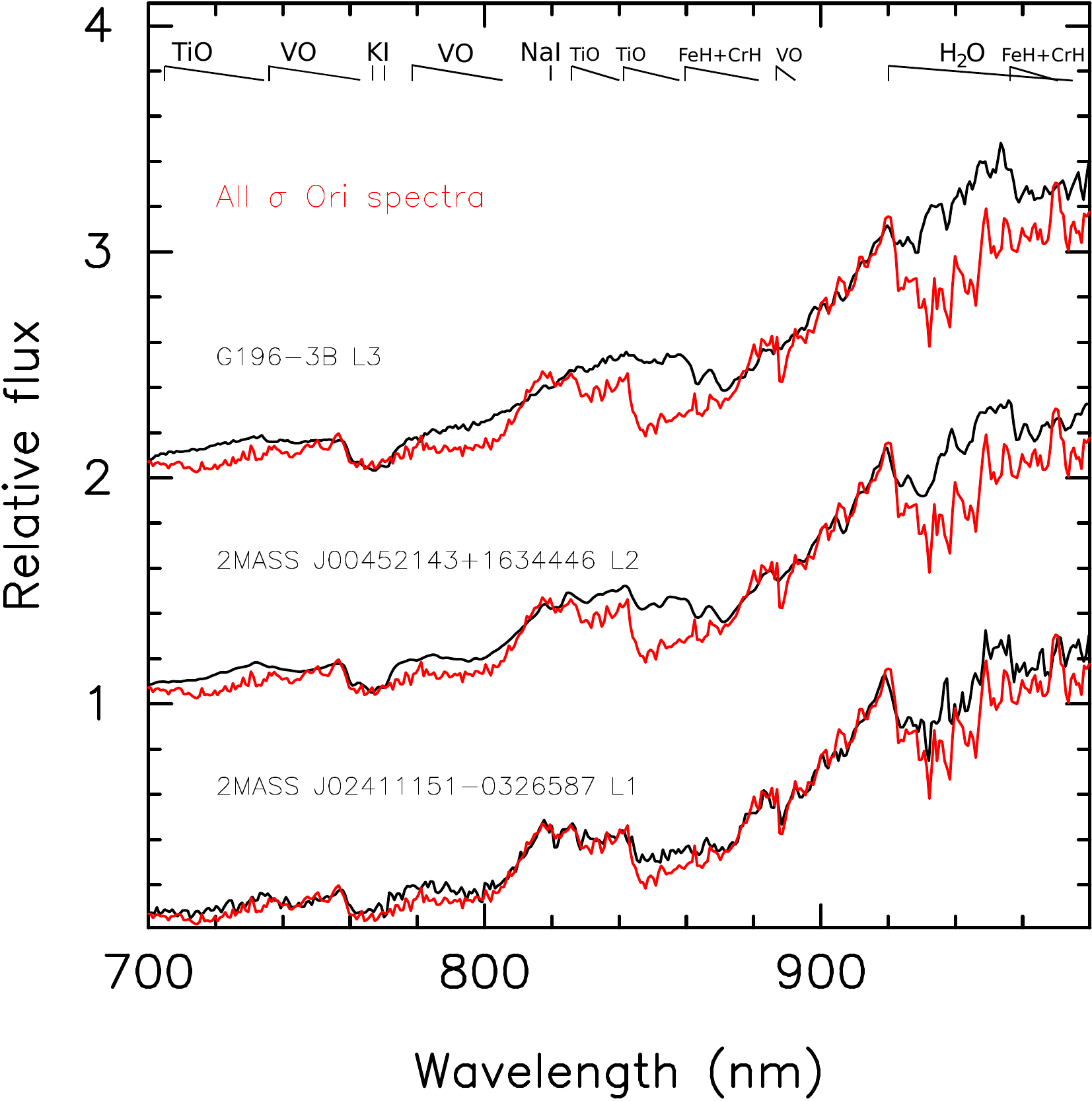}
\caption{The combination of all {\sc osiris} spectra of \so~members is shown in red. In black,
there are three field dwarfs (2MASS J02411151$-$0326587, L1; 2MASS J00452143$+$1634446, L2; and
G\,196$-$3B, L3; spectra from \citealt{osorio14a}) with  very low-gravity-score (VL-G) atmospheres 
assigned by \citet{allers13} and probable ages in the interval 10--500 Myr \citep{osorio14a,gagne14b}, 
i.e., older than the \so~cluster. All the spectra were acquired with the same instrumental setup and were
processed in the same manner. The most prominent features are labeled. The spectra are normalized 
to unity at $\approx$915 nm and are vertically shifted by a constant.
\label{allopt_with_young}}
\end{figure}

\begin{figure}        
\epsscale{0.5}
\plotone{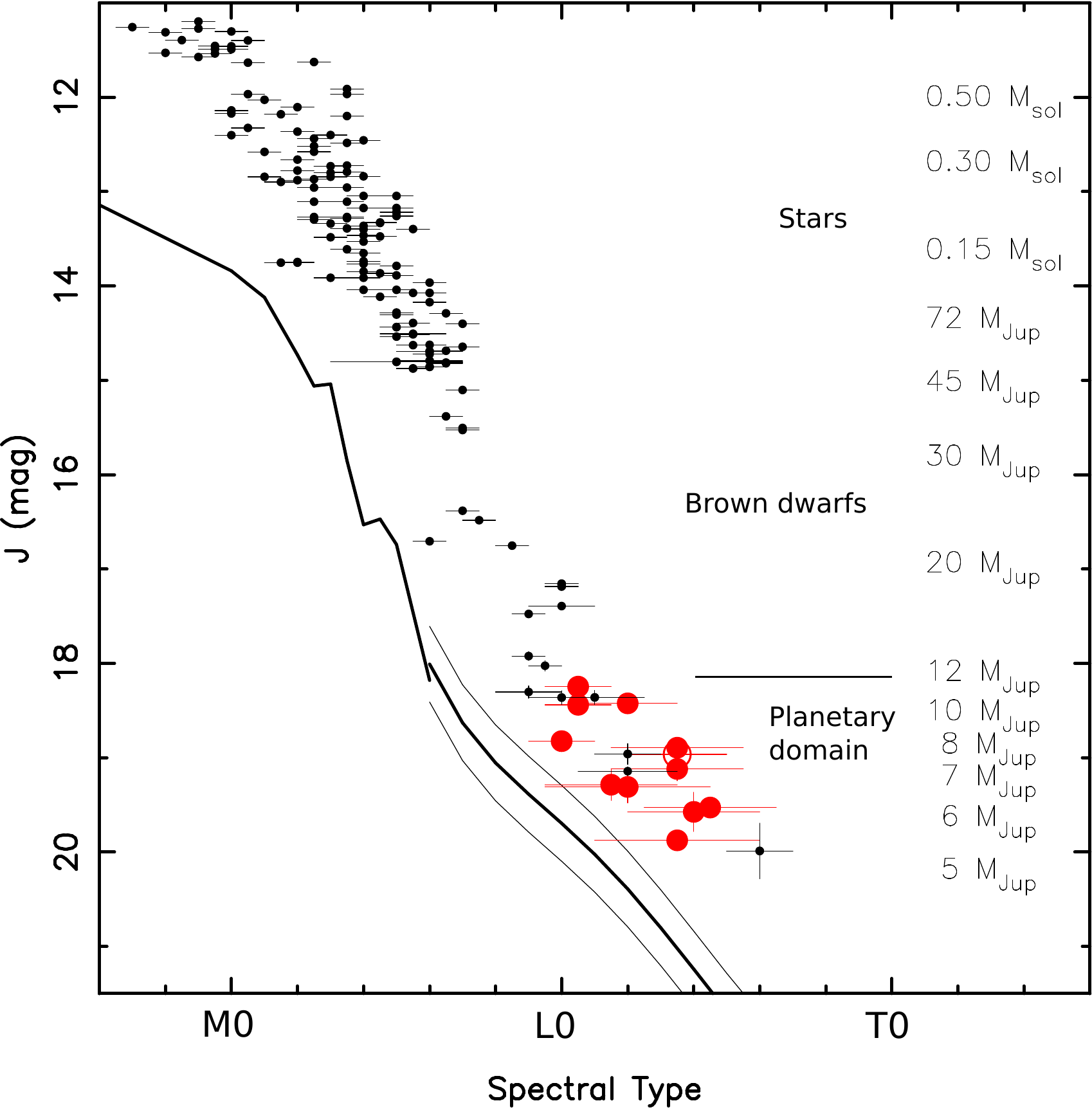}
\plotone{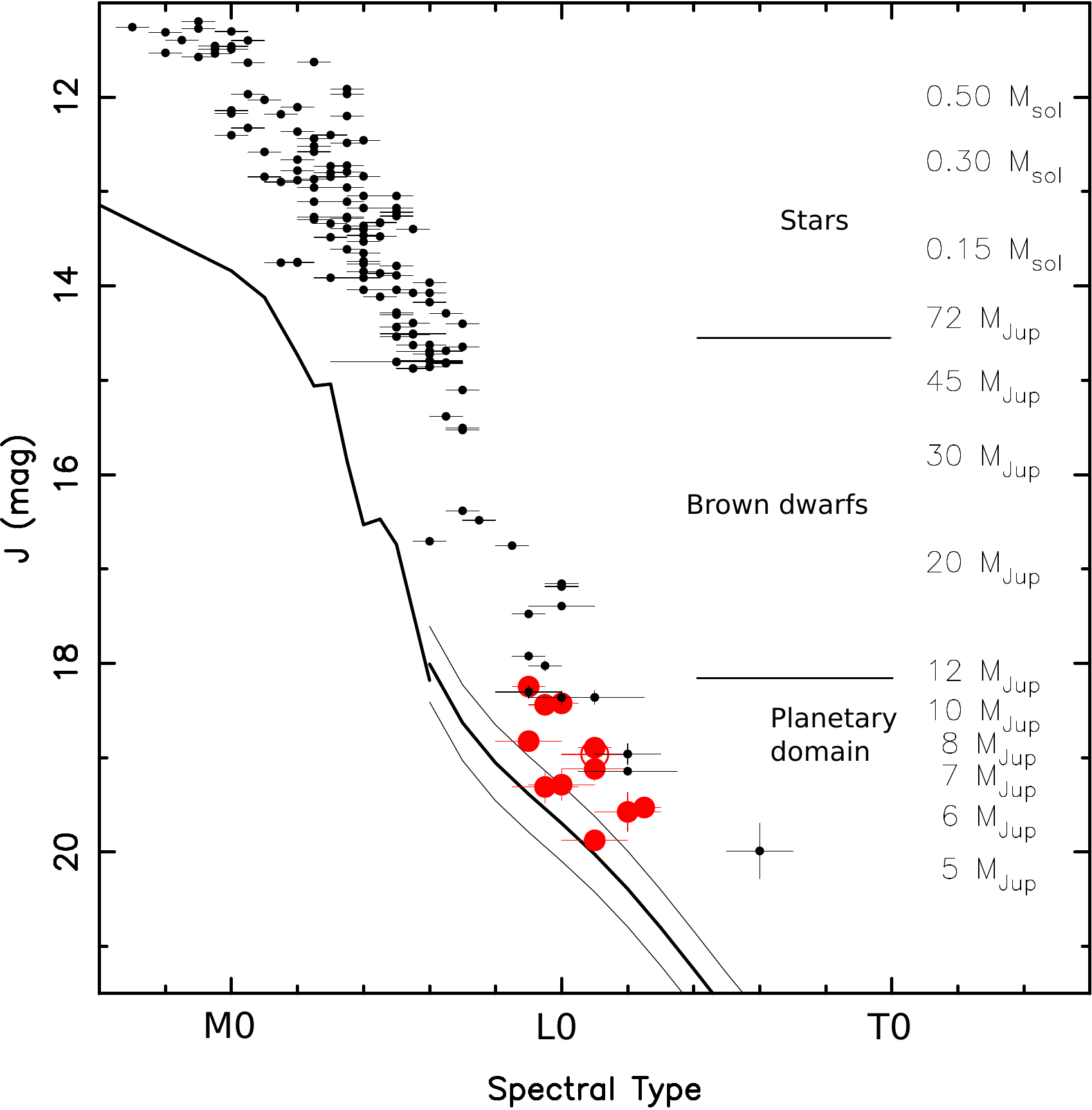}
\caption{Our spectroscopic targets (red symbols) along with other \so~candidates (black dots) 
from \citet{martin01}, \citet{barrado01}, \citet{sacco07}, and \citet{pena12} are shown in the 
$J$-band versus spectral type diagram. The top panel shows high-gravity-based spectral types
and the bottom panel shows the ``youth''-based spectral types derived for the program objects 
(red). The targets of this paper with confirmed spectroscopic signatures of youth are plotted as 
red filled circles, and J05385751$-$0229055 (see Section~\ref{membership}) is displayed with a red
open circle. The masses for solar metallicity, and the age and distance of \so~are labeled 
(Jupiter units for the substellar regime, and solar units for the stellar mass domain). The average 
location of the field sequence taken to the distance of the cluster is indicated by the thick solid
line: the thin solid lines represent the typical dispersion of the high-gravity field, as
discussed by \citet{faherty16}. As expected for their young age, \so~members are over-luminous as compared to the field.
\label{jspt}}
\end{figure}

\begin{figure}        
\epsscale{0.5}
\plotone{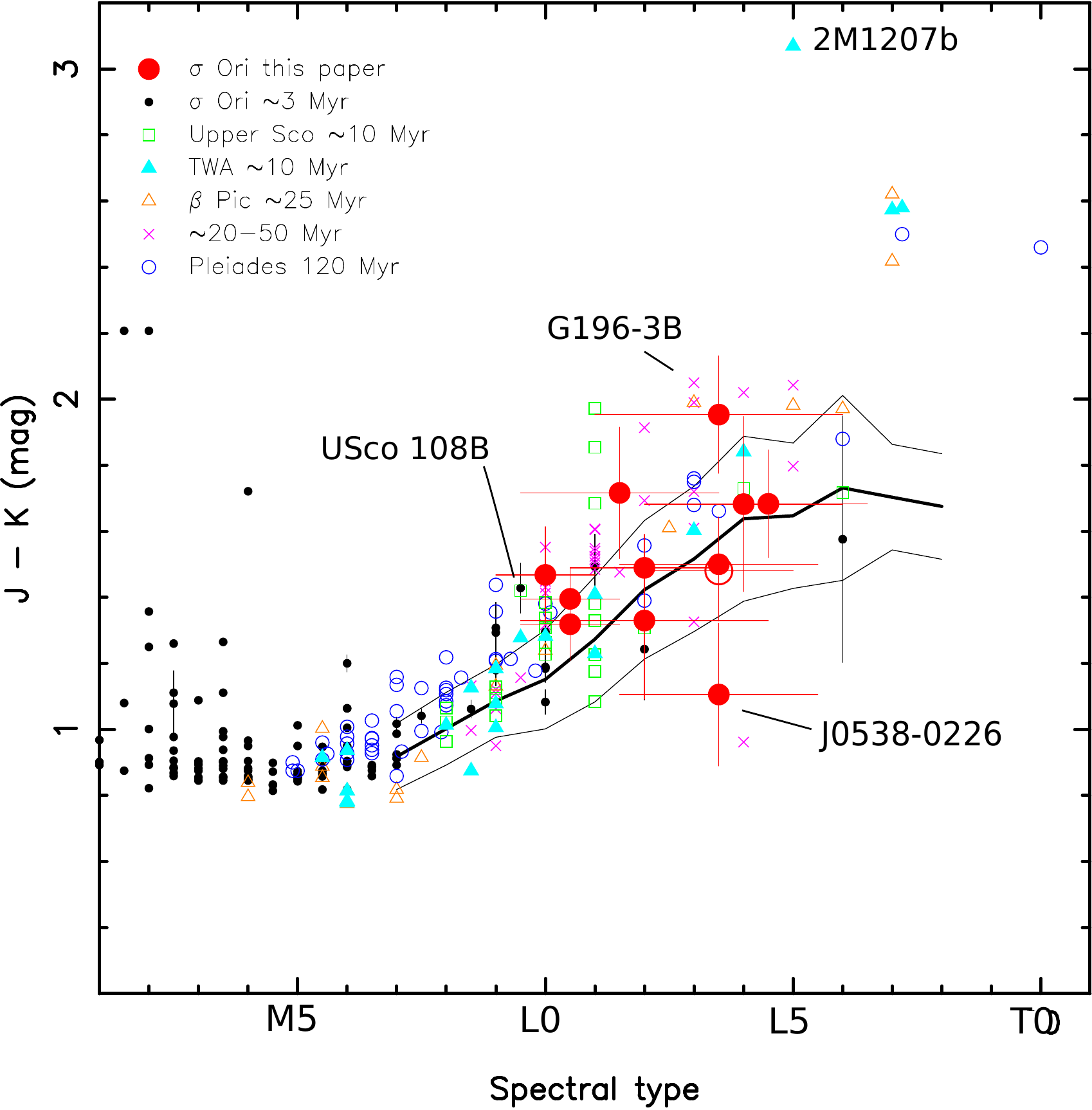}
\plotone{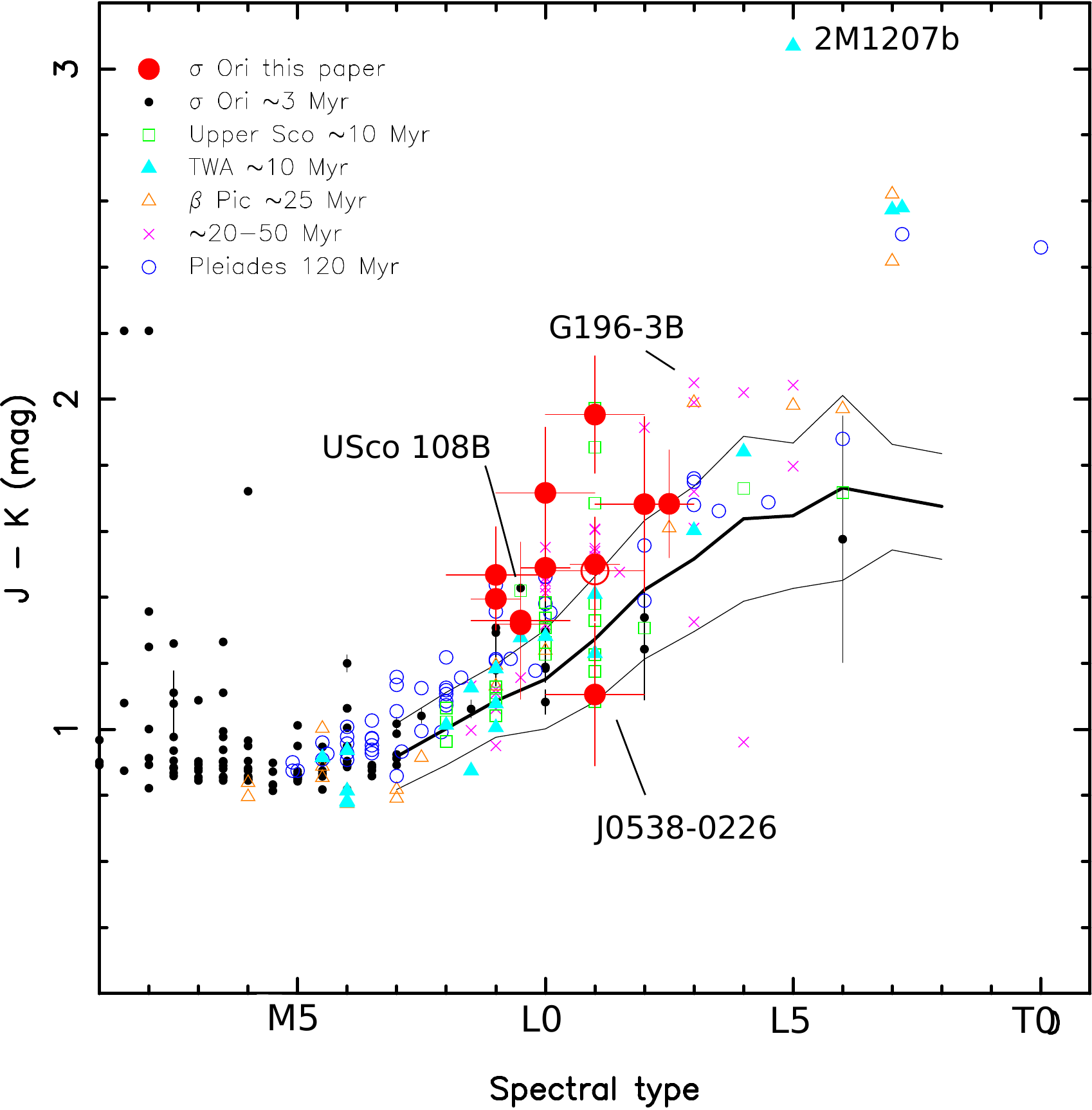}
\caption{The $J-K$ color as a function of spectral type for late-M and L dwarfs of different ages
(see legend). Our \so~targets are plotted as red symbols with their associated error bars: the 
top and bottom panels show the high-gravity- and ``youth''-based spectral types, respectively. For clarity, data 
corresponding to other regions come from the literature and do not have error bars. The average color of 
high-gravity field objects is shown by the thick solid line, and the 
color dispersion of the field is illustrated by the thin lines following \citet{faherty16}. Some objects
cited in the text are labeled.
\label{jkspt}}
\end{figure}

\begin{figure}        
\epsscale{1}
\plotone{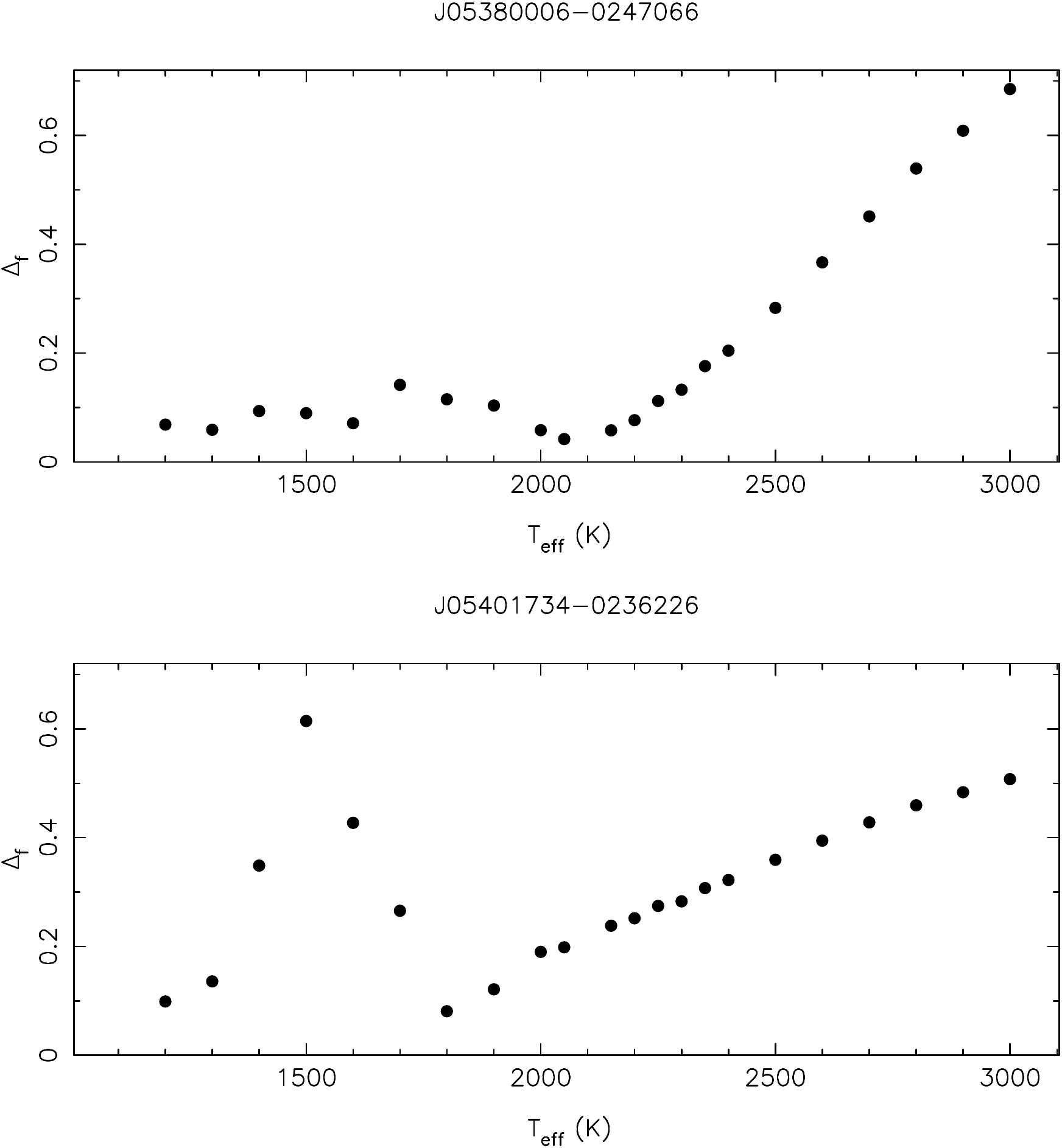}
\caption{The merit function of two \so~targets for determining the best-fit $T_{\rm eff}$ using the
BTSettl, log\,$g$ = 4.0 (cm\,s$^{-2}$) model atmospheres.
\label{goodness}}
\end{figure}

\begin{figure}        
\includegraphics[angle=-90,scale=0.85]{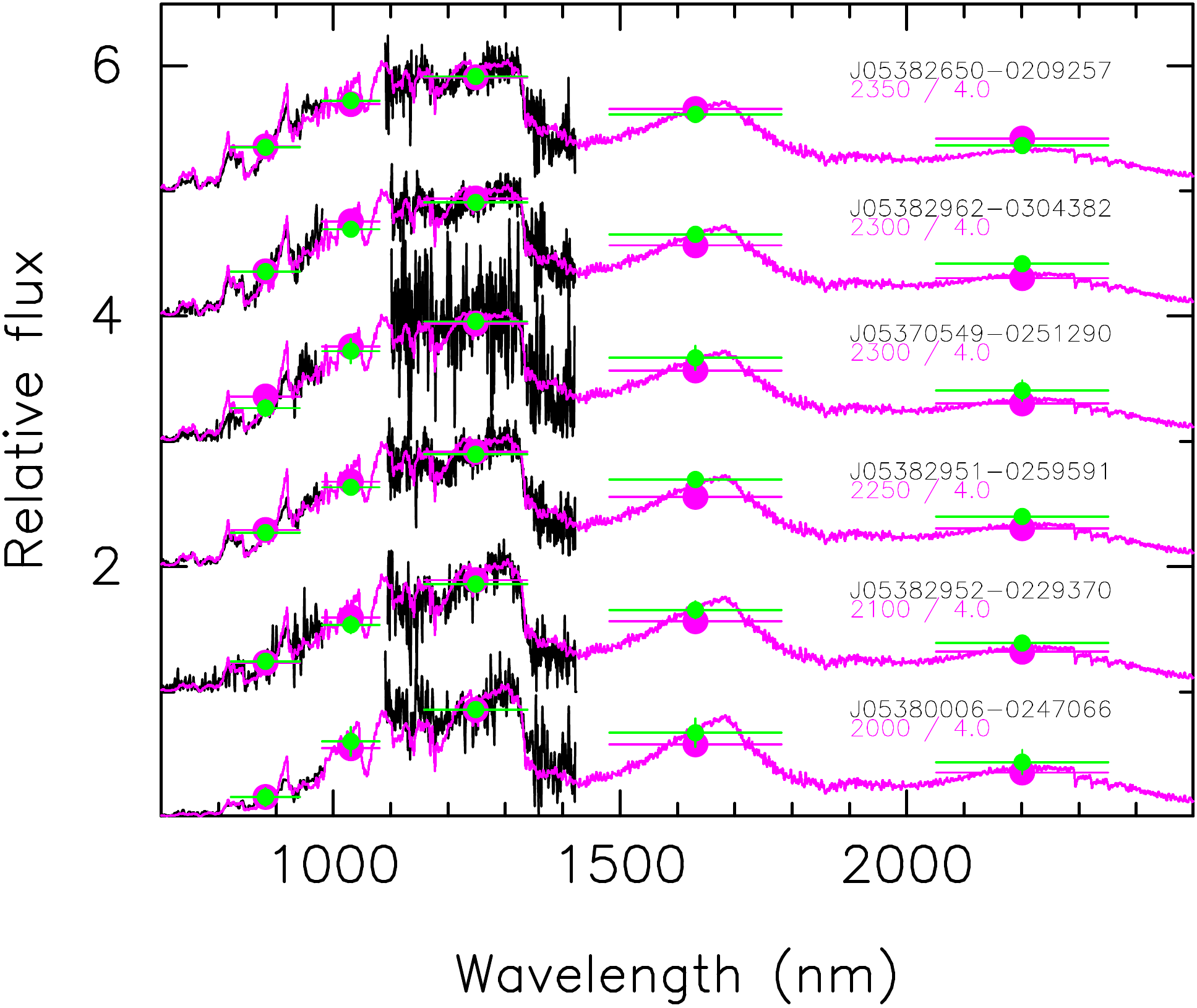}
\caption{Best-fit BT-Settl synthetic spectra and photometry using the {\sc vista} filters 
(magenta lines and dots) are plotted together with the observed optical ({\sc osiris}) and near-infrared
({\sc isaac}) spectra (black lines) and the {\sc vista} $ZYJHK$ photometry (green dots). All models are 
computed for solar metallicity and log\,$g$ = 4.0 (cm\,s$^{-2}$). Spectra and photometry are normalized 
to unity in the wavelength range 1.28--1.32 $\mu$m and offset vertically by a constant. The horizontal error bars of the 
photometry stand for the filter bandwidths. The temperatures are given in K.
\label{models1}}
\end{figure}

\begin{figure}        
\epsscale{1}
\plotone{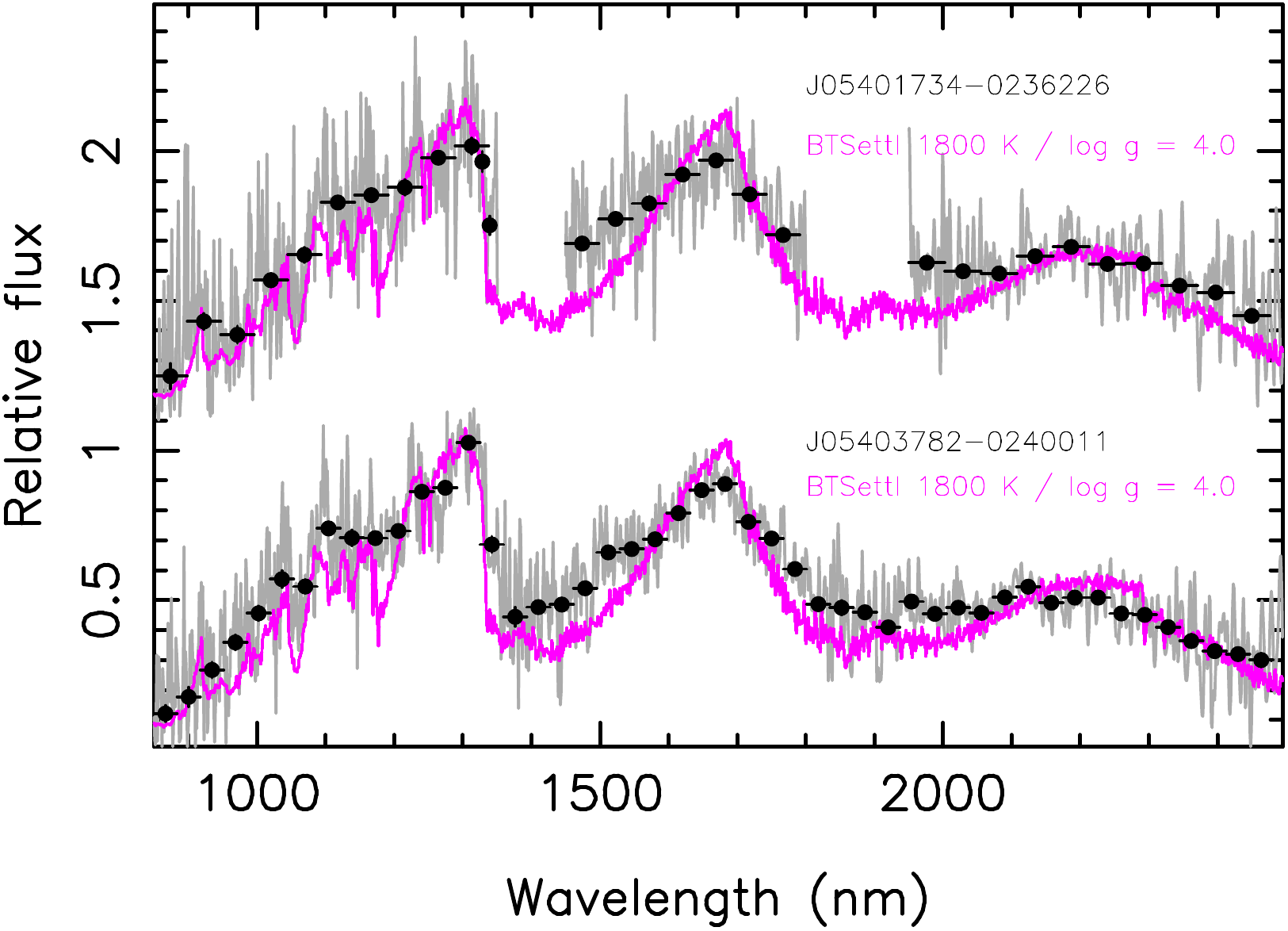}
\caption{Best-fit BT-Settl synthetic spectra (magenta lines) are plotted together with the observed
({\sc fire}) spectra (gray lines). The re-binned data are shown with black dots as in
Figure~\ref{fire}. All models are computed for solar metallicity and log\,$g$ = 4.0 (cm\,s$^{-2}$). 
Spectra are normalized to unity at 1.28--1.32 $\mu$m and offset vertically by a constant.
\label{models2}}
\end{figure}

With the low-resolution and modest S/N of our data, there are no obvious features in the 
{\sc osiris} spectrum of S\,Ori\,70 (Figure~\ref{opt}), except for a markedly increasing slope
from blue to red wavelengths typical of the T classification. This agrees with the determinations
of T5.5$\pm$1.0 and T6--T7 of \citet{osorio02b} and \citet{burgasser04}, respectively, based on
ground-based near-infrared spectra. More recently, \citet{pena15} employed high S/N data at $J$- 
and $H$-band wavelengths collected with the Wide Field Camera 3 of the Hubble Space Telescope 
(HST) to classify S\,Ori\,70. These authors found that the HST near-infrared spectrum better 
resembles T7$^{+0.5}_{-1.0}$ type and noted that the $K$-band spectrum of \citet{osorio02b} is 
better matched by earlier types, which agrees with the measured red $J-K$ color. No obvious 
signatures detectable at the low resolution of the HST spectrum ($R \approx 50$) have 
yet confirmed the 
low gravity atmosphere of S\,Ori\,70. 

The {\sl  Spitzer} colors of S\,Ori\,70 are also redder than those of the high-gravity spectral 
counterparts \citep{luhman08,scholz08,osorio08}. Because of the different pressure and density 
conditions at which H$_2$, CH$_4$, and CO absorptions are produced, low-gravity ultra-cool 
T-type objects tend to be brighter at $K$ and redder in all near-infrared and $[3.6]-[4.5]$ colors 
than comparable high-gravity objects (see discussions and figures in 
\citealt{knapp04,patten06,leggett07,liebert07}). As an example, \citet{naud14} reported on the 
discovery of GU\,Psc~b, a T3.5 dwarf, with a relatively strong $K$-band emission, that orbits 
 the 110--130-Myr AB Doradus moving group star GU\,Psc (70--130 Myr). And \citet{gagne15b} 
announced a T5.5 member of the same moving group with a mass of about 10--12 \mjup~and a red $J-K$ 
color for its spectral type. Other ``red'' T dwarfs have been reported in the literature (e.g., 
\citealt{luhman07b,goldman10,best15}), all of which probably have intermediate ages. Very recently, 
\citet{chiang15} published the finding of two methane dwarf candidates in the $\rho$~Oph 
star-forming region whose colors resemble those of S\,Ori\,70; and \citet{macintosh15} announced the
finding of the $\sim$20-Myr T-dwarf 51\,Eri\,b, which has a red $H-L_p$ color. This reddish nature
might provide a hint for the case of S\,Ori\,70, whose optical spectrum we explored to search for 
features that might reveal a low gravity atmosphere. 

\begin{figure}        
\epsscale{0.6}
\plotone{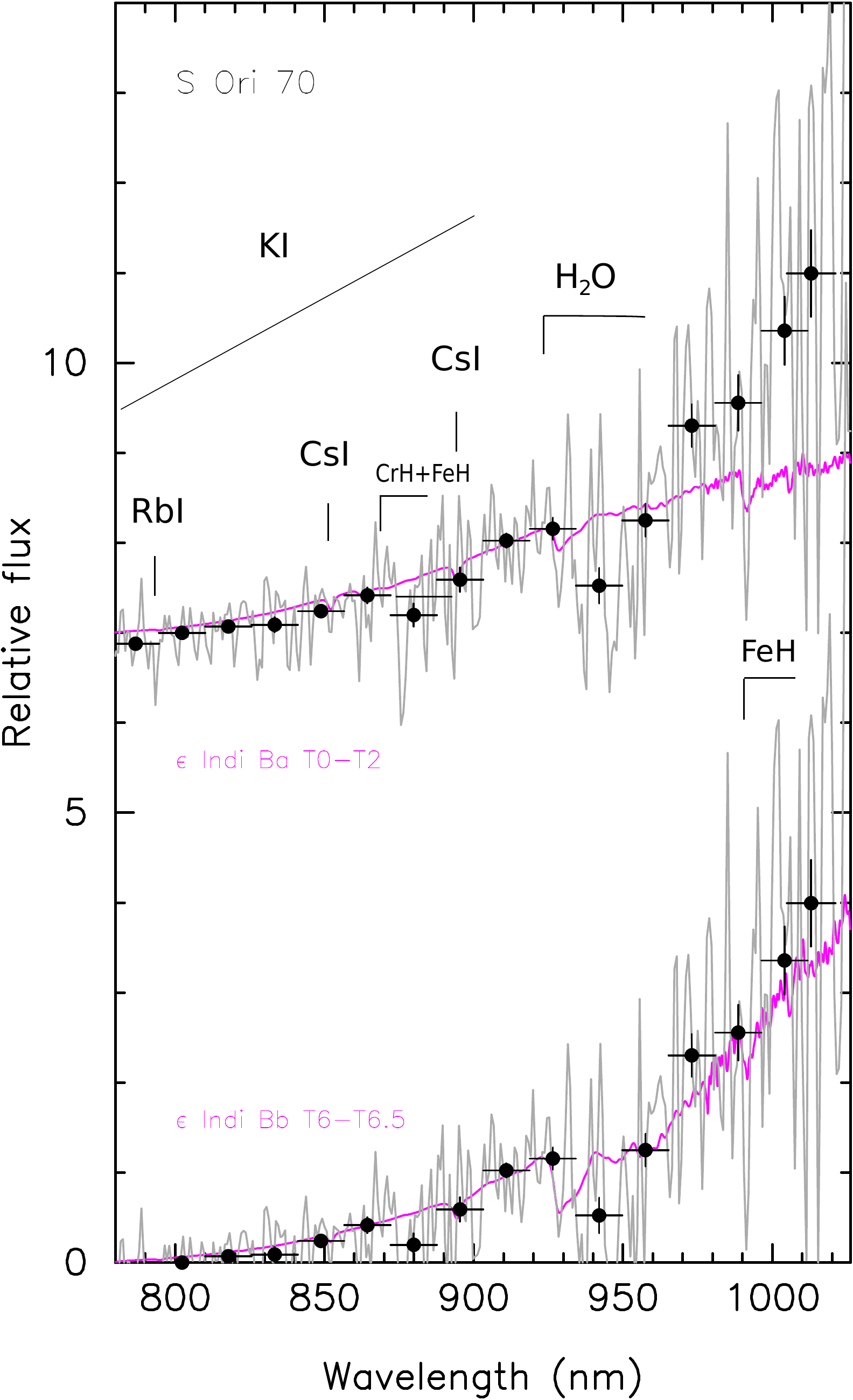}
\caption{The optical spectrum of S\,Ori\,70 (gray line and black dots as in Figure~\ref{opt}) is 
compared with the spectra (magenta) of $\epsilon$ Indi Ba (T0--T2, top) and $\epsilon$ Indi Bb (T6--T6.5, bottom) 
of \citet{king10}. The comparison data, normalized to unity at around 920 nm, were degraded 
to the resolution of S\,Ori\,70. They are shifted by a constant for clarity. Some molecular and atomic features are indicated. 
\label{sori70}}
\end{figure}

We compared the optical spectrum of S\,Ori\,70 with the low-resolution data for field T dwarf
spectral standards \citep{kirk10} with observed spectra available from the SpeX Prism 
Library\footnote{http://pono.ucsd.edu/~adam/browndwarfs/spexprism/}. The reference data were 
obtained with a resolution of $R$ = 75--200 and span roughly 0.65--2.55 $\mu$m. Using Equation~\ref{eq} 
and the observed T dwarf spectra instead of theoretical models, we searched for the field standard 
that minimized the $\Delta_f$ quantity in the wavelength interval 840--980 nm, where both the
{\sc osiris} and SpeX prism spectra have better quality. The best match was found to be T6
with an associated uncertainty of 1 subtype. The ``color-e'' index defined by \citet{burgasser03},
which measures the K\,{\sc i} spectral slope between 919 and 845 nm for late-L and T dwarfs,
also yields a T6 classification for S\,Ori\,70 (``color-e''\,=\,5.6). The visual comparison of
S\,Ori\,70's optical spectrum with other T dwarfs is provided in Figure~\ref{sori70}. The high
quality data of $\epsilon$\,Indi Ba (T0--T2) and Bb (T6--T6.5) published by \citet{king10} were 
used after a
suitable convolution with a Gaussian function for the spectral degradation to the resolution 
of $R$ = 200, which is close to that of the {\sc osiris} spectrum. Because the binary 
$\epsilon$\,Indi Bab \citep{mccaughrean04} is a companion to the nearby, well-known K4.5V star 
$\epsilon$\,Indi \citep{scholz03}, the system has been deeply studied in the literature 
(\citealt[and references therein]{king10}); the two T dwarfs can be used as templates. The lack of 
lithium absorption at 670.82 nm in their atmospheres provides solid evidence of their old ages and 
therefore high gravity nature.

From Figure~\ref{sori70}, it becomes apparent that the red slope of the optical spectrum of 
S\,Ori\,70 better matches that of T6--T7 high-gravity field dwarfs, which agrees with the
near-infrared classification. Water vapor absorption at $\lambda$935 nm appears prominent in the 
spectrum of S\,Ori\,70, even at the very low resolution of our data (we note that the spectrally 
degraded data of $\epsilon$\,Indi Ba and Bb also show strong water vapor features; see 
Figure~\ref{sori70}). As depicted in Figure~6 of \citet{burgasser03}, H$_2$O at 935 nm increases 
in strength with decreasing spectral type,  the strongest feature occurring at around T8. This supports the 
optical spectral classification of S\,Ori\,70 based on the red optical pseudo-continuum. At wavelengths 
blueward of $\sim$800 nm, there are practically no photon counts registered in the {\sc osiris} 
spectrum, thus preventing us from studying the blue wing of K\,{\sc i} and determining whether this signature
behaves differently in S\,Ori\,70. We determined an upper limit of pEW $\approx$ 20\,\AA~on the 
intensity of the Cs\,{\sc i} absorptions at 852.1 and 894.3 nm. This is not a very restrictive 
measurement since the great majority of field T dwarfs have pEWs ranging from 4 to 
9~\AA~(Cs\,{\sc i}), as indicated by \citet{burgasser03}. In summary, we found that the spectral 
classification of S\,Ori\,70 at optical and near-infrared wavelengths agrees to within one sutype. 
However, higher spectral resolution and better S/N data are required for a detailed quantitative 
analysis of its individual spectroscopic features for a proper assessment of its surface gravity. 

\clearpage



\clearpage


\begin{deluxetable}{lcllcccrc}
\tabletypesize{\scriptsize}
\tablecaption{Journal of spectroscopic observations\label{obslog}}
\tablewidth{0pt}
\renewcommand{\arraystretch}{1.02}
\setlength{\tabcolsep}{2.4pt}
\tablehead{
\colhead{Object} & \colhead{$J$\tablenotemark{a}} & \colhead{Obs$.$ date} & \colhead{Inst./Teles.} &
\colhead{Grating} & \colhead{Slit width} & \colhead{$\Delta \lambda$} &
\colhead{$t_{\rm exp}$} & \colhead{Air mass} \\
\colhead{ } & \colhead{(mag)} & \colhead{(UT)} & \colhead{ } &
\colhead{ } & \colhead{(\arcsec)} & \colhead{($\mu$m)} &
\colhead{(s)} & \colhead{} 
}
\startdata
 S\,Ori\,J05382650$-$0209257 & 18.25 &  2012 Dec 03 & {\sc isaac}/{\sc vlt}  & LR    & 1.0 & 1.097--1.339 &  8$\times$600  & 1.10--1.08 \\
                             &       &  2012 Dec 14 & {\sc osiris}/{\sc gtc} & R300R & 1.2 & 0.700--0.980 &  4$\times$1800 & 1.66--1.26 \\
 S\,Ori\,J05382962$-$0304382 & 18.40 &  2012 Dec 04 & {\sc isaac}/{\sc vlt}  & LR    & 1.0 & 1.097--1.339 &  8$\times$600  & 1.11--1.32 \\
                             &       &  2014 Nov 25 & {\sc osiris}/{\sc gtc} & R300R & 1.0 & 0.700--0.980 &  2$\times$1800 & 1.26--1.35 \\
 S\,Ori\,J05382951$-$0259591 & 18.42 &  2012 Dec 01 & {\sc isaac}/{\sc vlt}  & LR    & 1.0 & 1.097--1.339 & 10$\times$600  & 1.08--1.18 \\
                             &       &  2012 Dec 13 & {\sc osiris}/{\sc gtc} & R300R & 0.8 & 0.700--0.980 &  4$\times$1200 & 1.53--1.28 \\
 S\,Ori\,J05382471$-$0300283 & 18.64 &  2012 Dec 02 & {\sc isaac}/{\sc vlt}  & LR    & 1.0 & 1.097--1.339 &  8$\times$600  & 1.11--1.42\\
 S\,Ori\,J05382952$-$0229370 & 18.80 &  2013 Jan 26 & {\sc isaac}/{\sc vlt}  & LR    & 1.0 & 1.097--1.339 & 12$\times$600  & 1.37--2.60\\
                             &       &  2012 Dec 15 & {\sc osiris}/{\sc gtc} & R300R & 1.0 & 0.700--0.980 &  3$\times$1200 & 1.43--1.68 \\
 S\,Ori\,J05385751$-$0229055 & 19.04 &  2012 Dec 01 & {\sc isaac}/{\sc vlt}  & LR    & 1.0 & 1.097--1.339 &  8$\times$600  & 1.16--1.45\\
 S\,Ori\,J05370549$-$0251290 & 19.11 &  2013 Jan 27 & {\sc isaac}/{\sc vlt}  & LR    & 1.0 & 1.097--1.339 &  8$\times$600  & 1.54--2.62\\
                             &       &  2012 Dec 16 & {\sc osiris}/{\sc gtc} & R300R & 1.0 & 0.700--0.980 &  4$\times$1200 & 1.20--1.18 \\
 S\,Ori\,J05380323$-$0226568 & 19.26 &  2013 Jan 26 & {\sc isaac}/{\sc vlt}  & LR    & 1.0 & 1.097--1.339 & 16$\times$600  & 1.10--1.22\\
 S\,Ori\,J05400004$-$0240331 & 19.31 &  2012 Dec 15 & {\sc osiris}/{\sc gtc} & R300R & 1.0 & 0.700--0.980 &  4$\times$1800 & 1.17--1.24 \\
 S\,Ori\,J05380006$-$0247066 & 19.45 &  2012 Dec 02 & {\sc isaac}/{\sc vlt}  & LR    & 1.0 & 1.097--1.339 & 14$\times$600  & 1.14--1.09\\
                             &       &  2012 Dec 16 & {\sc osiris}/{\sc gtc} & R300R & 1.2 & 0.700--0.980 &  4$\times$1800 & 1.23--1.57 \\
                             &       &  2014 Nov 26 & {\sc osiris}/{\sc gtc} & R300R & 1.2 & 0.700--0.980 &  4$\times$1800 & 1.18--1.25 \\
 S\,Ori\,J05403782$-$0240011 & 19.53 &  2014 Dec 01 & {\sc fire}/Magellan    & LR    & 0.6 & 0.820--2.400 &  8$\times$180,8$\times$300 & 1.12--1.16\\
 S\,Ori\,70 & 19.83 &  2012 Dec 14 & {\sc osiris}/{\sc gtc} & R300R & 0.8 & 0.750--0.980 &  8$\times$1800 & 1.18--1.78 \\
 S\,Ori\,J05401734$-$0236226 & 19.88 & 2014 Dec 01 & {\sc fire}/Magellan    & LR    & 0.6 & 0.820--2.400 &  14$\times$300 & 1.22--1.55\\
\hline
2MASS\,J10170754$+$1308398\tablenotemark{b} & 14.10 & 2012 Dec 04 & {\sc isaac}/{\sc vlt}  & LR    & 1.0 & 1.097--1.339 &  2$\times$600  & 1.40--1.44 \\
2MASS\,J10584787$-$1548172\tablenotemark{c} & 14.16 & 2012 Dec 02 & {\sc isaac}/{\sc vlt}  & LR    & 1.0 & 1.097--1.339 &  2$\times$300  & 1.17--1.19 \\
                                            &       & 2012 Dec 15 & {\sc osiris}/{\sc gtc} & R300R & 0.8 & 0.700--0.980 &  2$\times$300  & 1.41--1.35 \\
2MASS\,J04234858$-$0414035\tablenotemark{d} & 14.47 & 2012 Dec 01 & {\sc isaac}/{\sc vlt}  & LR    & 1.0 & 1.097--1.339 &  2$\times$200  & 1.06--1.07 \\
\enddata
\tablenotetext{a}{{\sc vista} photometry given by \citet{pena12}.}
\tablenotetext{b}{Field L1 dwarf.}
\tablenotetext{c}{Field L3 dwarf.}
\tablenotetext{d}{Field L7.5 dwarf.}
\end{deluxetable}

\begin{deluxetable}{lccccr}
\tabletypesize{\scriptsize}
\tablecaption{Other spectral measurements. \label{pew}}
\tablewidth{0pt}
\renewcommand{\arraystretch}{1.02}
\setlength{\tabcolsep}{2.4pt}
\tablehead{
\colhead{Object} & \colhead{pEW\tablenotemark{a}}   & \colhead{$g$-class\tablenotemark{b}} & \colhead{$T_{\rm eff}$} & \colhead{log $L$/L$_\odot$} & \colhead{Mass} \\
\colhead{}       & \colhead{(\AA)}                  & \colhead{} & \colhead{(K)}          & \colhead{}                 & \colhead{(M$_{\rm Jup}$)}
}
\startdata
 S\,Ori\,51                  &   2.0$\pm$0.5\tablenotemark{c} & \nodata & \nodata & $-$2.60$\pm$0.16 &  20.8$\pm$3.7 \\
%
 S\,Ori\,J05382650$-$0209257 &   $\le$2.0      &  INT-G   & 2350$\pm$100 &  $-$3.02$\pm$0.17 &   12.9$\pm$2.7  \\
 S\,Ori\,J05382962$-$0304382 &   $\le$4.7      &  VL-G    & 2300$\pm$100 &  $-$3.05$\pm$0.17 &   12.4$\pm$2.6  \\
 S\,Ori\,J05382951$-$0259591 &    3.2$\pm$0.5  &  VL-G    & 2250$\pm$100 &  $-$3.07$\pm$0.17 &   12.2$\pm$2.6  \\
 S\,Ori\,J05382471$-$0300283 &   $\le$4.0      &  VL-G    & \nodata      &  $-$3.17$\pm$0.18 &   10.8$\pm$2.4  \\
 S\,Ori\,J05382952$-$0229370 &    4.5$\pm$0.7  &  INT-G   & 2100$\pm$100 &  $-$3.21$\pm$0.18 &   10.2$\pm$2.3  \\
 S\,Ori\,J05385751$-$0229055 &   $\le$8.0      &  VL-G    & \nodata      &  $-$3.27$\pm$0.18 &    9.5$\pm$2.2  \\
 S\,Ori\,J05370549$-$0251290 &   $\le$4.0      &  VL-G    & 2300$\pm$100 &  $-$3.36$\pm$0.19 &    8.6$\pm$2.0  \\
 S\,Ori\,J05380323$-$0226568 &   $\le$3.5      &  VL-G    & \nodata      &  $-$3.39$\pm$0.19 &    8.2$\pm$2.0  \\
 S\,Ori\,J05400004$-$0240331 &   \nodata       &  \nodata & \nodata      &  $-$3.37$\pm$0.19 &    8.4$\pm$2.0  \\
 S\,Ori\,J05380006$-$0247066 &   $\le$6.0      &  INT-G   & 2000$\pm$100 &  $-$3.40$\pm$0.19 &    8.1$\pm$2.0  \\
 S\,Ori\,J05403782$-$0240011 &   $\le$3.5      &  INT-G   & 1800$\pm$100 &  $-$3.47$\pm$0.20 &    7.4$\pm$2.0  \\
 S\,Ori\,J05401734$-$0236226 &   $\le$4.5      &  VL-G    & 1800$\pm$100 &  $-$3.55$\pm$0.21 &    6.7$\pm$1.8  \\
\enddata
\tablecomments{Temperatures are derived from spectral fitting of log\,$g$ = 4.0 (cm\,s$^{-2}$) and solar metallicity BTSettl model atmospheres.}
\tablenotetext{a}{Corresponding to K\,{\sc i} $\lambda$1.252 micron.}
\tablenotetext{b}{Gravity class as defined by \citet{allers13}.}
\tablenotetext{c}{Measure obtained from the spectrum published by \citet{mcgovern04}.}
\end{deluxetable}

\begin{deluxetable}{lcccccccccc}
\tabletypesize{\scriptsize}
\tablecaption{Spectral indices and spectral types. \label{spectypes}}
\tablewidth{0pt}
\renewcommand{\arraystretch}{1.0}
\setlength{\tabcolsep}{2.4pt}
\tablehead{
\colhead{Object} & \colhead{$Z-J$\tablenotemark{a}} & \colhead{$J-K$\tablenotemark{a}} & \colhead{$K-[4.5]$\tablenotemark{a}} & \colhead{PC3} & \colhead{SpT} & \colhead{$S1$} & \colhead{$S2$} & \colhead{$S3$} & \colhead{SpT ``youth''} & \colhead{SpT ``high $g$''} \\
\colhead{}       & \colhead{(mag)}                  & \colhead{(mag)}                  & \colhead{(mag)}                      & \colhead{}    & \colhead{opt} & \colhead{}     & \colhead{}     & \colhead{}     & \colhead{optNIR}        & \colhead{optNIR} 
}
\startdata
 S\,Ori\,51                  &  2.24$\pm$0.03 &  1.11$\pm$0.03 & 0.63$\pm$0.09 & \nodata &  M9.0\tablenotemark{b} &\nodata&\nodata&\nodata& \nodata & L0\tablenotemark{c} \\
%
 S\,Ori\,J05382650$-$0209257 &  2.21$\pm$0.06 &  1.19$\pm$0.07 & 1.02$\pm$0.19 & 2.74    &  L1.5    &  4.61 &  2.29 &  1.19 &  M9.0\,$\pm$\,0.5 & L0.5\,$\pm$\,1.0 \\
 S\,Ori\,J05382962$-$0304382 &  2.18$\pm$0.06 &  1.36$\pm$0.08 & \nodata       & 2.81    &  L1.5    &  4.74 &  2.19 &  1.23 &  M9.5\,$\pm$\,0.5 & L0.5\,$\pm$\,1.0 \\
 S\,Ori\,J05382951$-$0259591 &  2.46$\pm$0.07 &  1.32$\pm$0.07 & 0.76$\pm$0.16 & 2.32    &  L1.0    &  5.75 &  2.95 &  1.29 &  L0.0\,$\pm$\,0.5 & L2.0\,$\pm$\,1.5 \\
 S\,Ori\,J05382471$-$0300283 &  2.41$\pm$0.08 &  1.26$\pm$0.09 & 0.72$\pm$0.19 & \nodata &  \nodata &\nodata&\nodata&  1.16 &  M9.0\,$\pm$\,1.0 & L0.0\,$\pm$\,1.0 \\
 S\,Ori\,J05382952$-$0229370 &  2.53$\pm$0.10 &  1.34$\pm$0.09 & 1.30$\pm$0.16 & 4.07    &  L2.5    &  6.93 &  3.33 &  1.35 &  L1.0\,$\pm$\,0.5 & L3.5\,$\pm$\,2.0 \\
 S\,Ori\,J05385751$-$0229055 &  2.90$\pm$0.14 &  1.53$\pm$0.10 & 0.94$\pm$0.19 & \nodata &  \nodata &\nodata&\nodata&  1.35 &  L1.0\,$\pm$\,1.0 & L3.5\,$\pm$\,1.5 \\
 S\,Ori\,J05370549$-$0251290 &  2.55$\pm$0.12 &  1.26$\pm$0.13 & \nodata       & 2.77    &  L1.5    &  6.85 &  2.99 &  1.11 &  L0.0\,$\pm$\,1.0 & L1.5\,$\pm$\,2.0 \\
 S\,Ori\,J05380323$-$0226568 &  2.56$\pm$0.13 &  1.38$\pm$0.13 & 0.97$\pm$0.25 & \nodata &  \nodata &\nodata&\nodata&  1.31 &  L1.0\,$\pm$\,1.0 & L3.5\,$\pm$\,2.0 \\
 S\,Ori\,J05400004$-$0240331 &  2.42$\pm$0.12 &  \nodata       & \nodata       & 3.14    &  L2.0    &\nodata&\nodata&\nodata&  \nodata          & \nodata          \\
 S\,Ori\,J05380006$-$0247066 &  2.99$\pm$0.19 &  1.47$\pm$0.15 & 0.74$\pm$0.33 & 2.92    &  L1.5    &  12.0\tablenotemark{d} &  5.27 &  1.33 &  L2.0\,$\pm$\,1.0 & L4.0\,$\pm$\,2.0 \\
 S\,Ori\,J05403782$-$0240011 &  2.76$\pm$0.18 &  1.19$\pm$0.20 & \nodata       & \nodata &  \nodata &\nodata&  5.08 &  1.40 &  L2.5\,$\pm$\,0.5 & L4.5\,$\pm$\,2.0 \\
 S\,Ori\,J05401734$-$0236226 &  3.04$\pm$0.50 &  1.47$\pm$0.21 & \nodata       & \nodata &  \nodata &\nodata&  2.91 &  1.39 &  L1.0\,$\pm$\,1.0 & L3.5\,$\pm$\,2.5 \\
\enddata
\tablenotetext{a}{{\sc vista} and {\sl Spitzer} photometry given by \citet{pena12}.}
\tablenotetext{b}{Spectral type published by \citet{barrado01}.}
\tablenotetext{c}{Spectral type published by \citet{canty13}.}
\tablenotetext{d}{Large uncertainty.}
\end{deluxetable}

\begin{deluxetable}{crrrrccc}
\tabletypesize{\scriptsize}
\tablecaption{Polynomial coefficients of spectral indices. \label{coef}}
\tablewidth{0pt}
\renewcommand{\arraystretch}{1.0}
\setlength{\tabcolsep}{2.4pt}
\tablehead{
\colhead{Index} & \colhead{$c_0$} & \colhead{$c_1$} & \colhead{$c_2$} & \colhead{$c_3$} & rms & Index range & SpT range }
\startdata
$S1$ high-g &   $-$5.56 &    1.45 & \nodata    & \nodata & $\pm$0.7 & 1.69--7.92 & M9--L7 \\
$S2$ high-g &  $-$17.42 &   18.21 &    $-$6.71 &   0.87  & $\pm$0.4 & 1.29--4.35 & M9--L7 \\
$S3$ high-g & $-$815.28 & 2088.18 & $-$1791.97 & 515.21  & $\pm$1.0 & 1.01--1.36 & M9--L7 \\
$S1$ young  &   $-$3.20 &    0.59 & \nodata    & \nodata & $\pm$1.2 & 1.95--8.01 & M7.5--L4 \\
$S2$ young  &   $-$4.30 &    1.38 & \nodata    & \nodata & $\pm$0.8 & 1.36--5.22 & M7.5--L4 \\
$S3$ young  &  $-$14.85 &   11.94 & \nodata    & \nodata & $\pm$0.8 & 1.01--1.48 & M7.5--L4 \\
\enddata
\tablecomments{The rms is given in units of spectral subtype.}
\end{deluxetable}

\begin{deluxetable}{lccccc}
\tabletypesize{\scriptsize}
\tablecaption{UKIDSS and VISTA photometry of our targets. \label{ukidssphot}}
\tablewidth{0pt}
\renewcommand{\arraystretch}{1.02}
\setlength{\tabcolsep}{2.4pt}
\tablehead{
\colhead{Object} & \colhead{VISTA $J$} & \colhead{VISTA $K$} & \colhead{UKIDSS $J$}   & \colhead{UKIDSS $K$} & \colhead{UKIDSS $J-K$} \\
\colhead{}       & \colhead{(mag)}     & \colhead{(mag)}     & \colhead{(mag)}        & \colhead{(mag)}      & \colhead{(mag)}               
}
\startdata
  S\,Ori\,J05382650$-$0209257 &  18.25$\pm$0.03  &  17.06$\pm$0.06  &   18.25$\pm$0.07  &  16.85$\pm$0.06  &  1.40$\pm$0.10  \\ 
  S\,Ori\,J05382962$-$0304382 &  18.40$\pm$0.03  &  17.05$\pm$0.06  &   18.44$\pm$0.07  &  17.12$\pm$0.09  &  1.32$\pm$0.11  \\ 
  S\,Ori\,J05382951$-$0259591 &  18.42$\pm$0.03  &  17.09$\pm$0.06  &   18.42$\pm$0.07  &  16.94$\pm$0.07  &  1.49$\pm$0.10  \\ 
  S\,Ori\,J05382471$-$0300283 &  18.64$\pm$0.03  &  17.39$\pm$0.08  &   18.82$\pm$0.10  &  17.36$\pm$0.11  &  1.47$\pm$0.15  \\ 
  S\,Ori\,J05382952$-$0229370 &  18.80$\pm$0.04  &  17.46$\pm$0.09  &   18.89$\pm$0.11  &  17.39$\pm$0.09  &  1.50$\pm$0.14  \\ 
  S\,Ori\,J05385751$-$0229055 &  19.04$\pm$0.04  &  17.52$\pm$0.09  &   18.96$\pm$0.11  &  17.48$\pm$0.10  &  1.48$\pm$0.15  \\ 
  S\,Ori\,J05370549$-$0251290 &  19.11$\pm$0.04  &  17.84$\pm$0.12  &   19.29$\pm$0.17  &  17.57$\pm$0.11  &  1.72$\pm$0.20  \\ 
  S\,Ori\,J05380323$-$0226568 &  19.26$\pm$0.05  &  17.88$\pm$0.17  &   19.12$\pm$0.13  &  18.01$\pm$0.18  &  1.11$\pm$0.22  \\ 
  S\,Ori\,J05400004$-$0240331 &  19.31$\pm$0.05  &  17.76$\pm$0.11  &   19.31$\pm$0.17  &  17.98$\pm$0.17  &  1.33$\pm$0.24  \\ 
  S\,Ori\,J05380006$-$0247066 &  19.45$\pm$0.05  &  17.98$\pm$0.14  &   19.57$\pm$0.21  &  17.89$\pm$0.16  &  1.68$\pm$0.27  \\ 
  S\,Ori\,J05403782$-$0240011 &  19.53$\pm$0.06  &  18.34$\pm$0.19  &   \nodata         &  17.84$\pm$0.15  &  \nodata        \\ 
  S\,Ori\,J05401734$-$0236226 &  19.88$\pm$0.07  &  18.41$\pm$0.20  &   \nodata         &  17.92$\pm$0.16  &  \nodata        \\ 
\enddata
\end{deluxetable}

\end{document}